%% file: paper.tex
\documentclass{aa-old}           

\usepackage{amsmath}
\usepackage{natbib}
\usepackage{epsfig}
\usepackage{txfonts}

\input{definitions}

\bibpunct{(}{)}{;}{a}{}{,}

\begin{document}

\bibliographystyle{aa}

\title{Multidimensional supernova simulations with
       approximative neutrino transport}
\subtitle{I. Neutron star kicks and the anisotropy of
             neutrino-driven explosions \\
             in two spatial dimensions}

\author{
        L. Scheck     \and
        K. Kifonidis  \and
        H.-Th. Janka  \and
        E. M{\"u}ller
       }
        
\offprints{L. Scheck} 
\mail{scheck@mpa-garching.mpg.de}

\date{received; accepted}      
    
\institute{Max-Planck-Institut f\"ur Astrophysik,
           Karl-Schwarzschild-Stra{\ss}e 1, 
           D-85741 Garching, Germany}

\abstract{ We study hydrodynamic instabilities during the first
           seconds of core collapse supernovae by means of
           two-dimensional (2D) simulations with approximative
           neutrino transport and boundary conditions that
           parameterize the effects of the contracting neutron star
           and allow us to obtain sufficiently strong neutrino heating
           and, hence, neutrino-driven explosions. Confirming more
           idealised studies as well as supernova simulations with
           spectral transport, we find that random seed perturbations
           can grow by hydrodynamic instabilities to a globally
           asymmetric mass distribution in the region between the
           nascent neutron star and the accretion shock, leading to a
           dominance of dipole ($l=1$) and quadrupole ($l=2$) modes in
           the explosion ejecta, provided the onset of the supernova
           explosion is sufficiently slower than the growth time scale
           of the low-mode instability. By gravitational and
           hydrodynamic forces, the anisotropic mass distribution
           causes an acceleration of the nascent neutron star, which
           lasts for several seconds and can propel the neutron star
           to velocities of more than 1000$\,$km$\,$s$^{-1}$.  Because
           the explosion anisotropies develop chaotically and change
           by small differences in the fluid flow, the magnitude of
           the kick varies stochastically. No systematic dependence of
           the average neutron star velocity on the explosion energy
           or the properties of the considered progenitors is found.
           Instead, the anisotropy of the mass ejection and thus the
           kick seems to increase when the nascent neutron star
           contracts more quickly, and thus low-mode instabilities can
           grow more rapidly. Our more than 70 models separate into
           two groups, one with high and the other with low neutron
           star velocities and accelerations after one second of
           post-bounce evolution, depending on whether the $l=1$ mode
           is dominant in the ejecta or not. This leads to a
           bimodality of the distribution when the neutron star
           velocities are extrapolated to their terminal values.
           Establishing a link to the measured distribution of pulsar
           velocities, however, requires a much larger set of
           calculations and ultimately 3D modelling. }

\authorrunning{Scheck et al.}
\titlerunning{Multidimensional supernova simulations}

\maketitle  


\section{Introduction}
\label{sec:introduction}

Spectropolarimetry (\citealt{Leonard+06,Wang+03,Wang+01}, and
references therein) indicates that global anisotropies are a common
feature of many core-collapse supernovae (SNe). Since the asymmetry
inferred from the observed polarisation grows with the depth one can
look into the expanding supernova ejecta, the origin of the anisotropy
seems to be intrinsically linked to the mechanism of the explosion.
Recent high-resolution imaging of SN 1987\,A with the Hubble Space
Telescope \citep{Wang+02} as well as large values for the measured
space velocities of young galactic pulsars may be interpreted as a
support of such a link.  The average pulsar velocities are as high as
200-500 km/s, and some neutron stars (NSs) move through interstellar
space with more than 1000 km/s
\citep[e.g.][]{Cordes+93,LL94,HP97,Zou+05,Chatterjee+05}.  Claims of a
bimodality of the pulsar velocity distribution are still
controversial. While some authors have obtained evidence for such a
bimodality \citep{CC98,Fryer+98,ACC02,Brisken+03}, others found that a
simple Maxwellian fit works best \citep{HP97,Hobbs+05}.

Binary disruption, e.g. as a consequence of the SN explosions which
give birth to the NSs, does not lead to sufficiently high velocities.
Furthermore, the orbital parameters of many binary systems imply an
intrinsic acceleration mechanism of the pulsars, probably linked to
their creation (see \citealt{Lai01,Lai+01} for reviews). Quite a
number of explanations have been suggested, mostly involving
anisotropic mass ejection in the SN explosion or anisotropic neutrino
emission of the cooling, nascent NS. The former suggestion might be
supported by the fact that some pulsars seem to propagate in a
direction opposite to mass distribution asymmetries of their
associated SN remnants.  But a clear observational evidence is
missing, and hydrodynamic simulations have previously either produced
rather small recoil velocities \citep{JM94}, or started from the
assumption that a dipolar asymmetry was already present in the
pre-collapse iron core and thus gave rise to a large anisotropy of the
SN explosion \citep{BH96}. The origin of such big pre-collapse
perturbations, however, is not clear \citep{Murphy+04}. 

Suggestions that a ``neutrino rocket engine'' boosts NSs to large
velocities
(e.g. \citealt{Chugai84,Dorofeev+85,Burrows_Woosley86,Woosley87}; for
reviews see \citealt{Lai01} and \citealt{Lai+01}) make use of the fact
that the huge reservoir of gravitational binding energy released
during the collapse of the stellar core is mostly carried away by the
neutrinos. Creating a global emission anisotropy of these neutrinos of
even only 1\% -- which is sufficient to obtain a NS recoil of about
300$\,$km$\,$s$^{-1}$ --, however, turned out to be very difficult.
Most ideas refer to unknown neutrino properties (e.g.,
\citealt{Fuller+03,Fryer_Kusenko06} and refs.\ therein) and/or require
the presence of a very strong magnetic field with a large dipole
component (instead of being randomly structured and variable with
time) in the newly formed NS
\citep[e.g.][]{Arras_Lai99a,Arras_Lai99b,Socrates+06}.  Such
assumptions are not generally accepted and are not the result of
self-consistent calculations but put into the models ``by hand''.

If the observed high pulsar velocities indeed go back to the early
moments of the SN explosion, the simplest explanation would certainly
be a common origin of explosion asymmetries and pulsar acceleration.
In this case anisotropic ejection of mass would lead to a recoil (or
``kick'') of the NS due to (linear) momentum conservation.  Various
kinds of hydrodynamic instabilities might in fact be responsible for a
large-scale deformation of the ejecta and globally aspherical
explosions. Perturbation analysis of volume-filling thermal convection
in a fluid sphere by \cite{Chandra61}, found largest growth rates for
the $l=1,\,m=0$ mode (in terms of an expansion in spherical harmonics
$Y_l^m$ of order $l,\,m$). This is supported by (full $4 \pi$)
three-dimensional simulations of convection in red giant and
non-rotating main sequence stars
\citep{Woodward+03,Kuhlen+03}. Motivated by \citeauthor{Chandra61}'s
analysis, \cite{Herant95} speculated about the formation of a stable
$l=1,\,m=0$ convective mode in the neutrino-heated layers between the
gain radius and the supernova shock. In this configuration there
exists only a single buoyant bubble (outflow), and a single accretion
funnel (inflow), which reaches from the postshock region down to the
NS. \cite{Herant95} suggested the potential importance of such a
convective pattern for NS kicks up to nearly 1000$\,$km/s.
Instability of the accretion shock to a global Rayleigh-Taylor mode
which could lead to asymmetric shock expansion and a net recoil of the
NS of several 100$\,$km/s was also predicted by
\cite{Thompson00}. However, according to the linear analysis by
\cite{Foglizzo+06}, advection tends to stabilise the growth of
long-wavelength perturbations in the neutrino-heated accretion flow
behind the standing shock. A convective trigger of such instabilities
therefore requires the local growth rate to exceed a critical
threshold value.

Non-radial, low-mode instability of shocked accretion flows can also
be caused by the ``advective-acoustic cycle''. In the astrophysical
context, this instability was first discussed by
\cite{Foglizzo_Tagger00} and \cite{Foglizzo01,Foglizzo02} in
application to Bondi-Hoyle accretion of black holes. It was more
recently considered for supernova-core like conditions by
\cite{Galletti_Foglizzo05}. This instability relies on the fact that
the infall of entropy and vorticity perturbations produces acoustic
waves that propagate outward and create new entropy and vorticity
perturbations when reaching the shock, thus closing an amplifying
feedback cycle which eventually results in a dominant $l=1$ or $l=2$
mode. The advective-acoustic cycle can even operate under conditions,
in which convective instabilities are hampered, e.g. if the advection
of matter out of the convectively unstable region is too fast to allow
for a significant convective growth of small perturbations.

\cite{Blondin+03} investigated numerically an idealised setup for the
stalled shock in a supernova core and showed that a spherical shock is
dynamically unstable to non-radial perturbations, even without
neutrino heating and convection. The authors referred to this as the
``standing accretion shock instability'', or SASI, which reveals a
preferred growth of $l = 1$ mode deformation and was explained by
\cite{Blondin_Mezzacappa06} as a consequence of the propagation of
sound waves in the volume enclosed by the shock.

While these investigations lacked the use of a detailed description of
the neutrino physics and of the equation of state of the supernova
medium, \cite{Scheck+04,Janka+04a,Janka+04b,Janka+05},
\cite{Ohnishi+06}, \cite{Buras+06b}, and \cite{Burrows+06}
provided results which demonstrate that the instability of the
accretion shock also occurs in models which include the relevant
microphysics with more realism.  \cite{Scheck+04} suggested a link of
these low-mode instabilities of the supernova shock during the
neutrino-heating phase to global explosion asymmetries (see in
particular \citealt{Kifonidis+06}) and pulsar kicks.  Most previous
two-dimensional (2D) simulations of successful neutrino-driven
explosions \citep{HBC92,HBFC94,BF93,BHF95} failed to see the
development of $l=1,\,2$ modes (such an anisotropy, however, showed up
in one of the weakly exploding models of \citealt{JM96}) because most
of the simulations were done with limited computational wedges of only
$90^{\circ}$ to $120^{\circ}$ latitudinal width, or because very rapid
explosions were obtained. In these cases the low-mode asymmetries were
excluded by constraining boundary conditions, or they could not grow
in the time available between shock stagnation and revival. The latter
effect may have been the reason why low-mode instabilities were not
found to be dominant in the 3D simulations of \cite{FW02,Fryer+04},
which developed explosions on rather short time scales after bounce. It
is also possible that these 3D simulations were not evolved
sufficiently far in time to observe the formation of $l=1,\,2$
modes. Without a sufficiently strong contribution of the $l=1$ mode,
the neutron star recoil velocities remain low (typically less than
about 200\,km/s, see \citealt{JM94}).

The main goal of the present paper (the first in a series) is to show
that global anisotropies and large NS kicks can be obtained naturally
in the framework of the neutrino-driven SN explosion mechanism due to
the symmetry breaking by non-radial hydrodynamic instabilities,
without the need to resort to rapid rotation
\citep[e.g.][]{Kotake+03}, large pre-collapse perturbations in the
iron core \citep{BH96,Goldreich+96,Lai+00}, strong magnetic fields
\citep{Wheeler+02,Kotake+04}, anisotropic neutrino emission associated
with exotic neutrino properties
\citep[e.g.][]{Fryer_Kusenko06,Fuller+03}, or jets
\citep{Cen98,Khokhlov+99,Lai+01}. To this end we present a
comprehensive 2D parameter study of supernova dynamics that can be
considered as a significant improvement and extension of the earlier
calculations of \cite{JM96} with respect to the treatment of neutrino
transport, the assumed characteristics of the neutrino emission from
the neutron star core, the inclusion of rotation, the influence of the
initial seed perturbations, the spatial resolution, and the covered
evolutionary time of the supernova explosions.

Parts of the present work were already presented in a Letter by
\cite{Scheck+04}, but a detailed description of both our methods and
results will be given here. We proceed by summarising our numerical
algorithms and computational approach in Sect.~\ref{sec:numsetup}, and
our boundary conditions and initial data in
Sect.~\ref{sec:init_bounds}. We then give an overview of our
simulations in Sect.~\ref{sec:overview}, discussing two representative
neutrino-hydrodynamic calculations in some detail. In
Sect.~\ref{sec:correlations} we explore the dependence of our
simulations on the properties of the stellar progenitors and on the
assumed core neutrino fluxes, and establish correlations between
explosion parameters and neutron star
kicks. Section~\ref{sec:rotation} is devoted to the effects of
rotation. In Sect.~\ref{sec:recoil_robustness} we return to the
neutron star recoils and investigate their robustness with respect to
the approximations and assumptions that we have employed.
Furthermore, we investigate the long-time evolution of the recoil
velocities for a few models beyond the time interval of one second
after core bounce, for which we have evolved most of our models.
Estimating the terminal values of the NS velocities by a calibrated
extrapolation procedure, we will speculate about the possible
implications of our results for the velocity distribution of neutron
stars in Sect.~\ref{sec:kick_distribution}. A summary of this work and
our conclusions can be found in Section~\ref{sec:conclusions}. In
Appendix~\ref{app:definitions} we define and tabulate some physical
quantities of interest which characterise the different runs of our
large set of simulations. Furthermore, we describe the post-processing
procedures that we applied to the numerical calculations to compute
these characteristic quantities. Appendix~\ref{app:hydro_accel}
discusses the solution of the hydrodynamics equations in an
accelerated frame of reference. In Appendix~\ref{app:eexp} we analyse
the explosion energetics of our neutrino-driven
supernovae. Appendix~\ref{app:transport} finally details our new
neutrino transport scheme.

\section{Computational approach and numerical methods}
\label{sec:numsetup}

\subsection{Hydrodynamics and gravity}
\label{sec:hydrograv}

The basic version of the computer program that we employ for this
study is described in \cite{Kifonidis+03}. It consists of a
hydrodynamics module which is based on the direct Eulerian version of
the Piecewise Parabolic Method (PPM) of \cite{CW84} (augmented by the
HLLE solver of \citealt{Einfeldt88} to avoid the odd-even-decoupling
instability), and a module that computes the source terms for energy
and lepton number which enter the hydrodynamic equations due to
neutrino absorption, scattering, and emission processes (see below).
The equation of state is that of \cite{JM96}. In contrast to
\cite{Kifonidis+03,Kifonidis+06} we do not follow explosive
nucleosynthesis in this work. This allows us to save a considerable
amount of computer time, which is mandatory for carrying out an
extended parameter study like the one presented here. For the same
reason, the neutrino transport in our simulations is described
approximately (see Sect.~\ref{sec:transport}) and the dense NS core is
replaced by a moving inner boundary (usually a Lagrangian shell) whose
contraction mimics the shrinking proto-neutron star.

We include self-gravity with relativistic corrections by first
solving the Newtonian two-dimensional Poisson equation using a
Legendre expansion according to \cite{Mueller_Steinmetz95}, and by
subsequently replacing the ``spherical part'' of the resulting
gravitational potential of the 2D mass distribution by the ``effective
relativistic potential'' of \cite{RJ02} (for details, see
\citealt{Marek+06}). For describing the gravity of the central ``point
mass'' (i.e., the mass enclosed by our inner boundary) we use the
baryonic mass where Eq.~(53) in \cite{RJ02} requires the gravitational
mass. This turned out to yield very good agreement with the improved
version of the effective relativistic potential developed by
\cite{Marek+06}.

\subsection{Neutrino transport and neutrino source terms}
\label{sec:transport}

The original code version of \cite{Kifonidis+03} made use of a simple
light-bulb approximation \citep{JM96} in which luminosities of
neutrinos and antineutrinos of all flavours were imposed at the inner
boundary (which is usually below the neutrinospheres) and kept
constant with radius. These luminosities were typically \emph{not}
chosen to give accurate values for the fluxes prevailing below the
neutrinospheric layers, but their choice was guided by the asymptotic
luminosities that emerge from the contracting and accreting nascent
neutron star at {\em large} radii. This was necessary in order to cope
with the main problem of a light-bulb approach, namely that it
neglects changes of the neutrino fluxes and spectra that result from
the interactions of neutrinos with the stellar matter, thus ignoring,
for example, the contributions of the neutrino emission from accreted
matter to the neutrino luminosity.

In this work we considerably improve upon this former approach by
explicitly including these effects. We achieve this by abandoning the
light bulb in favour of a gray, characteristics-based scheme which can
approximate neutrino transport in the transparent and semi-transparent
regimes. The approach is not particularly suited to handle also the
regime of very large optical depths, $\tau$. Therefore we still
perform our simulations with an inner grid boundary at $\tau \approx
10 \dots 100$. However, the luminosities prescribed there have no
relation to those used in the older light-bulb calculations. We have
chosen them to reproduce qualitatively the evolution of the
luminosities in a Lagrangian mass shell below the neutrinospheres as
obtained in recent Boltzmann transport calculations (see also
Sect.~\ref{sec:neutrino_bounds} and especially
Appendix~\ref{sec:app_neutrino_bounds} for details).

The transport scheme itself solves the zeroth order moment equation of
the Boltzmann equation. The transport of neutrino number and energy is
accounted for separately, by integrating two such moment equations for
neutrinos and antineutrinos of all flavours
($\mathrm{e},\mu,\tau$). This allows us to adopt a non-equilibrium
description with the assumption that the spectral form is Fermi-Dirac,
but the neutrino temperatures $T_{\nu_i}$ are not necessarily equal to
the gas temperature $T$. Solving transport equations for neutrino
number and energy, we can determine locally neutrino number and energy
densities and thus the spectral temperatures $T_{\nu_i}$ from the mean
neutrino energies. A detailed description of our approximative
solution of the non-equilibrium transport problem and the exact
expressions for the employed interaction kernels can be found in
Appendix~\ref{app:transport}. While giving qualitatively similar
results as Boltzmann-solvers in spherical symmetry (cf.
Sect.~\ref{sec:impact_transport}), the computational cost of this
approximative transport scheme is two orders of magnitude lower.

\subsection{Numerical grid and frame of reference}
\label{sec:grid}

We adopt 2D spherical coordinates $(r,\theta)$ and assume axisymmetry.
Unless noted otherwise, the calculations presented in the following
are carried out in the full sphere, i.e. for $0 \leq \theta \leq \pi$,
with a grid that is equidistant in the lateral direction. A
non-equidistant grid is employed in the radial direction whose local
spacing, $\Delta r$, is chosen such that square-shaped cells are
obtained in the convective region, i.e.  $\Delta r \approx r\Delta
\theta$. Typically 400 radial and 180 lateral zones are used.

The outer boundary of the computational domain is typically located at
$\rob \approx 2 \times 10^4$\,km, while the inner boundary is placed
within the forming neutron star after core bounce, at a
\emph{Lagrangian} mass shell somewhat below the electron
neutrinosphere. The spacing of the zones near and below the
neutrinospheres is chosen such that variations of the optical depth
per zone remain smaller than a few. The baryonic matter of the neutron
star interior to the inner boundary, $M_{\rm core}$ (which is
typically $\sim 1.1\,\Msol$), is removed and its gravitational
attraction is taken into account by assuming a point mass at $r=0$
(see Sect.~\ref{sec:hydrograv}).

Since we are mainly interested in neutron star kicks in this paper, we
need to point out that the use of the inner boundary condition
(enclosing the NS ``core'') implies that the NS is attached to the
centre of our computational grid. It is therefore not free to move
relative to the ejecta during the simulation (unless special measures
are taken, see below). This is tantamount to assuming that the NS has
an infinite \emph{inertial} mass. Two implications result from this
approximation: A potential hydrodynamic feedback of a displacement of
the NS relative to the ejecta is neglected, and the neutron star
recoil velocity has to be determined indirectly in a post-processing
step by making use of the assumption of total momentum conservation
(see Appendix~\ref{app:definitions}).

The relative motion between neutron star and ejecta can, however, be
accounted for during a simulation by ``wagging the dog'', i.e. by
assuming that instead of the neutron star the ejecta move coherently
in the opposite direction of the neutron star's recoil. This can be
achieved technically by adding the velocity of the relative motion to
the gas velocity on the computational grid, which is tantamount to
performing (after every time step) a Galilei transformation to a new
inertial frame in which the neutron star core is at rest and centred
at $r=0$ (see Appendix~\ref{app:hydro_accel} for details). Simulations
including this procedure will be used to investigate potential
deficiencies of our standard assumption that the NS has an infinite
inertial mass and takes up momentum without starting to move (see
Sects.~\ref{sec:recoil} and \ref{sec:recoil_robustness}).

\section{Initial and boundary conditions}
\label{sec:init_bounds}

\subsection{Boundary conditions}

\subsubsection{Hydrodynamics}
\label{sec:hydro_bounds}

For solving the hydrodynamics equations reflecting boundary conditions
are imposed at the lateral boundaries at $\theta = 0$ and $\theta =
\pi$, while transmitting (i.e. zero gradient) boundary conditions are
employed at the outer radial boundary.

The inner boundary, which is located at the Lagrangian mass coordinate
where we cut our initial (i.e. immediate post-bounce) models, is taken
to be impenetrable. The contraction of this mass shell (and hence of
the neutron star core) is mimicked by moving the inner boundary of our
Eulerian grid from its initial radius, $R_{\rm ib}^{\rm i}$, inwards
to a final radius $R_{\rm ib}^{\rm f}$ according to the expression
\begin{equation}
  \Rib(t) =  \frac{R_{\rm ib}^{\rm i}}{1 \, + \, (1-\exp(-t/\tib)) \,
  (R_{\rm ib}^{\rm i}/R_{\rm ib}^{\rm f}-1)}
\label{eq:ribtimdep}
\end{equation}
of \cite{JM96}. The parameter $R_{\rm ib}^{\rm i}$ is typically in the
range $55\,{\rm km} < R_{\rm ib}^{\rm i} < 85\,{\rm km}$. 

For $R_{\rm ib}^{\rm f}$ and $\tib$ we use two alternative
prescriptions: In what we henceforth will call the ``standard boundary
contraction case'' -- as this is the original parametrization that was
employed by \cite{JM96} -- we set $R_{\rm ib}^{\rm f} = 15 \, {\rm
km}$ and $\tib = t_L$, where the time scale $t_L$ is connected to the
luminosity decay and is defined in
Appendix~\ref{sec:app_neutrino_bounds}. In the second prescription,
the so-called ``rapid boundary contraction case'', we set $R_{\rm
ib}^{\rm f} = 10.5 \, {\rm km}$ and $\tib = 0.25$\,s.

\begin{figure}[tpb!]
\includegraphics[width=8.5cm]{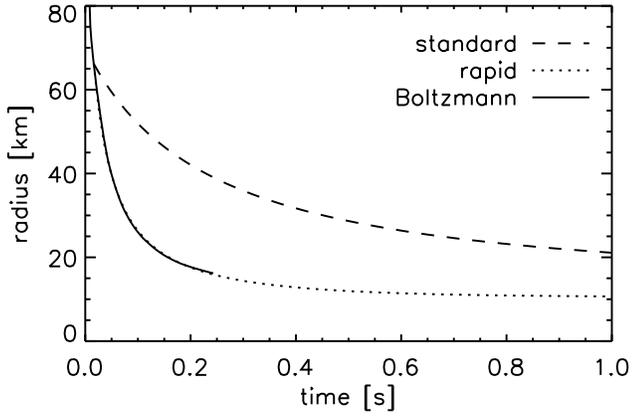}
\caption{Evolution after core bounce of the radius corresponding to a
  mass coordinate of $1.1 \Msol$ from a supernova simulation with
  Boltzmann neutrino transport \citep{Buras+03}, compared to the
  motion of the inner boundary radius as defined by
  Eq.~(\ref{eq:ribtimdep}) for the standard boundary contraction case
  (with $t_L=1$\,s) and the rapid boundary contraction case.}
\label{fig:boltz_rib}
\end{figure}

Figure~\ref{fig:boltz_rib} compares $\Rib(t)$ for both parameter
choices with each other and with data from a supernova simulation with
the nuclear equation of state of \cite{Lattimer_Swesty} and with
Boltzmann neutrino transport \citep{Buras+03} for one of our initial
models. The standard boundary contraction results in a larger final
radius and a slower contraction of the neutron star. The rapid
boundary contraction gives results which are almost indistinguishable
from the Boltzmann calculation. Although this latter parametrization
is potentially more realistic than the former, we have performed the
simulations with the ``standard'' case unless noted otherwise. On the
one hand this reduces the differences compared to our previous work to
only the treatment of the neutrino transport.  On the other hand, the
more compact neutron stars obtained with the ``rapid'' case require
the use of a much finer zoning for adequately resolving the steep
density drop in the neutron star ``atmosphere''. This leads to
computing times which are at least a factor of five larger than those
of the standard case. Exclusive use of the rapidly contracting
boundary would thus have severely reduced the number of computed
models, leading to much poorer statistics. For this reason we have
chosen the rapidly contracting boundary only for a limited set of
models to investigate whether the results obtained for the standard
case change qualitatively. Both cases together therefore provide
information how the results depend on the contraction behavior of the
forming neutron star {(see Sect.~\ref{sec:fastcon})}.

\subsubsection{Neutrinos}
\label{sec:neutrino_bounds}

The boundary conditions for the neutrino emission at the inner grid
boundary are chosen to be \emph{isotropic}. Luminosities and mean
energies for neutrinos and antineutrinos of all three flavours are
imposed there in order to solve the transport problem as described in
Appendix~\ref{app:transport}. These luminosities and energies are
chosen as time-dependent functions that are constrained by prescribed
and varied values for the total loss of energy and lepton number from
the core of the forming neutron star. For example, the lepton number
loss during the first second is of order 0.1--0.2 in all our
simulations, and the total (asymptotic) energy loss $\Delta
E^{\infty}_{\nu,\mathrm{core}}$ does not exceed the gravitational
energy
\begin{equation}
E \approx 3 \times 10^{53} \left({\Mns \over \Msol}\right)^2
\left({\Rns \over {\rm 10\,km}} \right)^{-1}~{\rm ergs},
\end{equation}
which can be released during the birth of a neutron star (see
Tables~\ref{tab:restab_b}--\ref{tab:restab_movns}).

\subsection{Initial models and initial perturbations}
\label{sec:inimod}

Our calculations are started at $\sim 15-20$\,ms after core bounce
from detailed post-collapse models. We make use of four such models
which are based on three different SN progenitors. The first was
calculated by \cite{Bruenn93} with a general relativistic,
one-dimensional (1D), Lagrangian hydrodynamics code coupled to
neutrino transport by multi-group, flux-limited diffusion (see his
Model WPE15 LS 180). It employs the $15\,\Msol$ progenitor of
\cite{WPE88}.  Simulations based on this model will henceforth be
called the ``B-series''.

Our second 1D post-collapse model, provided by M.~Rampp (priv. comm.),
uses a $15\,\Msol$ progenitor star of \cite{Limongi+00} and was
computed with the P{\sc rometheus} PPM hydrodynamics code coupled to
the V{\sc ertex} multi-group variable Eddington factor/Boltzmann
neutrino transport solver \citep{RJ02}. Our ``L-series'' of
simulations makes use of that model.

We also consider two post-bounce models that were computed for the
s15s7b2 progenitor of \cite{WW95} with P{\sc rometheus}/V{\sc ertex}
by \cite{Buras+03,Buras+06} (see their Models s15/1D and s15r). The
first of these models is from a one-dimensional simulation and gives
rise to the ``W-series'' of runs, while the second is a rotating,
two-dimensional (axisymmetric) model, which we use for our
``R-series'' of calculations. This latter model has been described in
detail in \cite{Mueller+04}.

The level of numerical noise in our hydrodynamics code is so low that
a one-dimensional, isotropic initial configuration remains isotropic,
even in the presence of a convectively unstable stratification.
Therefore we need to explicitly add random perturbations to trigger
the growth of non-radial hydrodynamic instabilities in the post-shock
flow. The portable, high-quality random number generator RANLUX due to
\cite{James94,James97} and \cite{Luescher94} is employed. We apply the
perturbation to the velocity field and typically use an amplitude of
0.1\%. To break the equatorial symmetry of the rotating 2D model of
\cite{Buras+03,Buras+06}, we have to add perturbations with an
amplitude of several per cent, since in this model the initial
perturbations have already grown to such a level by the time we map
the model to our full $180^{\circ}$ grid (see Sect.~\ref{sec:grid}).

\section{Overview of the simulations}
\label{sec:overview}

\subsection{The computed models}
\label{sec:the_models}

Tables~\ref{tab:restab_b}--\ref{tab:restab_movns} give an overview of
all our simulations (which were performed with the ``standard boundary
contraction'') in terms of some characteristic quantities which are
defined in Appendix~\ref{app:definitions}.

The naming convention we have chosen for the models is the following:
The first letter denotes the initial model (i.e. the
progenitor/post-bounce data), followed by a two-digit code which
corresponds to the chosen value for the total asymptotic neutrino
energy loss of the neutron star core, $\Delta
E^{\infty}_{\nu,\mathrm{core}}$, in units of $\frac{1}{100}\,\Msol
c^2$. Thus B18, for example, refers to a simulation based on the
\cite{WPE88}/\cite{Bruenn93} initial data with an assumed release of
gravitational binding energy of the core of $\Delta
E^{\infty}_{\nu,\mathrm{core}} = 0.18\,\Msol c^2$. The second
fundamental model parameter, the luminosity time scale $t_L$, is not
taken into account in the model names, because it has the same value
for all models of a series. The chosen value in each case is given in
the captions of Tables~\ref{tab:restab_b}--\ref{tab:restab_movns}.

Simulations performed on a larger grid (with an outer boundary radius
of $10^{10}$\,cm and 500 radial zones) are indicated by the letter
``g'' appended to the model name, e.g. B18-g, simulations which
account for the recoil motion of the neutron star contain the letter
``m'' in the model name, and model series started from different
random seed perturbations are denoted by numbers appended to the model
names. Hence Model B18-1 differs from Model B18 (and from Models
B18-2, B18-3 etc.)  only in the random perturbations imposed on the
initial velocity distribution (with the perturbation amplitude being
the same in all cases).

\begin{figure*}[tpb!]
\centering
\begin{tabular}{cc}
\vspace{-0.1cm}
\includegraphics[angle=0,width=7.0cm]{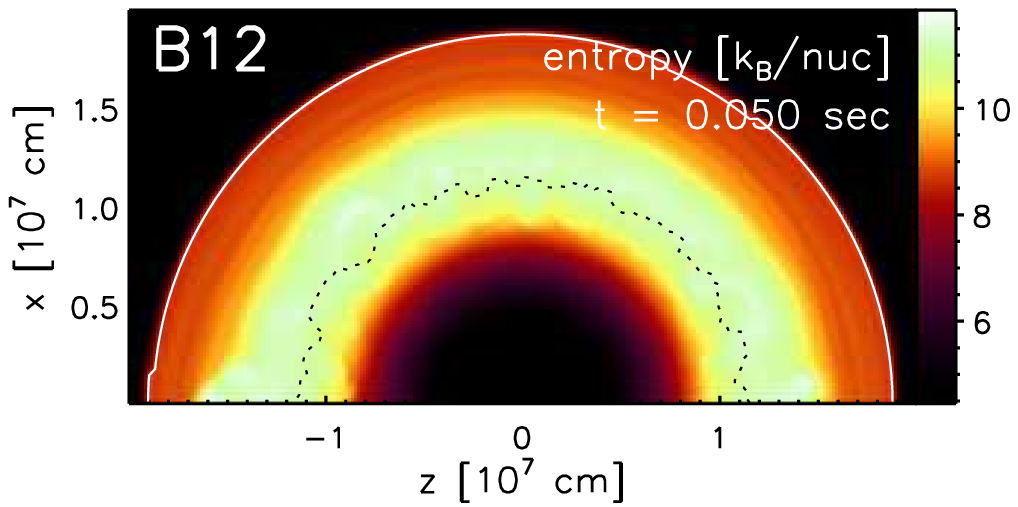}   &   
\includegraphics[angle=0,width=7.0cm]{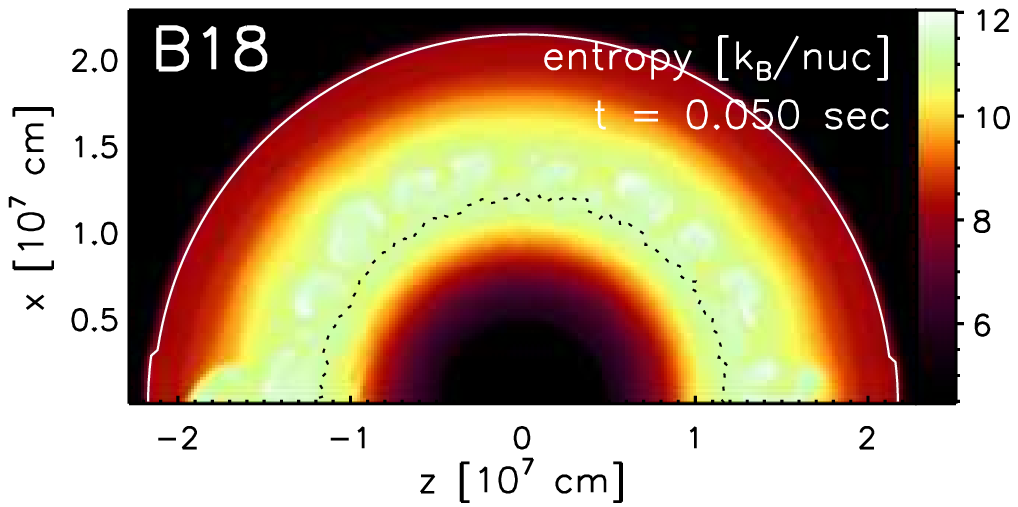}   \\  
\vspace{-0.1cm}
\includegraphics[angle=0,width=7.0cm]{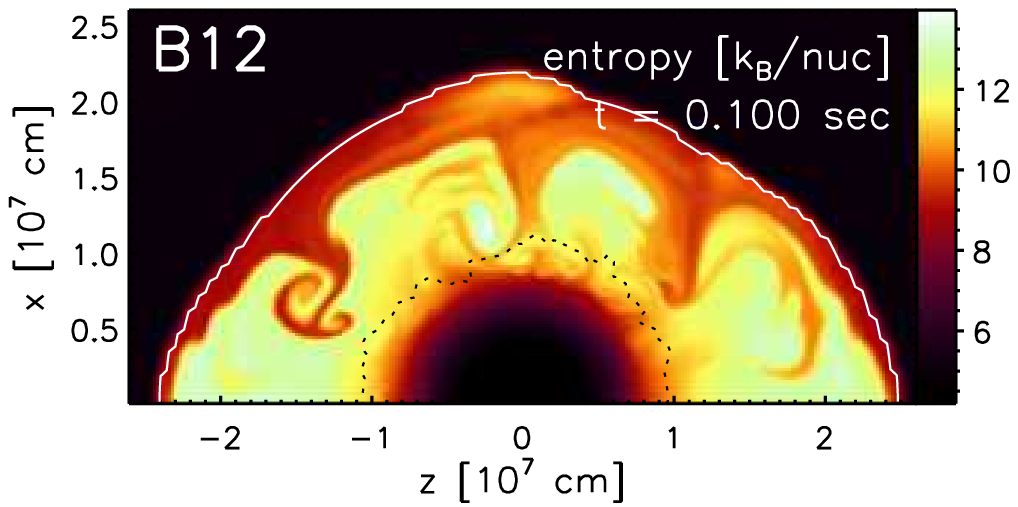}  &   
\includegraphics[angle=0,width=7.0cm]{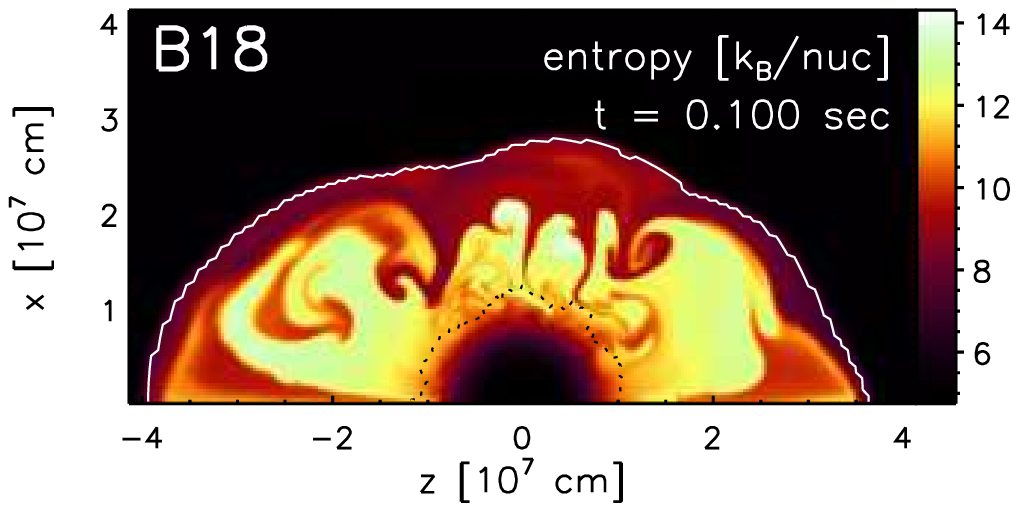}  \\  
\vspace{-0.1cm}
\includegraphics[angle=0,width=7.0cm]{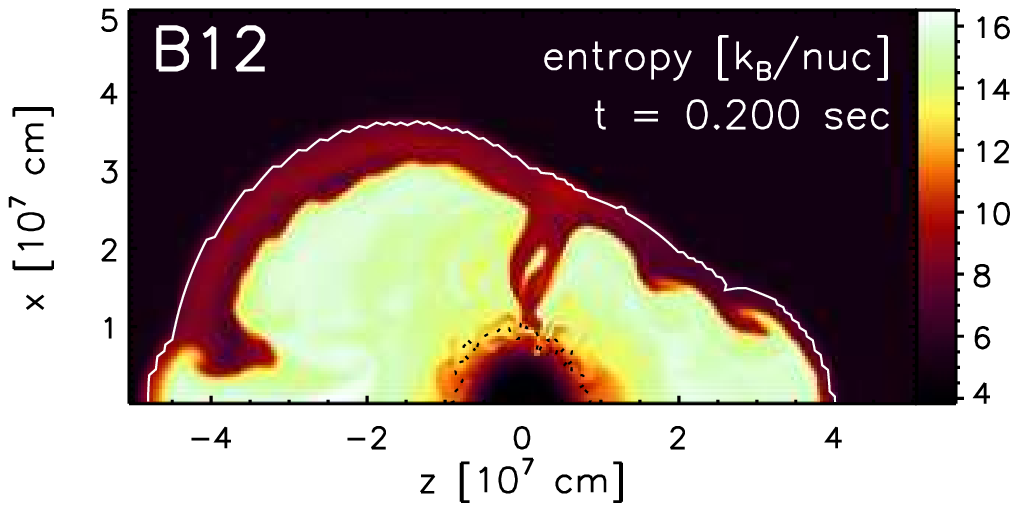}  &   
\includegraphics[angle=0,width=7.0cm]{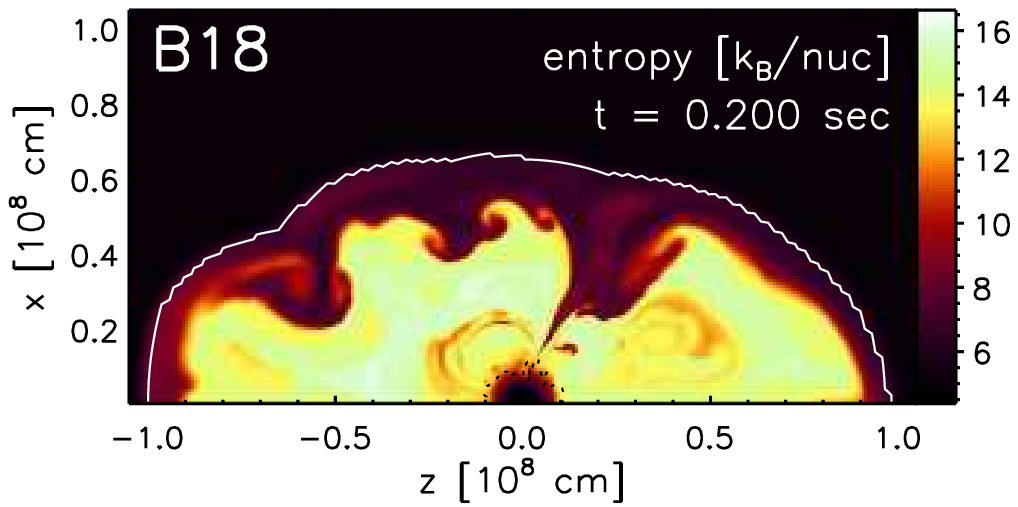}  \\  
\vspace{-0.1cm}
\includegraphics[angle=0,width=7.0cm]{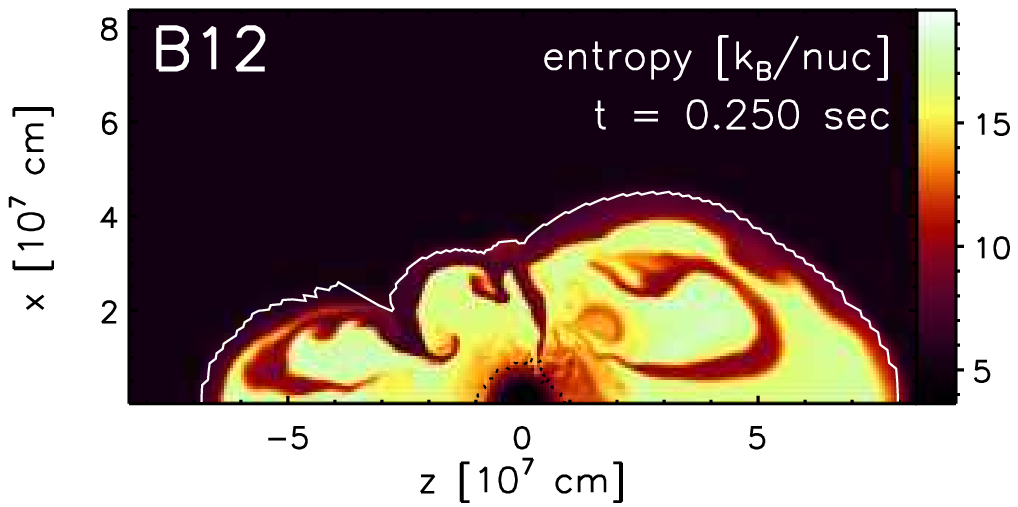}  &   
\includegraphics[angle=0,width=7.0cm]{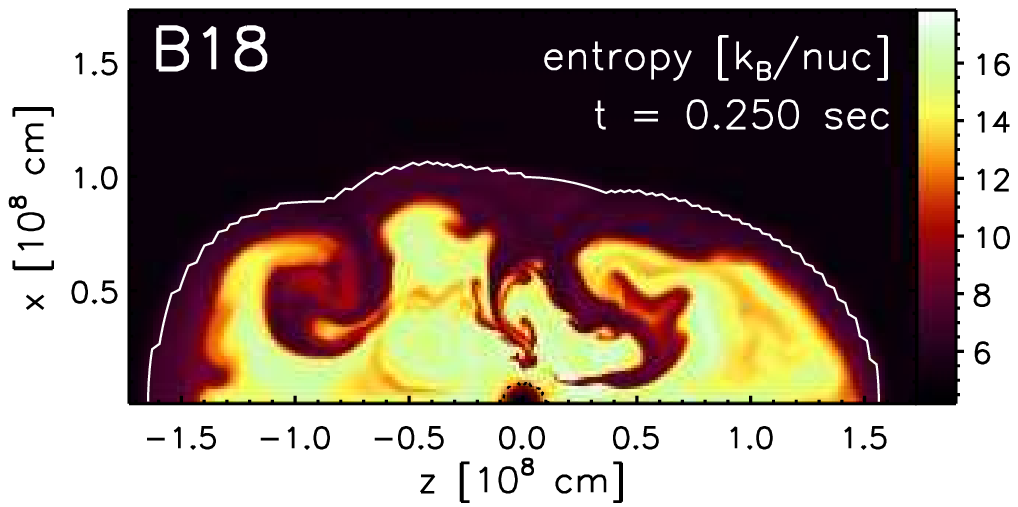}  \\  
\vspace{-0.1cm}
\includegraphics[angle=0,width=7.0cm]{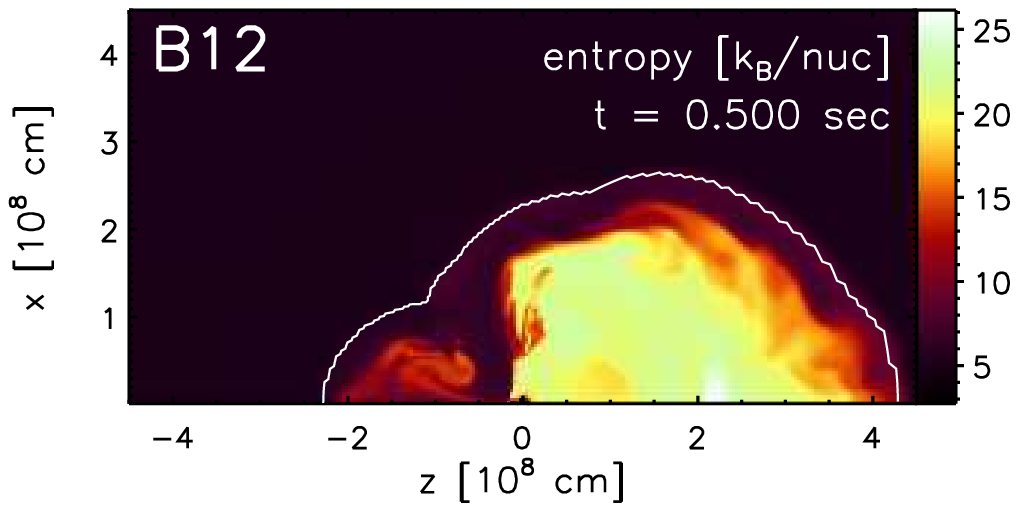}  &   
\includegraphics[angle=0,width=7.0cm]{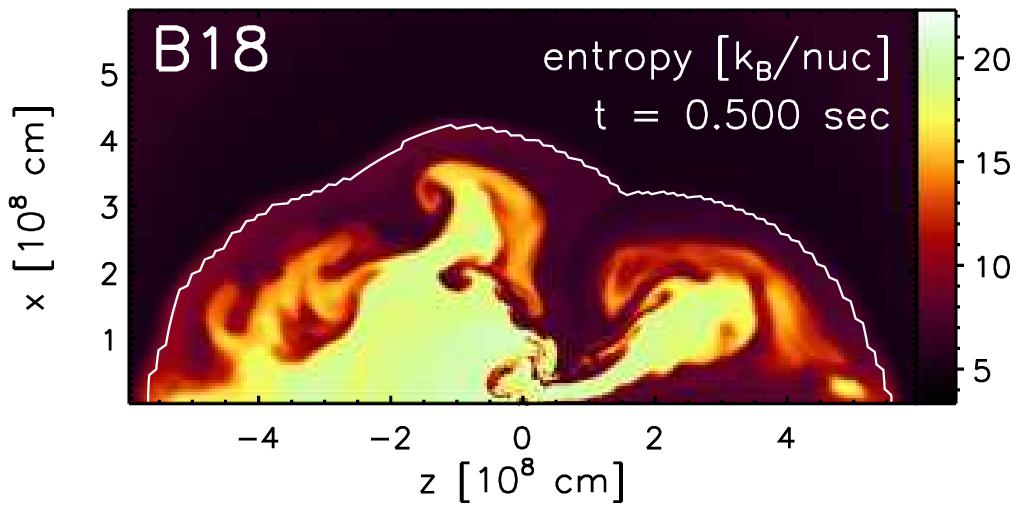}  \\  
\vspace{-0.1cm}
\includegraphics[angle=0,width=7.0cm]{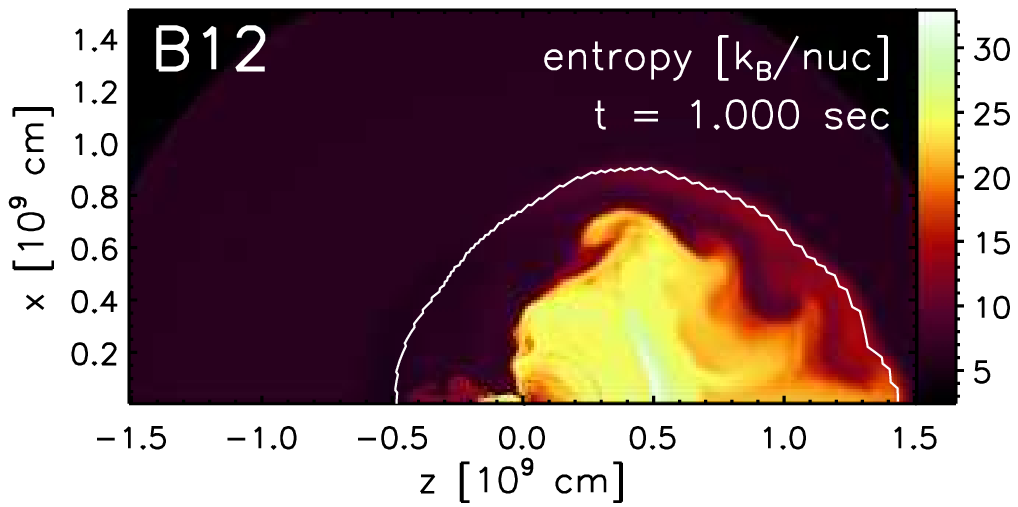} &   
\includegraphics[angle=0,width=7.0cm]{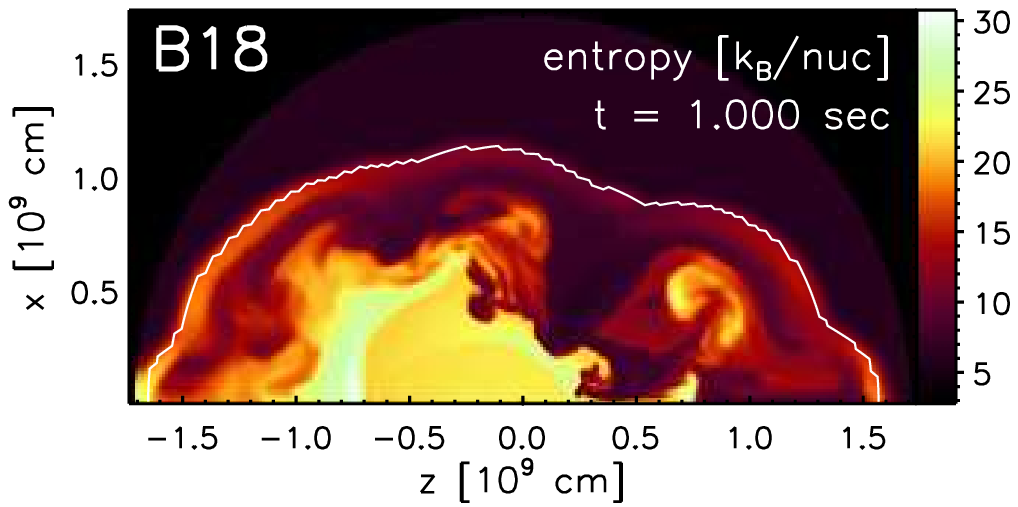}     

\end{tabular}
\caption{Entropy distributions in Model B12 (left column) and Model
  B18 (right column) for different times.  The figures are plotted
  such that the polar axis is oriented horizontally with ``south''
  ($\theta=\pi$) on the left and ``north'' ($\theta=0$) on the right.
  Dotted black lines mark the gain radius and white lines the
  supernova shock.  Note that the scales differ between the plots.
  Convective activity starts with small Rayleigh-Taylor structures
  ($t=50\,$ms) which then grow and merge to larger cells and global
  anisotropy. In contrast to Model B18, the low-energy model B12
  develops pronounced bipolar oscillations (compare the plots for
  $t=200\,$ms and $t=250\,$ms between both cases). After the explosion
  has set in, the convective pattern ``freezes out'' and the expansion
  continues essentially self-similarly (see the plots for $t=500\,$ms
  and $t=1000\,$ms). At 1$\,$s after bounce Model~B18 shows the
  emergence of an essentially spherical neutrino-driven wind expanding
  away from the neutron star surface (region around the coordinate
  center in the bottom right panel).}
\label{fig:stot}
\end{figure*}

\begin{figure*}[tpb!]
\centering
\begin{tabular}{cc}
\includegraphics[angle=0,width=8.5cm]{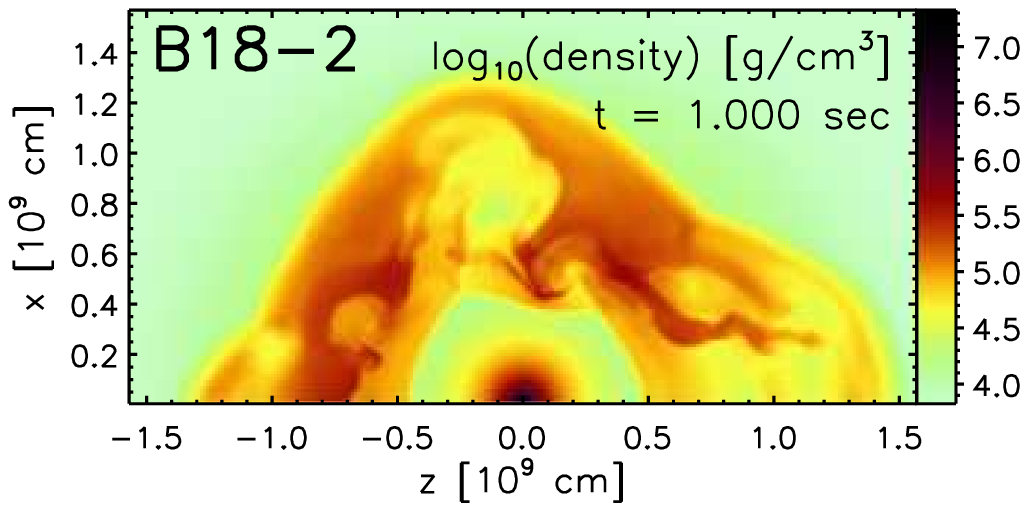} &
\includegraphics[angle=0,width=8.5cm]{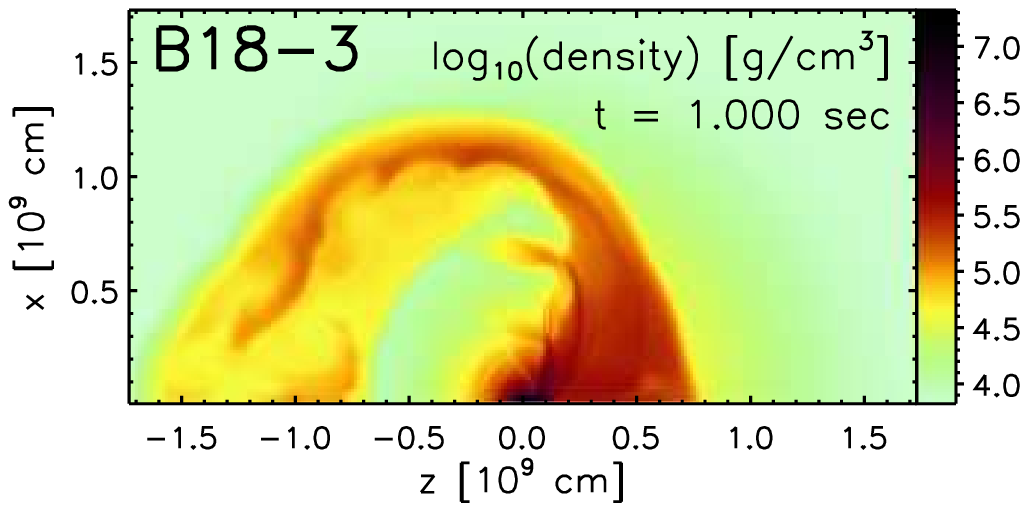} \\
\includegraphics[angle=0,width=8.5cm]{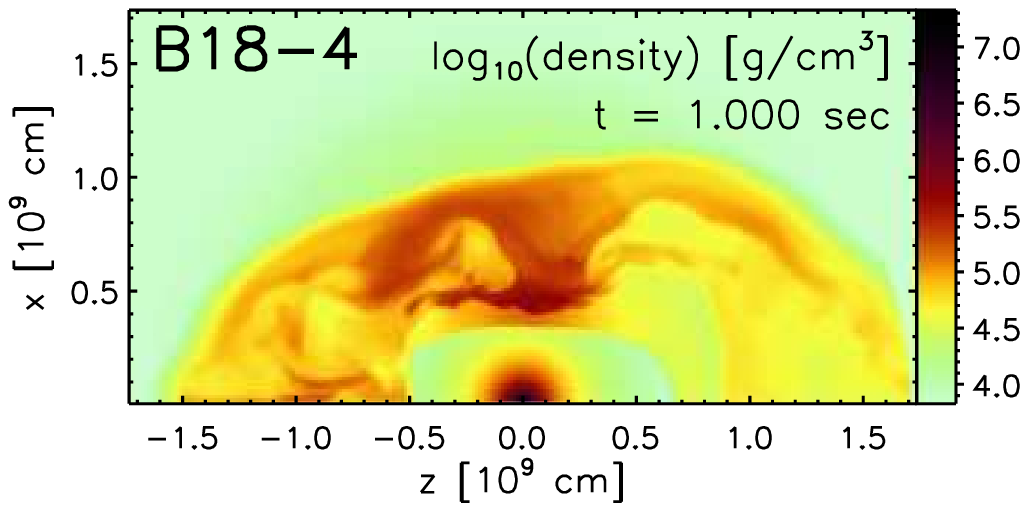} &
\includegraphics[angle=0,width=8.5cm]{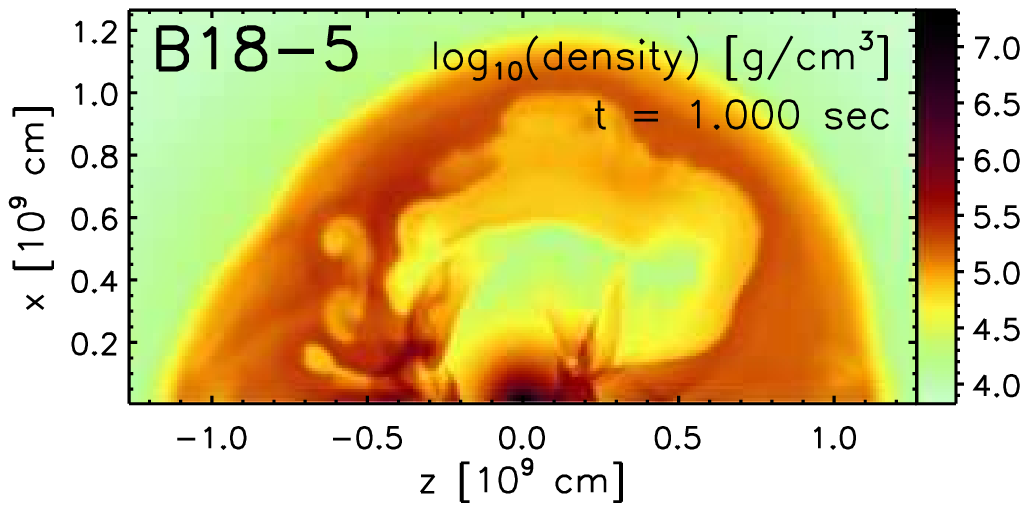}
\end{tabular}
\caption{Density distributions one second after core bounce for four
  simulations with the same initial and boundary conditions as Model
  B18, but different patterns of the random seed perturbations imposed
  on the velocity field of the initial model. The amplitudes of the
  perturbations ($10^{-3}$) are the same in all cases. The morphology
  of the explosion depends in a chaotic way on the initial
  perturbations.}
\label{fig:b18_rseed}
\end{figure*}

Note that in Tables~\ref{tab:restab_b}--\ref{tab:restab_movns} the
total lepton number and energy loss of the neutron star core,
$\dYecoretone$ and $\Delta E^{\rm tot}_{\nu,\mathrm{core}}$,
respectively, the time-integrated energy loss in $\nue$ and $\nuebar$,
$\Delta E_{500}$, the explosion energy, $\Eexp$, the anisotropy
parameter, $\alphag$, the shock deformation, $d_{\rm shock}$, the
neutron star mass and recoil velocity, $\Mns$, and $\vzns$,
respectively, as well as the correction of the latter due to the
``neutrino rocket effect'', $\vznsnu$, are all given at the time
$t=1$\,s, at which we usually stop our simulations.

We need to point out here that the listed neutron star velocities are
\emph{not} the final ones, but that even at the end of our simulations
the neutron stars can still experience a large acceleration. We
therefore also give this acceleration, $\azns=\mathrm{d}v_z^{\rm
ns}/\dt$ (averaged over the last $100\,$ms and without neutrino
effects), and will attempt to estimate the final neutron star
velocities in Sect.~\ref{sec:longterm}.

\subsection{The character of the flow}
\label{sec:flow_character}

Giving an accurate qualitative description of the flow that
establishes in our calculations is a difficult endeavour, as the
evolution that we observe during the first $\sim 300-400$\,ms is
wildly time-dependent and extremely nonlinear. One may even
characterise it as chaotic. The layer between the proto-neutron star
and the supernova shock is Ledoux-unstable, because a negative entropy
gradient is established due to neutrino heating within $\sim 50$\,ms
after bounce. In all simulations discussed here, it is consequently
Ledoux convection which breaks the initial spherical symmetry: small
Rayleigh-Taylor mushrooms grow from the imposed random seed
perturbations and start rising towards the shock. They merge quickly
and grow to fewer but larger bubbles that deform the shock and push it
outward (Fig.~\ref{fig:stot}).

Due to the violent motions of the rising high-entropy plumes the shock
gets bumpy and deformed, and caustic-like kinks of the shock emerge
where two such bubbles approach each other and collide. Downstream of
the shock, decelerated and compressed matter forms a high-density
(low-entropy) shell, which sits atop high-entropy material that boils
vigorously as it is heated by neutrinos from below. The interface
between these layers is Rayleigh-Taylor unstable \citep{Herant95} and
gives therefore rise to narrow, low-entropy downflows of matter, which
penetrate from the postshock layer to the neutron star with supersonic
velocities. When they reach surroundings with entropies lower than
their own, the downflows are decelerated and their material spreads
rapidly around the neutron star. The evolution of these downflows is
highly dynamic. They form, merge with other accretion funnels, or are
blown away by the rising buoyant matter on a time scale of
$10$--$20\,$ms, while their number decreases with time. The most
massive of these downflows originate from the kinks at the shock
surface, where the deceleration of the infalling matter is weaker due
to the (local) obliqueness of the shock.

During this phase of violent ``boiling'' the shock develops a strong,
time-dependent deformation and expands slowly outward. In Model B12,
whose evolution is shown in the sequence of plots on the left side of
Fig.~\ref{fig:stot}, pronounced bipolar hemispheric oscillations
become visible after about $150\,$ms. Such bipolar oscillations (and
the consequent ``sloshing'') of the shock have been found to be
typical of $l=1$ mode instabilities as associated with the
advective-acoustic instability or the SASI. Hence these instabilities
are likely to dominate the evolution of this model in this strongly
nonlinear phase. Note that Model B12 differs from Model B18 (on the
right side of Fig.~\ref{fig:stot}) by lower neutrino luminosities at
the inner boundary and, correspondingly, by a later onset of the
explosion (at $\texp=220\,$ms compared to $\texp=152\,$ms) and a lower
explosion energy of $0.37 \times \foe$ versus $1.16 \times \foe$
(measured at $t=1$\,s). Model B18 shows also violent convective
activity, but no bipolar oscillations. This fact might indicate that
in this model the convective instability might play a more important
role. A detailed investigation of the growth of different kinds of
non-radial instabilities in the postshock flow and their competition
will be presented in a subsequent paper of this series (Scheck et al.\
2006, in preparation). Here we only note that their combined
effects can have a decisive impact on the explosion mechanism of
supernovae, since one-dimensional counterparts of both Models B12 and
B18 failed to explode.

The highly dynamic phase of the evolution comes to an end around
$300-400$\,ms after bounce. At that time the explosion is well
underway, and the overall flow settles into a state of
quasi-self-similar expansion, which is remarkably stable (compare the
lower panels of Fig.~\ref{fig:stot}, and see also
\citealt{Herant95}). Yet the consequences of the dynamic phase are
felt long thereafter, since the strongly nonlinear boiling motions
lead to a final morphology that is sensitive to even tiny initial
differences in the flow. These may not only result from the influence
of different boundary conditions, as in case of Models B12 and
B18. The late-time morphology is even sensitive to the seed
perturbation that we apply to trigger the non-radial
instabilities. Figure~\ref{fig:b18_rseed} illustrates this for the
B18-series of models in which the seed perturbation was varied as
described in Sect.~\ref{sec:the_models}. It is obvious that the
dominant mode in the flow is unpredictable. It can be $l=2$, with two
bubbles which are separated by a single accretion funnel, and which
occupy roughly a hemisphere each (as in Model B18-4 and the original
Model B18). Yet the bubbles may also differ significantly in size
resulting in a dominance of the $l=1$ anisotropy as in Model B18-3
(top right panel of Fig.~\ref{fig:b18_rseed}) and Model B12 (left
panels of Fig.~\ref{fig:stot}). This sensitivity to the seed
perturbation is so extreme that the system may be described as
exhibiting symmetry breaking in a chaotic manner. In fact even the
same model computed on different machines (with slightly different
64-bit round off behaviour) may actually end up with a different
morphology.

For sufficiently high core luminosities, accretion of matter onto the
neutron star is eventually superseded by the onset of a nearly
spherically symmetric neutrino-driven wind (see the region around the
coordinate center in the lowermost right panel of Fig.~\ref{fig:stot}
and in the left panels of Fig.~\ref{fig:b18_rseed}; cf.\ also
\citealt{BHF95,JM96}). If the wind is strong enough, as in Model B18
where the mass-loss rate of the nascent neutron star by the wind is
$\dotMns = -5.1\times10^{-2}\,\Msol/{\rm s}$, it blows away the
accretion funnel and establishes a high-entropy shell or cavity of
rapidly expanding low-density material around the neutron star, which
is separated from the ejecta by a strong reverse shock.  Otherwise
accretion through the funnel continues until more than about 1\,s
after bounce, as in Model B12. In this case the accreted material
reaches infall velocities of about 1/4 of the speed of light, while
the accretion rate at $t = 1$\,s has decreased to $\dot M_{\rm accr}
\approx 4\times10^{-2}\,\Msol/{\rm s}$.  Since at the same time the
neutron star mass changes at a rate of $\dotMns \approx
1.1\times10^{-2}\,\Msol/{\rm s}$, only a fraction of $\sim 25\%$ of
the infalling matter is actually integrated into the neutron star. The
remaining $75\%$ are heated and reejected with high velocity in a
neutrino-driven wind that inflates a buoyant bubble of neutrino-heated
gas in the northern hemisphere opposite to the remaining accretion
funnel on the other side (see Fig.~\ref{fig:stot}, lowermost left
panel). The resulting flow, which is characterised by a strong dipole
mode, can be conveyed only incompletely with plots such as
Fig.~\ref{fig:stot} and is much more impressively captured by movies
that we have produced from our data\footnote{A collection of movies is
  provided as online material of this article.}.

These movies also show that the impact and rapid deceleration of the
accretion streams in the vicinity of the forming NS create acoustic
and weak shock waves that emanate from the neutron star's surface. As
was recently also noted by \cite{Burrows+06}, these waves propagate
predominantly into the hemisphere opposite to the accretion funnel.
While traversing the low-density cavity, also the acoustic waves
steepen into shocks, dissipate energy, and heat the expanding
material. However, in accordance with the results of \cite{JM96}, we
observe only a modest production of entropy due to these waves when
they propagate outward in the rapidly expanding neutrino-driven wind.
In case of the \cite{Burrows+06} simulation the acoustic energy input
from neutron star g-mode oscillations was found to be crucial for the
explosion of an 11$\,M_\odot$ model.  Nonspherical accretion was found
to lead to the excitation of core g-mode oscillations at late times
($\ga\,$300--500$\,$ms) after bounce, whose sonic damping transfers a
significant amount of acoustic power to the surrounding medium and
supernova shock. G-mode oscillations are in fact also present in the
outer layers of our neutron stars -- i.e.\ in the layers that are
included on our grid -- but their amplitudes are modest and thus they
do not lead to the strong consequences reported by \cite{Burrows+06}.
It is possible, however, that our simulations underestimate such
effects, which would require the inclusion of the whole neutron star
without excising the central core, and the ability to follow the
excitation of deep modes due to a self-consistent coupling between
accretion, core motion, and core-mode generation. On the other hand,
our models are characterized by explosions due to convectively
supported neutrino heating (whereas the 11$\,M_\odot$ model of Burrows
et al.\ seemingly did not explode in that way). After the explosion
has been launched, our models reveal the development of a strong
neutrino-driven wind, in which the dissipation of acoustic waves may
have a different effect than in a more or less static gas around the
oscillating neutron star.  We point out that numerical simulations of
this neutrino-driven outflow require very high radial resolution of
the steep density gradient near the neutron star surface. We are not
sure whether sufficiently fine grid zoning was guaranteed in the
simulation by \cite{Burrows+06}.  A more detailed investigation of
such questions is in progress.

\subsection{The influence of the neutrino transport}
\label{sec:impact_transport}

\begin{figure}[tpb!]
\centering
\includegraphics[angle=0,width=8.5cm]{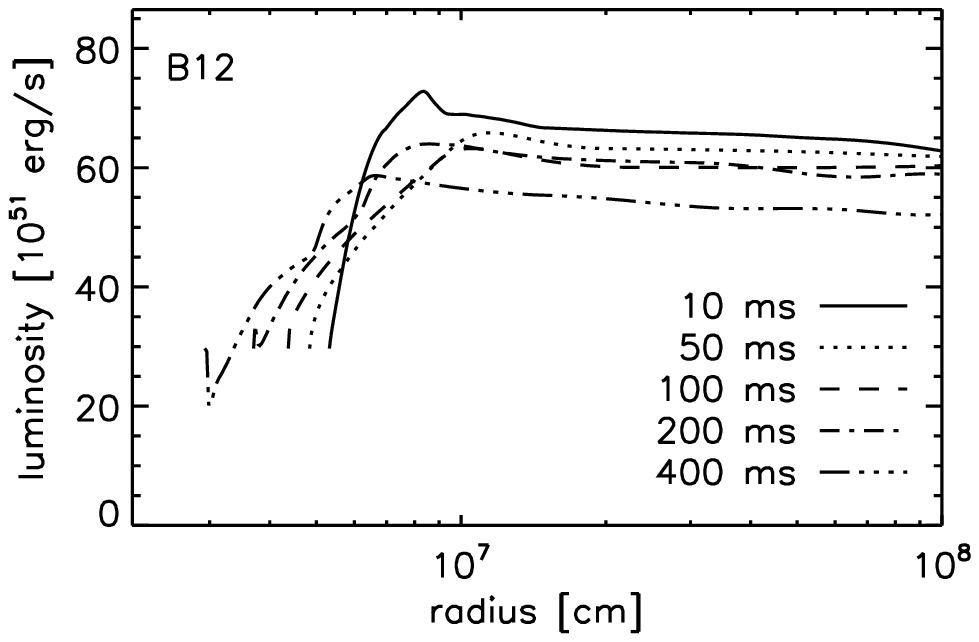} 
\includegraphics[angle=0,width=8.5cm]{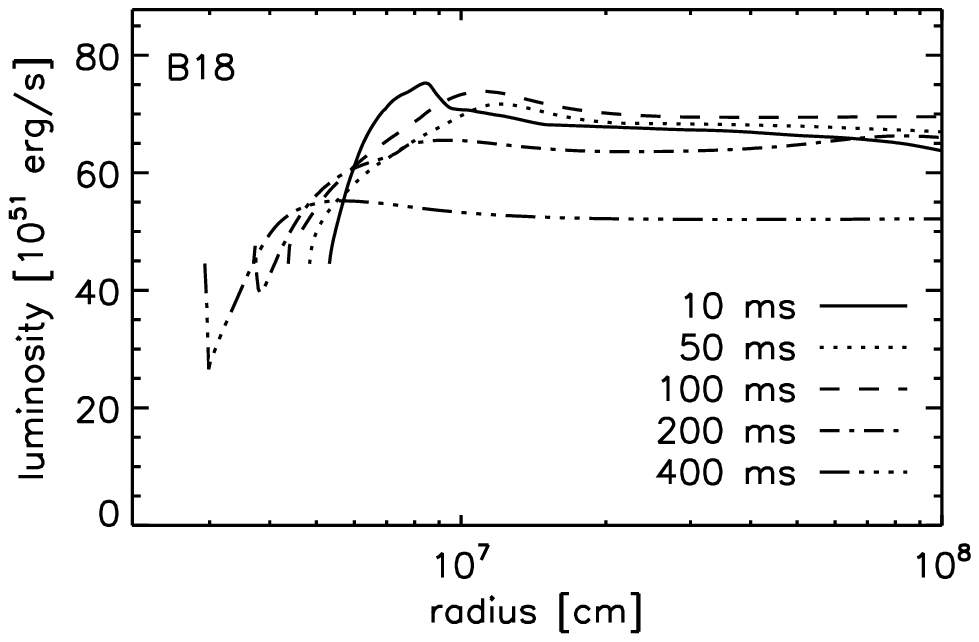}
\caption{Radial profiles of the sum of the $\nue$ and $\nuebar$
  luminosities for Models B12 and B18 at different times after the
  start of the simulations.}
\label{fig:lum_profiles}
\end{figure}

\begin{figure}[tpb!]
\centering
\includegraphics[angle=0,width=8.5cm]{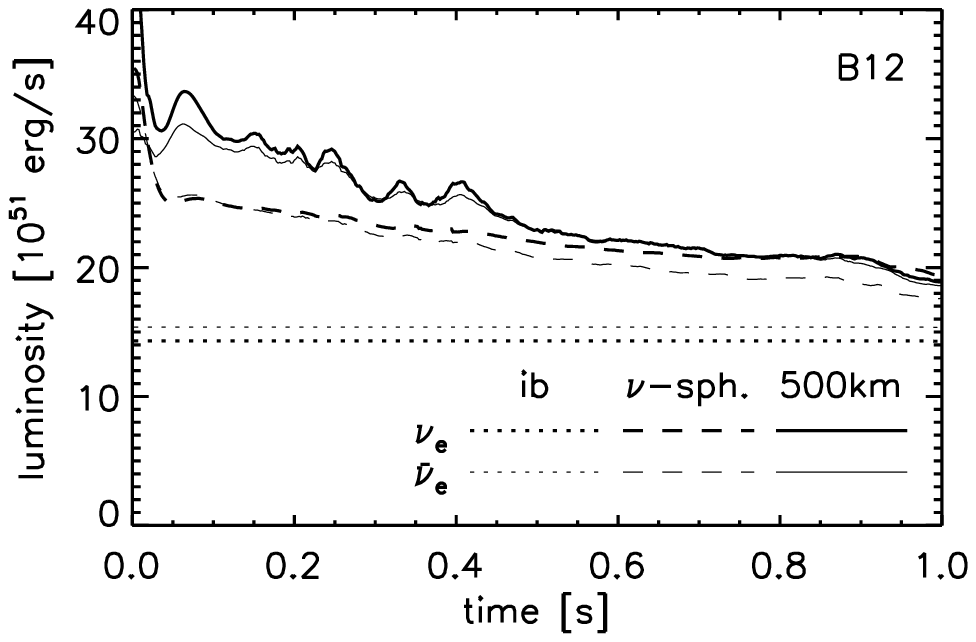}
\includegraphics[angle=0,width=8.5cm]{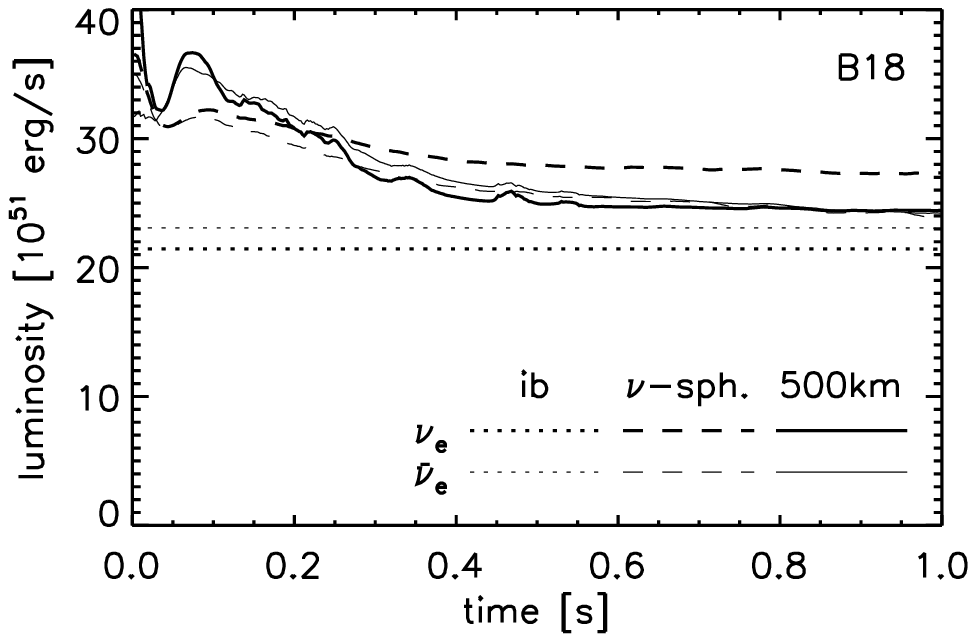}
\caption{Evolution of the luminosities of $\nue$ and $\nuebar$ at the
         inner boundary, at the $\nue$-sphere, and at a radius of
         $500$\,km. Note the different importance of the accretion
         contribution to the luminosity in the low-energy explosion
         (Model B12) compared to the high-energy explosion (Model B18)
         and the rapid decay of the accretion luminosity after the
         onset of the explosion in the latter model.}
\label{fig:lightcurves}
\end{figure}

The fact that earlier 2D simulations, which were performed with a
neutrino light-bulb description \citep{JM96,Kifonidis+03}, were not
dominated by low-order modes, poses the question to which extent the
development of such global asphericity in the flow is sensitive to the
treatment of the neutrino transport. Figure~\ref{fig:lum_profiles}
shows that our new neutrino transport description yields radial
profiles for the sum of the $\nue$ and $\nuebar$ luminosities which
deviate markedly from the radius-independent luminosities used in a
light-bulb approach: The luminosities are significantly modified
compared to the values imposed at the inner boundary. After some
adjustment to the local thermodynamic conditions, which takes place in
a few radial zones next to the inner boundary, the luminosities rise
steeply in the cooling region below the gain radius, and decline
slightly in the heating region farther out. The rise is caused by the
creation of neutrinos when gravitational energy is released during the
accretion and the contraction of the neutron star, while the slight
decline results from the absorption of the $\nue$ and $\nuebar$ in the
heating region.

\begin{figure}[tpb!]
\centering
\includegraphics[angle=0,width=8.5cm]{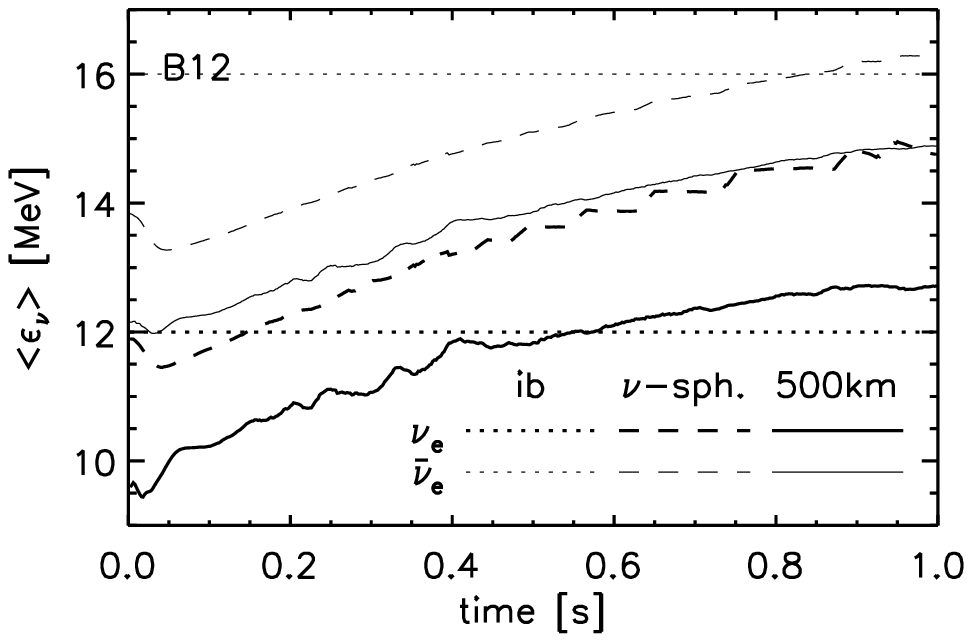}
\includegraphics[angle=0,width=8.5cm]{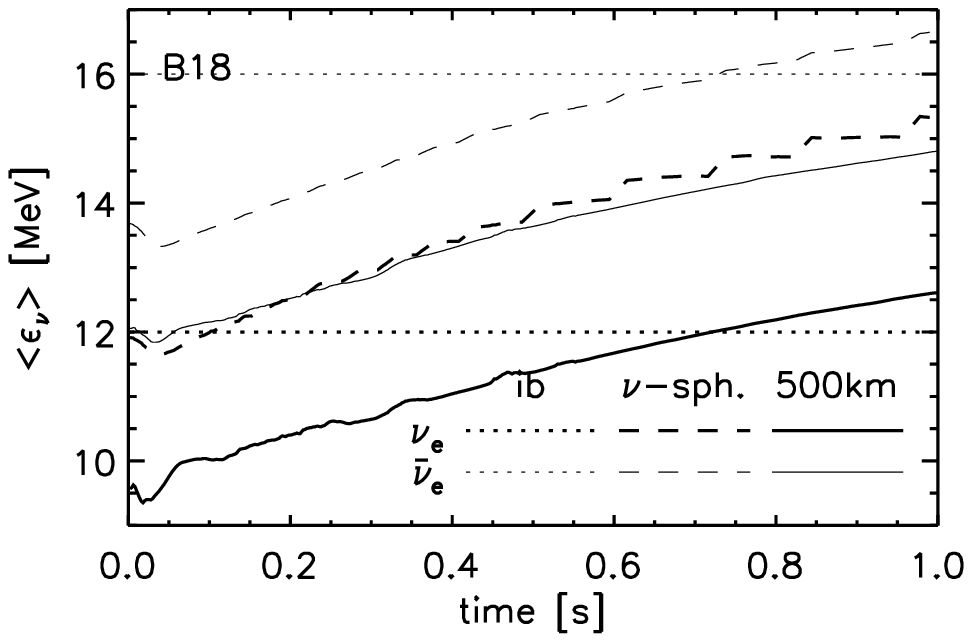}
\caption{Evolution of the mean $\nue$ and $\nuebar$ energies at the
  inner boundary (ib), at the $\nue$-sphere, and at a radius of
  $500\,$km. Note that due to the contraction and compressional
  heating of the nascent neutron star the average energies of the
  radiated neutrinos continue to rise until the end of our simulated
  evolution.}
\label{fig:nuenergies}
\end{figure}

The ``accretion'' luminosity that is produced on the grid is usually
of the same order as the luminosity emerging from the core.  In
low-energy models, like B12, the accretion component exceeds the core
component early on, while in high-energy models the core component is
dominant at all times (see the neutrino ``lightcurves'' for Models B12
and B18 shown in Fig.~\ref{fig:lightcurves}).

Our transport scheme can account qualitatively well for the evolution
of core and accretion components of the neutrino luminosities, for the
radial and temporal evolution of the luminosities and mean energies of
the radiated neutrinos, and for the relative sizes of the $\nue$ and
$\nuebar$ emission (see
Figs.~\ref{fig:lum_profiles}--\ref{fig:nuenergies}, and compare with
\citealt{Liebendoerfer+01}, \citealt{Liebendoerfer+05},
\citealt{RJ00}, \citealt{Buras+03,Buras+06,Buras+06b} and
\citealt{Thompson+03}). We therefore think that our current transport
treatment is a reasonably good method for performing parametric
explosion studies with the aim to better understand the role of
hydrodynamic instabilities during the shock-revival phase of
neutrino-driven supernova explosions.
 
Yet, we point out here that all the previously not modelled neutrino
transport effects are \emph{not} the reason why the development of
$l=1,2$ modes is seen here, whereas it was not visible in the
calculations of \cite{JM96} and \cite{Kifonidis+03}. Highly
anisotropic explosions can also be obtained with the light-bulb
assumption (see \citealt{Janka+02,Janka+04c} for an example). In other
words, the details of the functional form of $L(r)$, which are visible
in Fig.~\ref{fig:lum_profiles}, are \emph{not} decisive for the growth
of the $l=1,2$ modes. What is crucial, however, is that the explosions
in the current models start {\em slowly}. This was not the case in all
but one of the simulations of \cite{JM96} and \cite{Kifonidis+03},
where the neutrino luminosities were assumed to decay exponentially
with a time scale of typically 500--700$\,$ms (see Table~1 in
\citealt{JM96}) instead of varying slowly. The exponential,
``burst-like'' decline of the neutrino light bulb implied fairly high
initial luminosities (i.e., 4.5--5$\times 10^{52}\,$erg$\,$s$^{-1}$
for $\nu_{\mathrm e}$ plus $\bar\nu_{\mathrm e}$ at the inner grid
boundary) -- which were required in case of the exponential decay for
getting ``typical'' supernova explosion energies -- and thus strong
neutrino heating occurred at early times after bounce. This led to
rapid explosions, which in turn did not leave enough time for the
convective cells and bubbles to merge before the expansion became so
fast that it continued in a quasi self-similar way. Since these
bubbles are initially small, their early ``freezing out'' in the
rapidly expanding flow had the effect that small structures (i.e.
high-order modes) prevailed until very late times. The rapid
explosions also caused a quick end of accretion onto the proto-neutron
star, and therefore the neutron stars remained small.  In contrast,
the present transport description gives neutrino luminosities between
the neutrino spheres and a radius of 500\,km that vary much less
steeply than exponentially with time (see Fig.~\ref{fig:lightcurves}).
This leads to explosion time scales that are sufficiently long to
allow for the formation of low-order modes.

\begin{figure*}[tpb!]
\centering
\begin{tabular}{ccc}
\includegraphics[width=5cm]{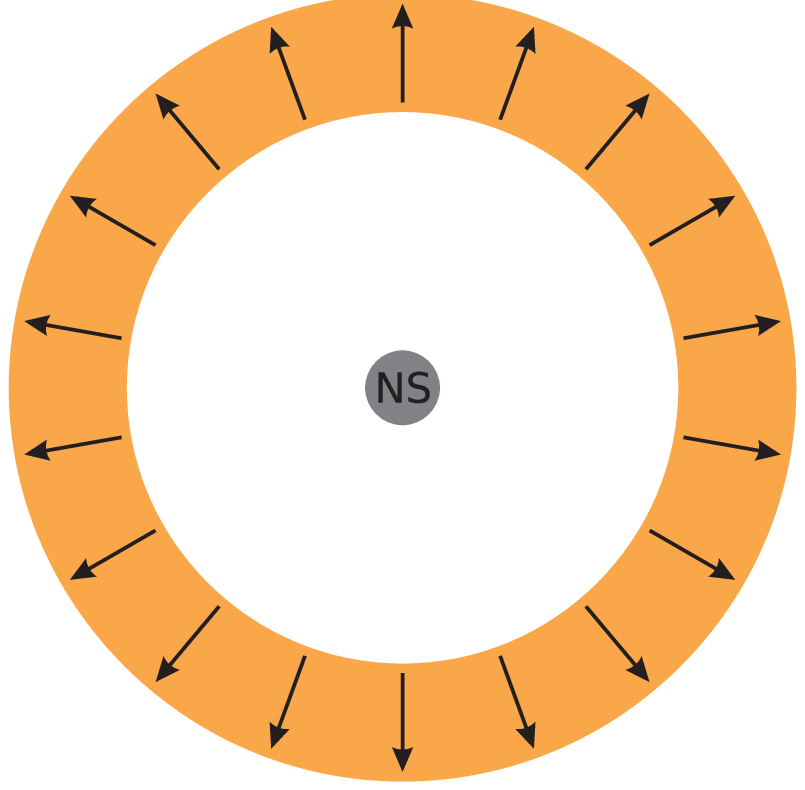} &
\includegraphics[width=5cm]{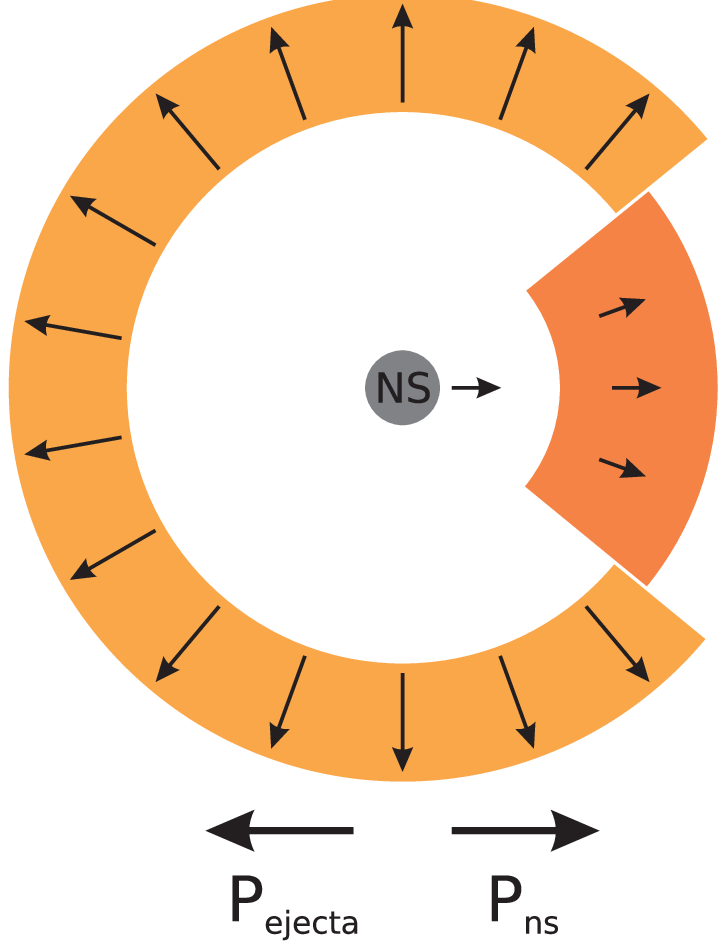} &
\includegraphics[width=5cm]{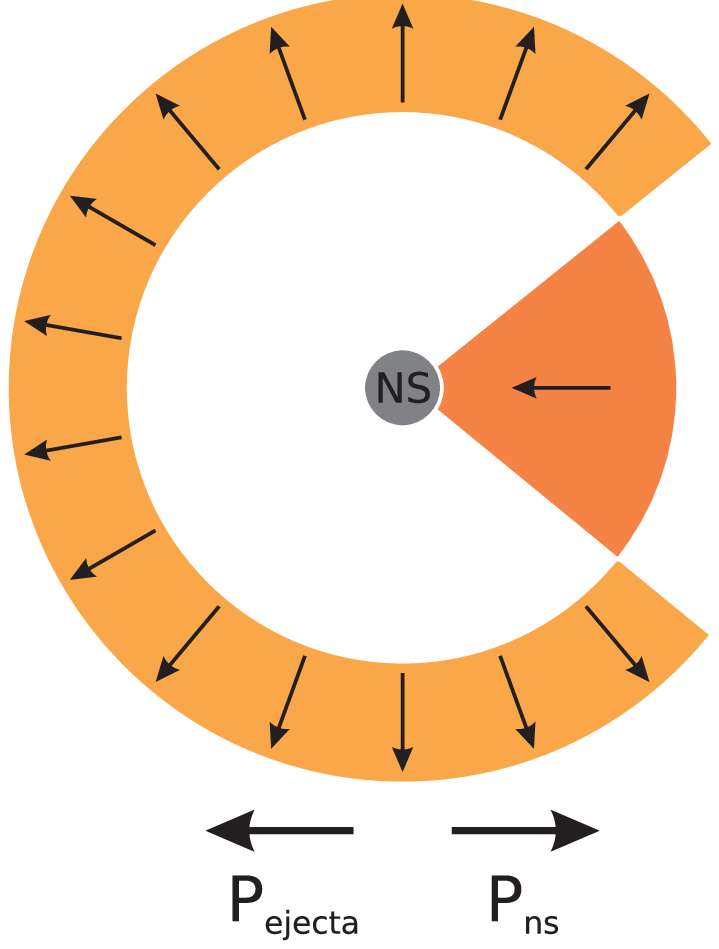}
\end{tabular}
\caption{Graphical illustration of the momentum balance between
  neutron star and ejecta. The largest fraction of the ejecta mass is
  concentrated in a dense shell behind the shock (bright coloured
  ring). For a spherical explosion (left panel) the momenta of the
  neutron star and the ejecta are zero. If the expansion of one
  hemisphere lags behind the other, the gas has a net momentum in the
  direction of the faster expanding hemisphere. The neutron star is
  always accelerated in the opposite direction, i.e. towards the slower
  moving gas (middle and right panels). This acceleration can be
  mediated by the gravitational attraction of the anisotropic ejecta
  (middle panel). In case accretion flows reach down to the neutron
  star surface (right panel), additional (hydrodynamic) forces may
  contribute, but the gravitational force, in general, remains
  dominant.}
\label{fig:mombal}
\end{figure*}

\subsection{Acceleration and recoil of the neutron star}
\label{sec:recoil}

As is detailed in Appendix~\ref{app:definitions}, in a 2D
axi\-symmetric calculation the neutron star recoil can only have a
component parallel to the $z$-axis. For its calculation only the
$z$-momenta of the gas in the northern and southern hemispheres need
to be considered (see Eq.~\ref{eq:mombal2d}). If the momentum density
of the ejecta, $p_z(r,\theta)$, is mirror symmetric with respect to
the equatorial plane, i.e., if
\begin{equation}
  p_z(r,\theta)=-p_z(r,\pi-\theta)
\end{equation}
holds, the sum of the $z-$momenta of the two hemispheres vanishes
\begin{equation}
P_{z,\mathrm{gas}} = P_{z,\mathrm{gas}}^{\rm N} + 
                     P_{z,\mathrm{gas}}^{\rm S} = 0,
\end{equation}
and thus also the neutron star remains at rest
(cf. Eq.~\ref{eq:vns_pgas}). The latter situation is given e.g. for an
$l=2$ mode, i.e. for two polar bubbles of equal size separated by a
single downflow which is located in the equatorial plane. However, in
general the expansion of the ejecta will proceed differently in the
two hemispheres, so that a net momentum $P_{z,\mathrm{gas}} \ne 0$
will result.

\begin{figure*}[tpb!]
\centering
\begin{tabular}{cc}
\includegraphics[angle=0,width=8.5cm]{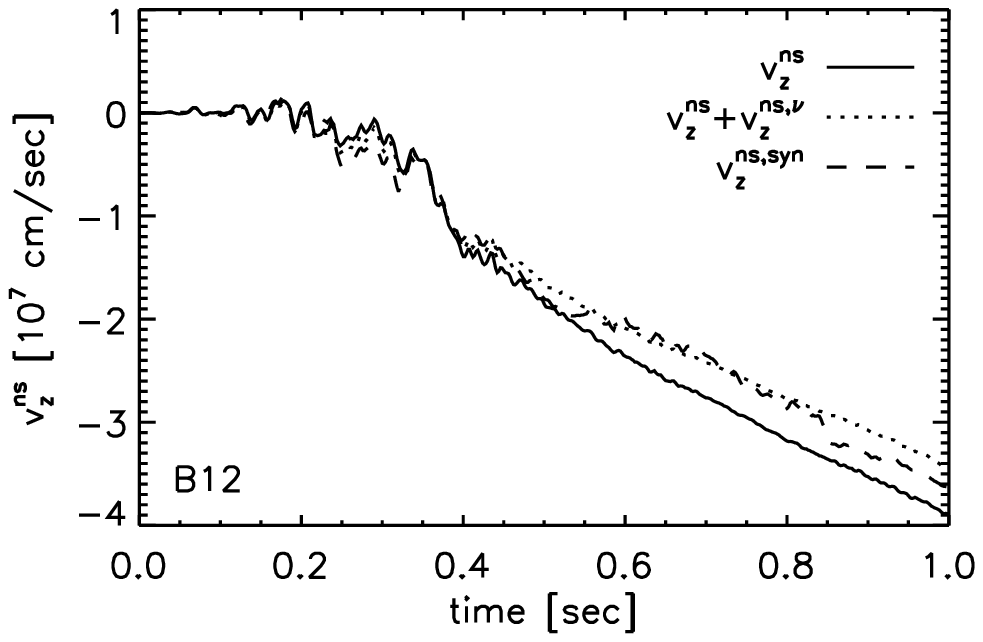}&
\includegraphics[angle=0,width=8.5cm]{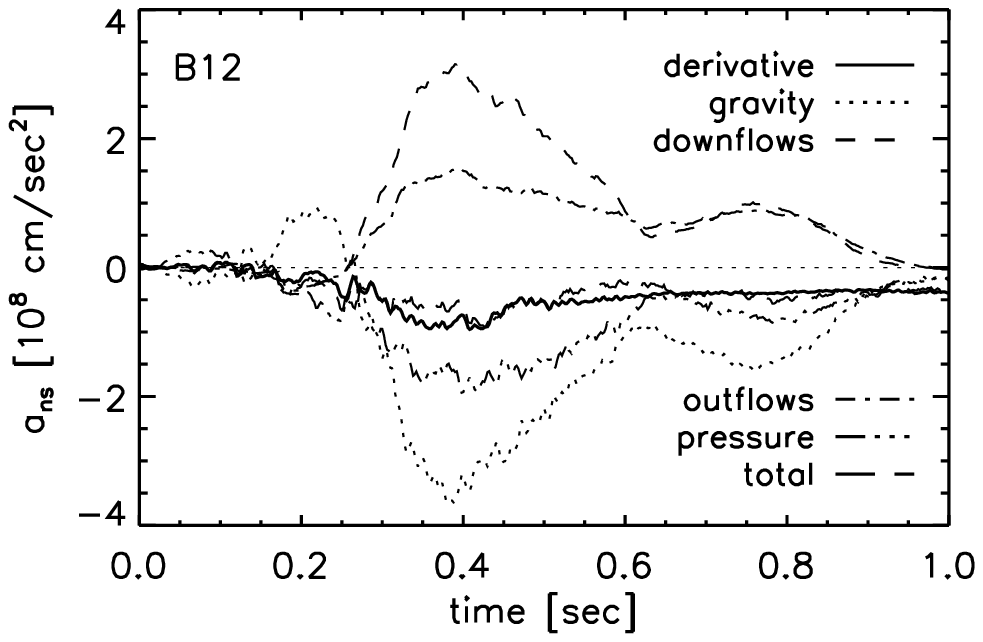}\\
\includegraphics[angle=0,width=8.5cm]{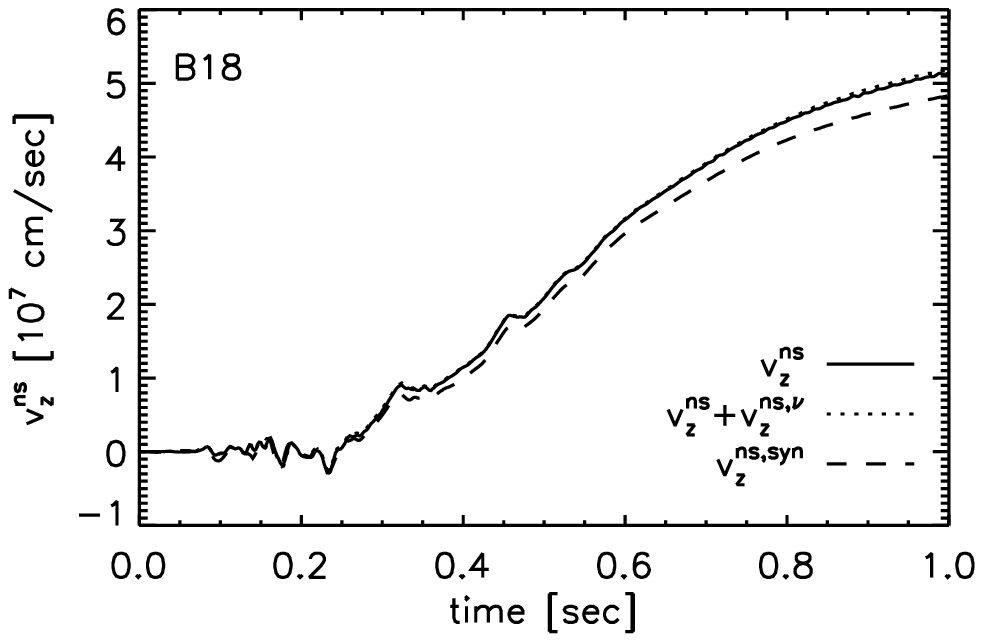}&
\includegraphics[angle=0,width=8.5cm]{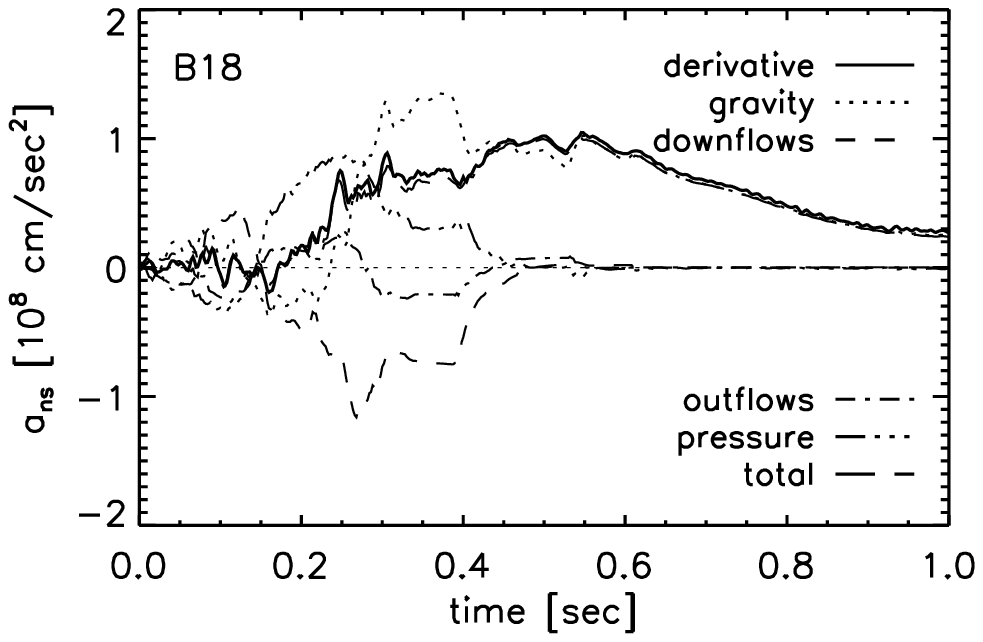}
\end{tabular}
\caption{{\em Left:}\/ Evolution of the neutron star velocities for
  Models B12 and B18. The solid curve is the neutron star recoil
  velocity derived from total momentum conservation of gas and neutron
  star (see Eq.~\ref{eq:vns_pgas}). The dotted curve includes
  corrections due to anisotropic neutrino emission. The dashed curve
  is an estimate obtained by integrating Eq.~(\ref{eq:ans_est3d}) over
  time. {\em Right:}\/ Evolution of the neutron star acceleration
  (solid curve), as computed from the (numerical) time derivative of
  the solid curve shown in the velocity plots on the left side. Also
  shown are the individual terms of Eq.~(\ref{eq:ans_est3d}),
  corresponding to momentum transfer due to downflows, outflows (e.g.
  in the neutrino-driven wind), anisotropic pressure distribution
  around the neutron star, and gravitational pull of the anisotropic
  ejecta.  The sum of these terms (the long-dashed curve labelled
  ``total'') agrees well with the (solid) curve obtained independently
  from total momentum conservation applied to the hydrodynamics
  results.}
\label{fig:nsvel}
\end{figure*}

If a single downflow has formed, e.g., in the southern hemisphere,
the expansion of the ejecta will be hampered there. On the other hand
it will proceed unaffected in the northern hemisphere, and thus
$|P_{z,\mathrm{gas}}^{\rm N}|$ will be larger than
$|P_{z,\mathrm{gas}}^{\rm S}|$. Hence, $\Pzgas$ will be dominated by
$P_{z,\mathrm{gas}}^{\rm N}$ (which is positive since all of the gas
in this hemisphere is moving outwards). According to
Eq.~(\ref{eq:vns_pgas}), the neutron star must then be accelerated in
the negative $z$-direction, i.e. towards the (southern) hemisphere
which contains the downflow. This is the situation that ultimately
establishes in Model B12 (compare Fig.~\ref{fig:stot} and
Table~\ref{tab:momb12b18}), and which is also illustrated in the right
panel of Fig.~\ref{fig:mombal}. In this case the neutron star has
attained a velocity of $\vzns=-389\,\mathrm{km/s}$ one second after
core bounce, and is still accelerated with
$\azns=-372\,\mathrm{km/s^2}$ (Fig.~\ref{fig:nsvel}).

\begin{table}
\centering
\caption{Integrated momenta of the ejecta in the northern
  ($\theta<\pi/2$) and southern hemispheres as well as their
  sum, $\Pzgas$, and the resulting neutron star recoil velocity,
  $\vzns$, at $t=1\,$s.}
\begin{tabular}{lrrrr}
\hline
\hline
Model & $P_{z,\mathrm{gas}}^{\rm N} \; [\frac{\rm g\,cm}{\rm s}]$  & $P_{z,\mathrm{gas}}^{\rm S} \; [\frac{\rm g\,cm}{\rm s}]$
       & $\Pzgas \; [\frac{\rm g\,cm}{\rm s}]$ & $\vzns \; [\frac{\rm km}{\rm s}]$ \\
\hline
B12    & $1.26 \times 10^{41}$ & $-0.20 \times 10^{41}$ & $ 1.06 \times 10^{41}$ & $-389.3$ \\
B18    & $1.77 \times 10^{41}$ & $-3.07 \times 10^{41}$ & $-1.30 \times 10^{41}$ & $ 515.1$ \\
\hline       
\end{tabular}
\label{tab:momb12b18}
\end{table}

Model B18 also develops a single downflow, which, however, is located
in the northern hemisphere, rather close to the equator. Although this
model is clearly less anisotropic than Model B12 (which is dominated
by an $l=1$ mode), the larger explosion energy and faster expansion
result in a larger $|\Pzgas|$ (Table~\ref{tab:momb12b18}). This leads
to a larger neutron star kick of $\vzns = 515\,\mathrm{km/s}$ at
$t=1\,$s, while the acceleration at this time is
$\azns=290\,\mathrm{km/s^2}$ (Fig.~\ref{fig:nsvel}). Note that these
values are comparable to the mean pulsar birth velocities derived from
observations (see Sect.~\ref{sec:longterm}), and that they are
considerably higher than those found in earlier simulations
\citep[e.g.][]{JM94}. This is mainly due to the low-order modes in our
calculations, which result in a larger gas momentum anisotropy,
$\alphag$ (cf.\ Eq.~\ref{eq:def_alphag} for a definition), compared to
previous work.

The neutron star velocities shown in Fig.~\ref{fig:nsvel} (left
panels), as well as their time-derivatives labelled with ``derivative''
and plotted with solid lines in the acceleration plots of
Fig.~\ref{fig:nsvel} (right panels), are calculated from the
simulation data with our standard post-processing approach by assuming
total momentum conservation (see Appendix~\ref{app:definitions}). The
use of this approach requires a justification, because, numerically,
energy and momentum might not be strictly conserved (i.e. up to
machine accuracy)\footnote{The energy and momentum conservation
properties of neutrino-hydrodynamic codes like the employed one are
discussed in much detail in \cite{RJ02} and
\cite{Marek+06}.}. Moreover, momentum conservation can be guaranteed
analytically only if the gravitational potential can be written as the
solution of a Poisson equation. This is, of course, the case for
Newtonian gravity. Yet, for the ``general relativistic potential'' of
\cite{RJ02} that we used in the simulations, an equivalent of the
Poisson equation cannot be derived \citep{Marek+06}.

Since the neutron star kicks discussed in this work depend on the
anisotropic distribution of the ejected gas, we do not expect that the
small general relativistic corrections or numerical errors of the
mentioned kind can seriously affect the results of our calculations to
an extent that unrealistically large values for the neutron star
recoil velocities are obtained. This expectation is supported by the
fact that we find similarly large neutron star kicks in simulations
with Newtonian gravity (see Sect.~\ref{sec:fastcon}).  Since the NS
recoil estimated from our simulations is a consequence of the
anisotropic ejection of mass in the explosion, it is also unlikely to
be linked to nonconservation of energy and/or momentum on a small
level. In order to provide additional and independent evidence that
the neutron star velocities estimated on grounds of the assumption of
total momentum conservation are reliable, we check them by verifying
the estimated neutron star acceleration as a sum of the different
forces which contribute to a momentum transfer to the neutron star.

For this purpose we consider a sphere of radius $r_0 \approx 1.1 \Rns$
that encloses the neutron star. The time-derivative of the neutron
star momentum (and hence the neutron star acceleration at a certain
time) can then be obtained by integrating the Euler equation over the
volume of that sphere, resulting in
\begin{equation}
  \dot{\vec{P}}_{\rm ns} \approx 
   -\oint_{r=r_0} {\cal P} \; {\rm d} \vec{S} -
    \oint_{r=r_0} \rho\vvec \, v_r \,{\rm d}S +
    \int_{r>r_0} G  \Mns \, \frac{\rvec}{r^3} \; {\rm d}m.
\label{eq:ans_est3d}
\end{equation}
Here the individual terms account for the varying pressure ${\cal P}$
around the sphere, the flux of momentum through the surface of the
sphere, and the gravitational acceleration due to the anisotropic
matter distribution outside the sphere. For the gravitational term we
assume that the matter distribution inside the sphere is spherically
symmetric and that the gravitational potential is \emph{Newtonian}.

The time evolution of the acceleration corresponding to these terms,
calculated from the data of Models B12 and B18, is shown in the right
panels of Fig.~\ref{fig:nsvel}. Here the second term has been split
into two components associated with momentum flux into (``downflows'')
and out of the sphere (``outflows''). Also displayed is the sum of all
terms (labelled by ``total''). Integration over time of the latter
quantity yields the dashed velocity curve for $v_z^{\rm ns,syn}$ in
the left panels of Fig.~\ref{fig:nsvel}. This should be compared to
the solid curve ($\vzns$) which was computed with our standard
post-processing approach of the gas momentum (and which includes the
effects due to general relativistic corrections). It is evident that
there are only small differences between both results, which are
significantly smaller than 10\%.  This demonstrates that the flow
morphology indeed produces an anisotropic momentum transfer to the
nascent NS, which is the cause of the estimated NS velocities.

\begin{figure*}[tpb!]
\centering
\begin{tabular}{cc}
\includegraphics[width=8.5cm]{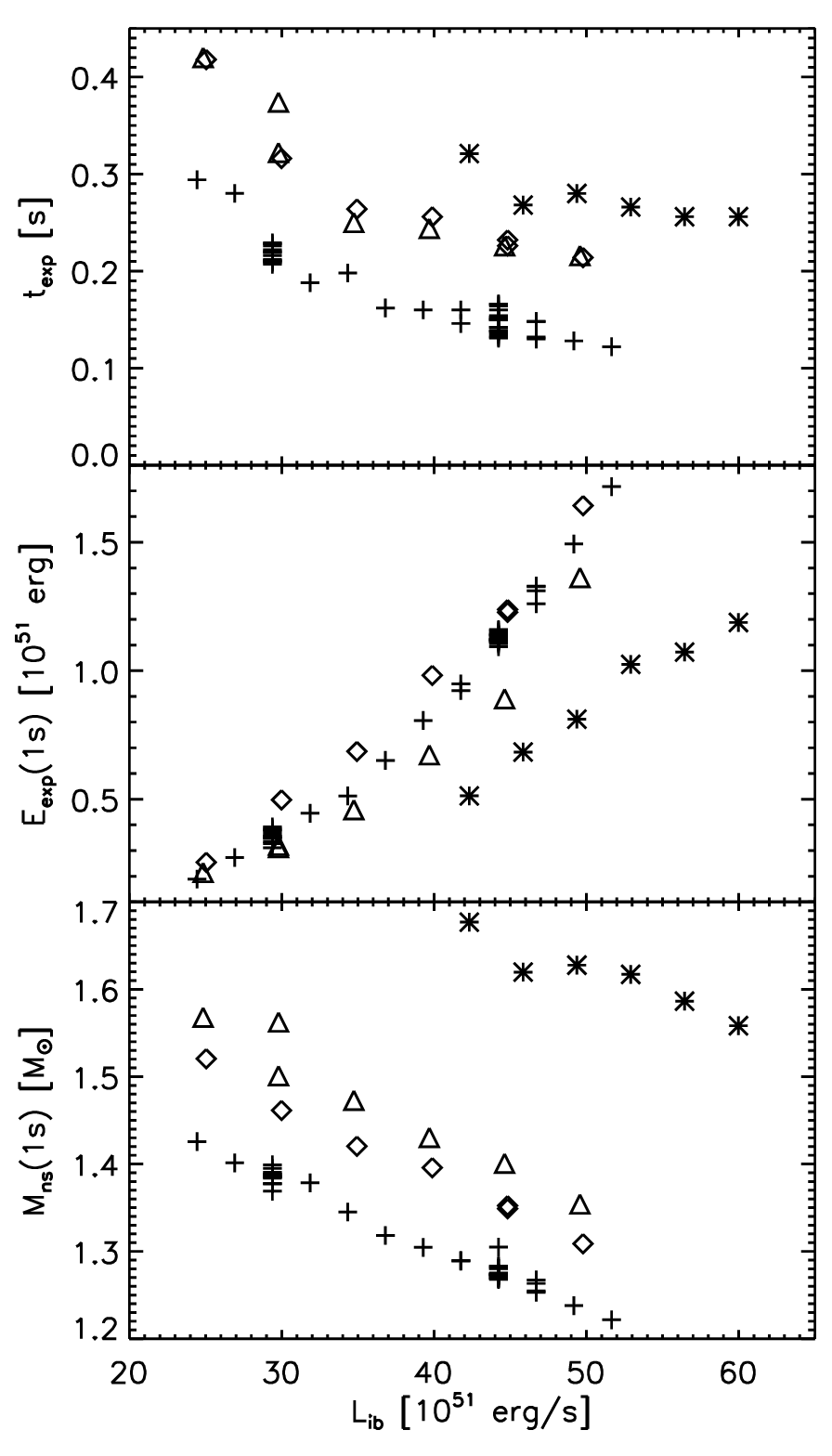} &
\includegraphics[width=8.5cm]{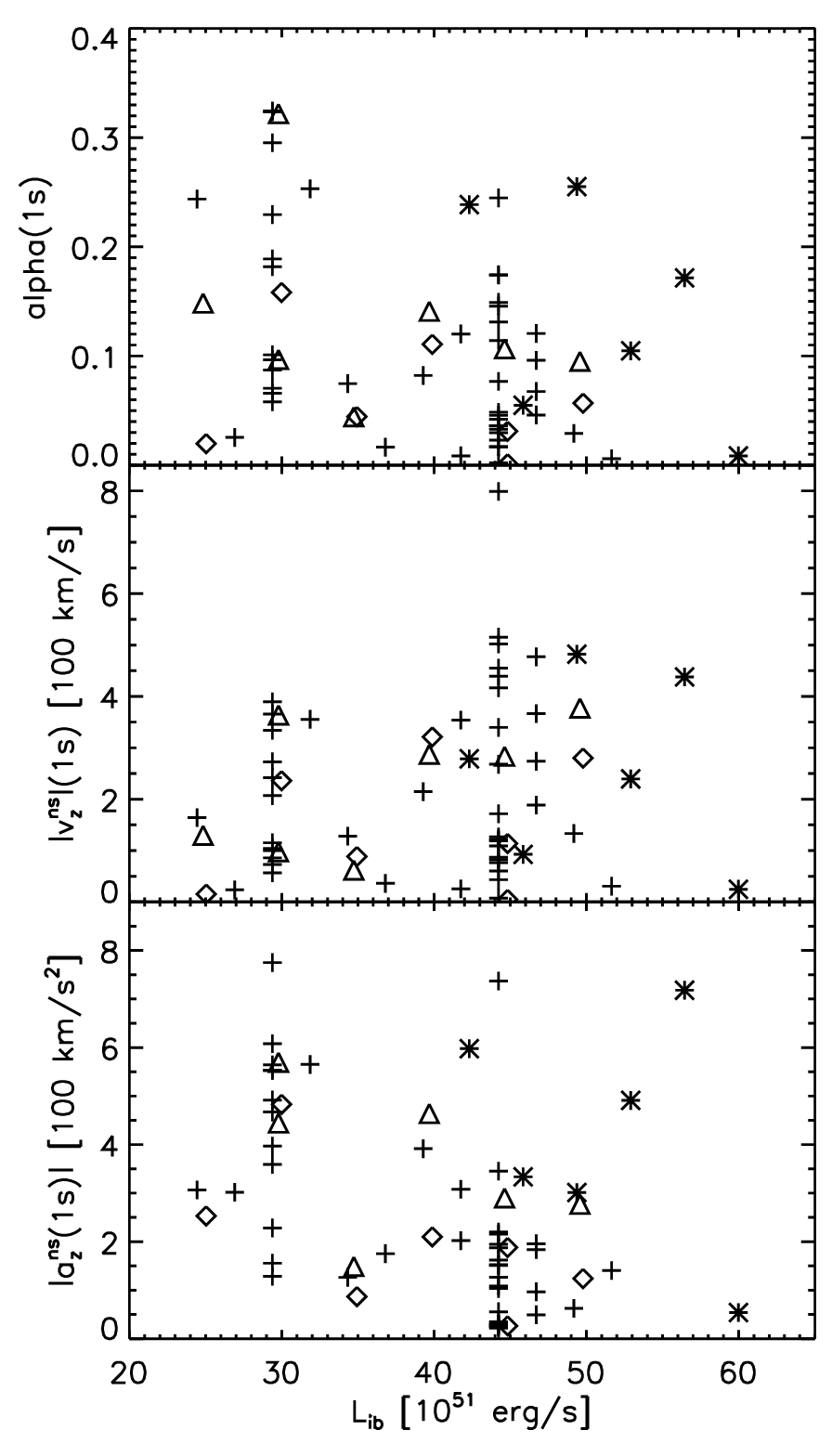}
\end{tabular}
\caption{Dependence of some global quantities on the inner boundary
  luminosity. The quantities in the left column (explosion time scale,
  explosion energy, and neutron star mass) depend only on the
  progenitor and the boundary conditions. The quantities in the right
  column (anisotropy, neutron star velocity, and acceleration) are
  strongly influenced by the initial perturbations. All time-dependent
  quantities are shown at $t=1s$. Crosses stand for the B-series of
  models, stars mark results for the L-series, triangles denote the
  W-series, and diamonds refer to the R-series of models (see
  Sect.~\ref{sec:inimod} for the differences between these models).}
\label{fig:all-over-l_ena_ib}
\end{figure*}

An interesting implication of Fig.~\ref{fig:nsvel} is the fact that
the largest contribution to the acceleration is, in general, due to
the gravitational term. In certain evolutionary phases also the other
terms may contribute significantly. Yet, the total acceleration points
nearly always in the same direction as the gravitational pull.
Momentum transfer by pressure and gas flow (the first and the second
term in Eq.~\ref{eq:ans_est3d}) are only important as long as the
inhomogeneous ejecta have sonic contact with the neutron star and thus
can exert hydrodynamic forces on the central object. This is the
situation found in Model B18 for times before $t\approx0.5\,$s.  After
that time the supersonic neutrino-driven wind, which is very strong in
this energetic model (due to the high neutrino luminosities) has blown
away the accretion downflows from the neutron star. Ongoing
acceleration is then exclusively caused by the gravitational pull of
the anisotropic ejecta and decreases slowly as the nearly spherically
symmetric wind clears the surroundings of the neutron star.
Hydrodynamic forces therefore do not contribute at later times in
Model B18. On the other hand, they are important at all times in Model
B12. The acceleration due to the momentum flux associated with the
narrow downflows that reach the neutron star is usually the second
most important term, and is directed opposite to the gravitational
acceleration.  Anisotropies in the pressure distribution and wind
outflow contribute on a smaller level.

Finally, we show in Fig.~\ref{fig:nsvel} (left) with dotted lines the
neutron star velocities corrected for the effects of anisotropic
neutrino emission (see Appendix~\ref{app:definitions}). These effects
turn out to be small. For Model B12 the neutron star kick is thus
reduced by about 10\%, which is unusually large. For most of our
models (including Model B18) the corrections due to anisotropic
neutrino emission are smaller than 5\% (for more details, see
Sect.~\ref{sec:anisotropic_emission}).

\section{Dependence on the initial model 
          and the core luminosity}
\label{sec:correlations}

In this section we discuss the variation of the quantities introduced
in Appendix~\ref{app:definitions} as functions of the initial model
and a systematic variation of the imposed core neutrino luminosity
$L_{\rm ib}$. Tables~\ref{tab:restab_b}--\ref{tab:restab_movns} give
an overview. To facilitate their interpretation, we also display the
most important quantities for all models as a function of $L_{\rm ib}$
graphically in Fig.~\ref{fig:all-over-l_ena_ib}.

The results plotted in that figure show that the neutrino-driven
mechanism as computed in our models is able to account for different
key observational aspects of supernovae and neutron stars
simultaneously, provided that sufficient time is available for
low-order unstable modes to form. Typical supernova explosion
energies of about $10^{51}$\,erg, typical baryonic neutron star masses
around $1.4\,\Msol$ (actually between $1.3$ and $1.6\,\Msol$ depending
on the progenitor) \emph{and} high neutron star recoils (with a
maximum of 800\,km/s in Model B18-3 after 1\,s of post-bounce
evolution, see Table~\ref{tab:restab_b}), are obtained at the same
time.

What is also apparent is that the quantities displayed in
Fig.~\ref{fig:all-over-l_ena_ib} can be grouped in two classes, those
which show a clear correlation with the core luminosity, $\Lib$, and
those which do not. Among the former are the explosion time scale,
$\texp$, the explosion energy, $\eexp$, and the neutron star mass,
$\Mns$. For a given initial model these integral quantities show a
systematic variation with the boundary luminosity with only little
scatter. In particular, these quantities (together with the mean shock
radius) are only weakly affected by varying the random seed
perturbations in the way described in Sect.~\ref{sec:the_models} (see
Tables~\ref{tab:restab_b}--\ref{tab:restab_movns}). Among the
quantities which do not correlate with $\Lib$ are the ones that depend
on the morphology of the explosion, i.e. the anisotropy parameter,
$\alphag$, the neutron star recoil velocity, $\vzns$, and the neutron
star acceleration, $\ans$. These show a strong sensitivity to small
differences in the flow (e.g. to the initial perturbations), and hence
essentially stochastic behaviour. The large scatter of neutron star
recoil velocities for Models B18-1 to B18-6 of between $\sim 80$\,km/s
and 800\,km/s (see Table~\ref{tab:restab_b}) illustrates this clearly.

A higher luminosity, $\Lib$, from the neutron star core causes the
explosion to develop faster, to become more energetic, and to leave
behind a neutron star with a smaller mass, because less material can
be accreted onto the core when the explosion occurs faster. The
monotonic correlation between $\Lib$ and the explosion energy $\eexp$
shows that our chosen approach to parameterise our simulations can
also be interpreted as one in terms of explosion energy. In this sense
$\Lib$ and $\eexp$ can be exchanged as governing parameters.  Note,
however, that the $\Lib$-$\eexp$ relation differs between the initial
models.

\begin{figure}[tpb!]
\centering
\includegraphics[width=8.5cm]{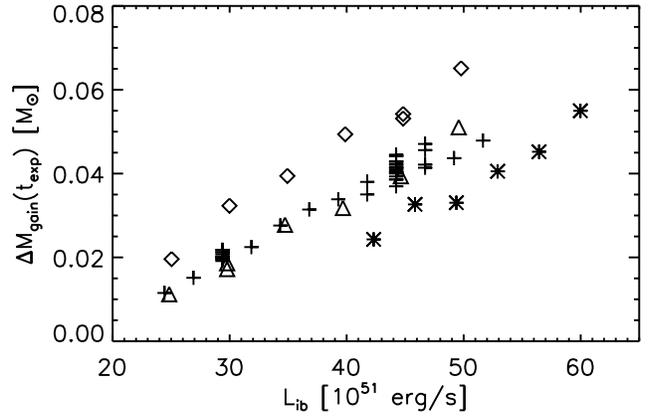}
\caption{Mass of the gain layer at the onset of the explosion
  ($\texp$) as a function of the boundary luminosity for the set of
  models displayed in Fig.~\ref{fig:all-over-l_ena_ib}. For every
  initial model there exists an approximately linear relation between
  $\Delta M_{\rm gain}$ and $\Lib$.}
\label{fig:mgain}
\end{figure}

A similar behaviour is also visible in Fig.~\ref{fig:mgain} for
$\dMg(\texp)$, the mass contained in the gain layer at time $\texp$,
as a function of $\Lib$ for all models. In fact, it is actually
$\dMg(\texp)$ which is responsible for the progenitor dependence of
the $\Lib$-$\eexp$ relation visible in
Fig.~\ref{fig:all-over-l_ena_ib}, mainly because the recombination of
free nucleons to $\alpha$ particles and nuclei in the expanding and
cooling ejecta from the gain layer yields a significant fraction of
the final explosion energy. This energy contribution increases with
more mass in the gain layer. The rest of the explosion energy is due
to the power of the neutrino-driven wind of the proto-neutron star
(see Appendix~\ref{app:eexp}). Since $\dMg(\texp)$ depends on the mass
accretion rate through the shock, there is a dependence on the density
profile of the progenitor star.  The different initial models reveal
significant differences in this respect. In particular, the
\citeauthor{Limongi+00} progenitor exhibits considerably higher
densities at the edge of the iron core and in the silicon shell than
the Woosley et al. models, but this progenitor explodes later and thus
at a time when the mass accretion rate has already decreased
significantly.

It should be noted that rotation will also affect $\dMg(\texp)$ (see
Sect.~\ref{sec:rotation}). The systematically larger mass of the gain
layer (Fig.~\ref{fig:mgain}), and the up to $\sim 50\%$ higher
explosion energies of the rotating models compared to the non-rotating
models of the s15s7b2 progenitor (Fig.~\ref{fig:all-over-l_ena_ib}),
though, are strongly affected by the larger initial perturbations that
we have used in the rotating case (see Sects.~\ref{sec:inimod} and
\ref{sec:rotation}).

A progenitor dependence is also visible in case of $\texp$ and $\Mns$
as a function of $\Lib$, as displayed in the left column of
Fig.~\ref{fig:all-over-l_ena_ib}. The simulations that are based on
the newer $15\,\Msol$ progenitor model s15s7b2 of \cite{WW95} give
explosion time scales that are systematically higher by $\sim 30\%$,
and final neutron star masses that are higher by $\sim 10\%$ than
those of the older \cite{WPE88} core. On the other hand, the results
belonging to the \cite{Limongi+00} progenitor again exhibit larger
systematic deviations from those for the Woosley et al. stars. The
higher mass accretion rate in simulations with the
\citeauthor{Limongi+00} progenitor delays the development of
convective motions, and thus the onset of the explosion ($\texp$)
compared to the other models. This prolongs the time that the revived
bounce-shock needs to reach a certain radius. It also reduces the
explosion energy, and leads to a larger neutron star mass, for a given
value of the boundary luminosity $\Lib$.

\begin{figure}[tpb!]
\centering
\includegraphics[width=8.5cm]{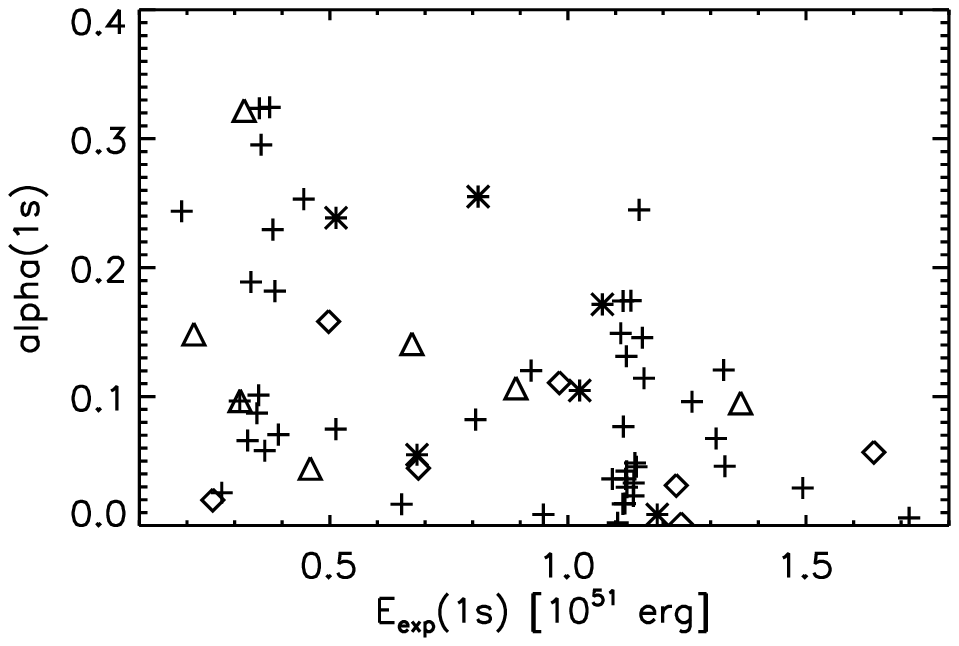}
\includegraphics[width=8.5cm]{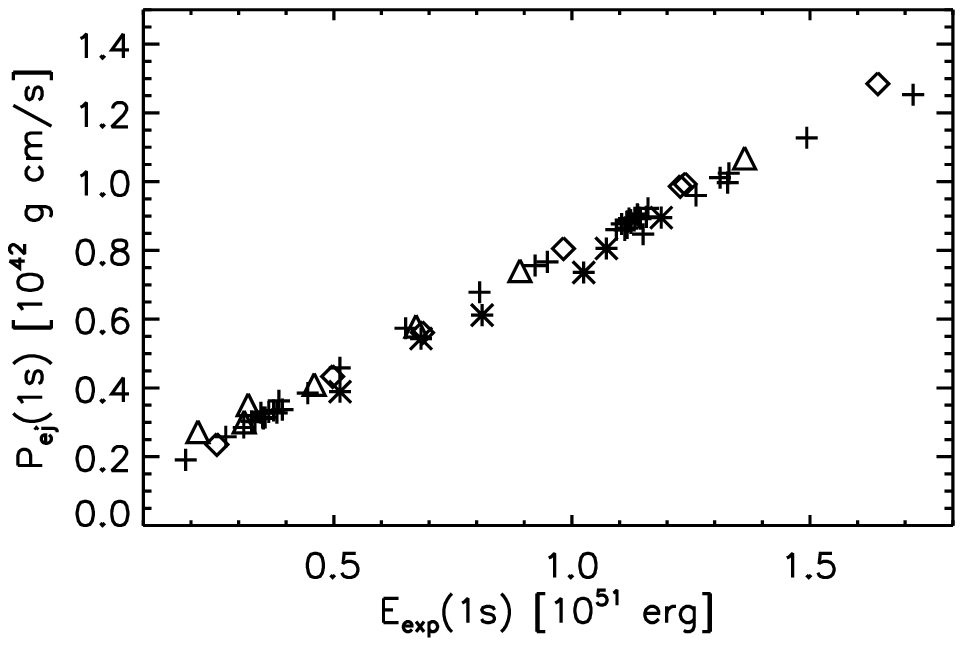}  
\caption{Anisotropy parameter $\alphag$ (upper panel) and (scalar)
  quasi-momentum of the ejecta, $\Pej$, (lower panel, see
  Eq.~\ref{eq:def_pej}) for a time of 1 second after core bounce as a
  function of the explosion energy. The different symbols have the
  same meaning as in Fig.~\ref{fig:all-over-l_ena_ib}. }
\label{fig:alpha-v_ns-e_exp}
\end{figure}

\begin{figure}[tpb!]
\centering
\includegraphics[width=8.5cm]{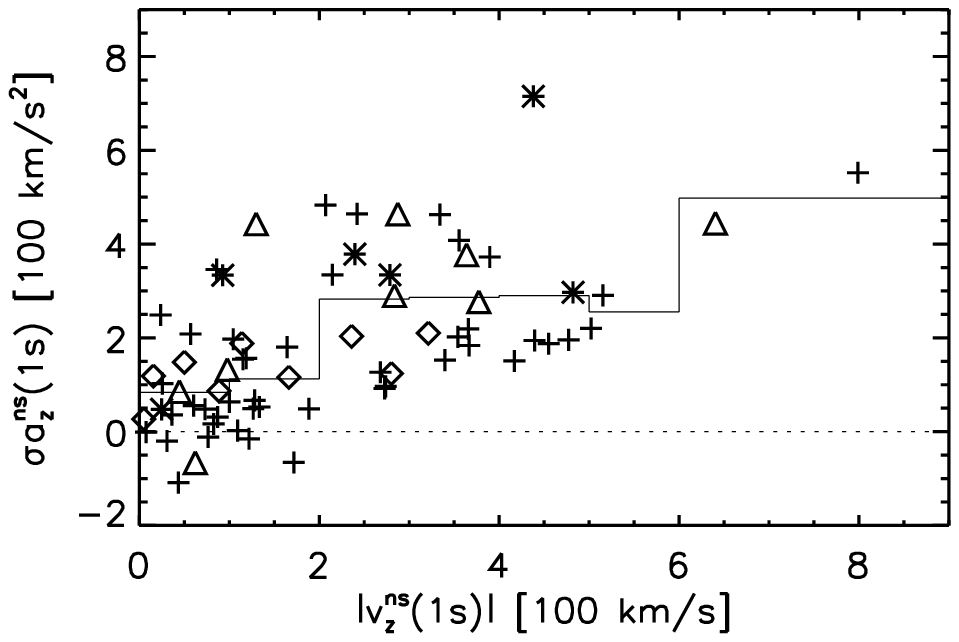}
\includegraphics[width=8.5cm]{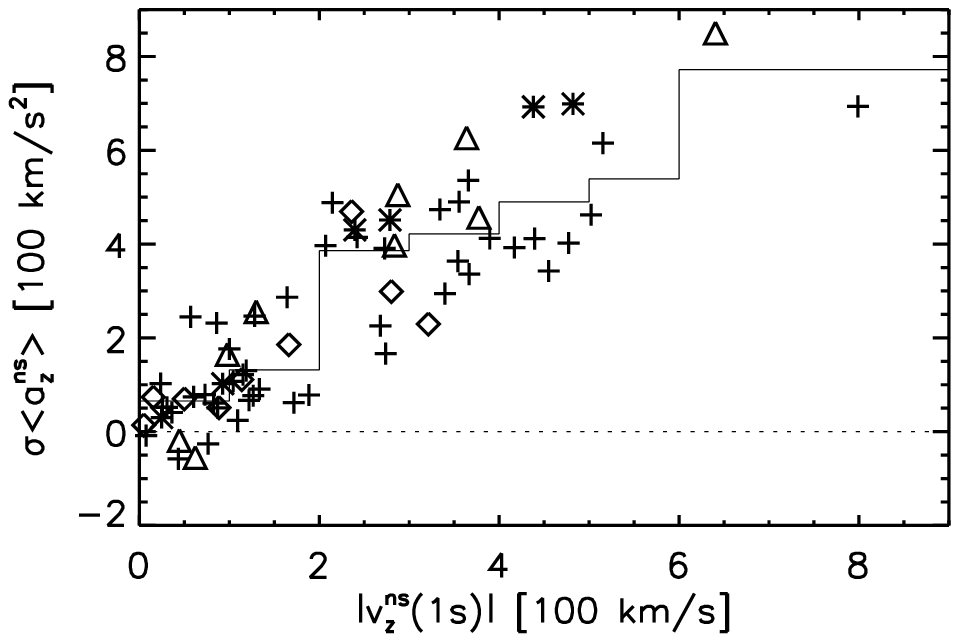}
\caption{{\em Top:}\/ Neutron star acceleration as a function of the
  neutron star velocity after one second. {\em Bottom:}\/ Acceleration
  computed as time-averaged value over the last half of a second of
  the simulations versus neutron star velocity.  The acceleration is
  multiplied by a factor $\sigma=\mathrm{sign}(\vzns)$, i.e.
  $\sigma\langle\azns\rangle<0$ corresponds to a deceleration of the
  neutron star.  The different symbols have the same meaning as in
  Fig.~\ref{fig:all-over-l_ena_ib}.  Typically, low values of the
  acceleration ($\sigma\langle\azns\rangle \lesssim
  250\,\mathrm{km/s^2}$) are associated with low velocities
  ($|\vzns|<200\,\mathrm{km/s}$), while much higher values of
  $\sigma\langle\azns\rangle$ are reached for higher velocities
  $|\vzns|$. This suggests two components of the distribution, one
  with low velocities and lower average acceleration values and one
  with both values being higher. The thin solid line indicates the
  mean values of $\sigma\langle\azns\rangle$, binned in velocity
  intervals of $100\,$km/s.}
\label{fig:vzns_azns_eexp2}
\end{figure}

We focus now on the right column of Fig.~\ref{fig:all-over-l_ena_ib}.
Recalling the highly nonlinear, chaotic hydrodynamic evolution in
response to a variation of the initial perturbations described in
Sect.~\ref{sec:flow_character}, one can understand that there is no
clear correlation between $\Lib$ and the quantities $\alphag$,
$\vzns$, and $\azns$, which depend on the explosion morphology. When,
however, $\alphag$ is plotted as a function of the explosion energy
(see Fig.~\ref{fig:alpha-v_ns-e_exp}), it becomes apparent that the
area near the upper right corner in the $\alphag$--$\eexp$ diagram,
satisfying
\begin{equation}
  \alphag \, / \, \alpha_0 \; + \; \eexp \, / \, \eexpn \; > \; 1
\end{equation}
with $\eexpn \approx 2 \times 10^{51}{\rm erg}$ and $\alpha_0 \approx
0.3$, is almost void. This indicates that high-energy explosions with
large anisotropies are disfavoured, which is plausible because there
is not sufficient time available for high-order modes to merge.  In
order to assess the impact of this result on the neutron star recoil
by virtue of Eq.~(\ref{eq:rel_alphag_vns}), we need to consider also
the scalar quantity $\Pej$, which is defined in
Eq.~(\ref{eq:def_pej}). Figure~\ref{fig:alpha-v_ns-e_exp} shows that
it is linearly increasing with the explosion energy.  Since $|\vzns|
\propto \alphag \Pej$, this increase of $\Pej$ with $\eexp$ will tend
to compensate the smaller values of $\alphag$ for higher explosion
energies. Therefore high neutron star velocities (up to
$800\,\mathrm{km/s}$ at $t=1\,$s) can result for a wide range of
explosion energies, or, equivalently, boundary luminosities
(cf. Fig.~\ref{fig:all-over-l_ena_ib}). Indeed we see that neither
bipolar oscillations nor the dominance of an $l=1$ mode are excluded
when the explosion energy is moderately large. We expect, however,
that for sufficiently large boundary luminosities the explosion time
scale, and correspondingly $\alphag$, will become so small that the
neutron star velocities will remain low for (very) large explosion
energies.

An important result of the present work is that neutron stars which
have attained high velocities at $t=1\,$s typically experience very
high accelerations, too (reaching up to more than
$700\,\mathrm{km/s^2}$). This becomes apparent in the panels of
Fig.~\ref{fig:vzns_azns_eexp2}, which display the acceleration at the
end of our simulations (top) or averaged over the last half of a
second, respectively, as a function of the neutron star recoil
velocity. In fact two populations of models may be discriminated, a
low-velocity, low-acceleration component and a second component
extending to much higher accelerations and velocities. The latter
contains simulations with a strong contribution of the $l=1$ mode,
whereas the former is made up of models in which $l=2$ or higher modes
are dominant. Since in many of the simulations the accelerations are
still high at $t=1$\,s, one can expect that their neutron star recoil
velocities will significantly increase at still later times. We will
discuss this in Sect.~\ref{sec:longterm}.

\section{The effects of rotation}
\label{sec:rotation}

We have shown that the magnitude of the neutron star recoil depends
sensitively on the convective mode. Here we will consider the
influence of rotation, which can have an effect on the pattern of
convection via the H{\o}iland condition, which states that the flow is
stable to convection if
\begin{align}
\label{eq:hoi1}
  \CHone & := \CS + \CL \\
         & =  \frac{1}{x^3} \frac{\djz^2}{\dx} + 
              \frac{1}{\rho} \vec{a} \cdot
              \left\{ \drhodS \vec{\nabla} S + 
              \drhodYe \vec{\nabla} Y_e \right\} > 0, \nonumber
\end{align}
holds \citep[see e.g.][]{Tassoul78}. Here $\vec{a}$ is the
total (gravitational and centrifugal) acceleration, and $j_z$ is the
specific angular momentum ($j_z = x \cdot v_{\phi}$, where $x = r \sin
\theta$ is the distance from the axis of rotation). In the
non-rotating case the condition of Eq.~(\ref{eq:hoi1}) reduces to the
familiar Ledoux criterion for stability, $\CL>0$, whereas for
negligible entropy- and $\Ye$-gradients Eq.~(\ref{eq:hoi1}) becomes
the Solberg-condition $\CS>0$.

In order to investigate how rotation changes the morphology, the
energetics of the explosion, and the neutron star recoil velocities,
we have computed the R-series of our models. These models start from a
post-bounce configuration with a perturbation amplitude of several
percent (cf. Sect.~\ref{sec:inimod}), which is more than an order of
magnitude larger than the standard perturbations that we employed in
our non-rotating models. Such a large increase of the perturbation
amplitude leads to noticeable changes in the explosion time scale and
energy. A clean discussion of rotationally induced effects therefore
requires recomputing some of the non-rotating models with a higher
amplitude of the initial random perturbations. We do this in case of
Models W12-c and W18-c (see Table~\ref{tab:restab_rw1218c}), in which
the same initial perturbations are applied as in Models R12-c and
R18-c, whose results are also listed in
Table~\ref{tab:restab_rw1218c}.

\subsection{Evolution of the rotation rate}
\label{sec:evoprot}

\begin{table*}
\begin{center}
  \caption{Rotating and non-rotating models with the same initial
    perturbations. For more details, see the caption of
    Table~\ref{tab:restab_b}.}
\label{tab:restab_rw1218c}
\input{restab-rw1218c}
\end{center}
\end{table*}

\begin{figure}[tpb!]
\centering
\includegraphics[angle=0,width=8.5cm]{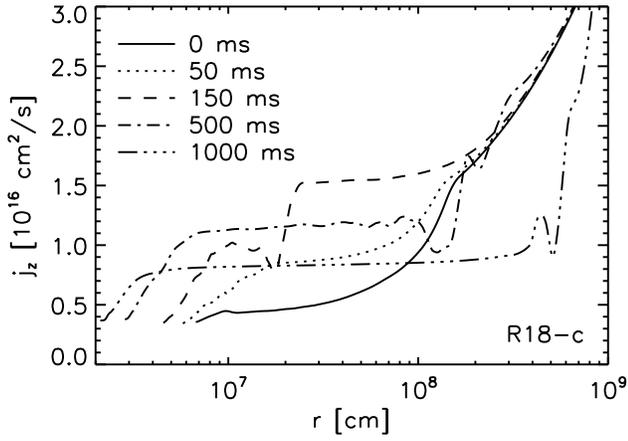}
\caption{Radial profiles of the specific angular momentum, $j_z$, at
         the equator of Model R18-c for several times after the start
         of the simulation.}
\label{fig:jz_profiles}
\end{figure}

The initial rotation profile that we employ was discussed in detail by
\cite{Mueller+04} (see also Fig.~1 there). Our choice of this angular
velocity profile on the one hand maximises rotational effects in view
of the most recent evolution calculations for magnetised rotating
massive stars: it yields rotation rates that are more than a factor of
two higher in the iron core, and on average a factor of ten higher in
the silicon shell than in the calculations of \cite{Heger+04}. On the
other hand, it avoids sub-millisecond rotation of the newly formed
neutron star, which would result for rotation rates that are
significantly higher than the 0.5 rad/s with which our iron core was
assumed to rotate prior to collapse. With our ``standard'' contraction
law the proto-neutron star {\bf spins up} due to angular
momentum conservation to a maximum angular velocity of about
$8\times10^2\,{\rm rad/s}$ at one second after core bounce. This
corresponds to a rotation period of several milliseconds at the end of
our simulations (and close to $1-2$\,ms after NS contraction to a
radius of 10\,km).

\subsection{Morphology}

\begin{figure}[tpb!]
\centering
\includegraphics[angle=0,width=8.5cm]{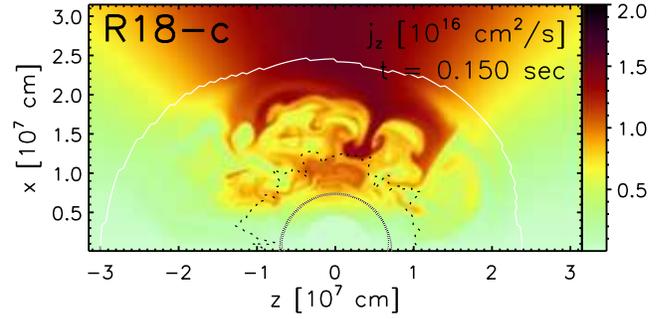}
\caption{Distribution of the specific angular momentum $j_z$ of the
  rotating model R18-c at $t=150$\,ms. Matter with larger and larger
  specific angular momentum has fallen through the shock (outer solid
  line), which leads to an overall positive gradient $\djz^2/\dx$ in
  the gain layer. However, due to convection, which is suppressed only
  near the poles, the $j_z$ stratification and its gradient are
  locally perturbed. The rotation axis is oriented horizontally.}
\label{fig:jz-dist-150}
\end{figure}

\begin{figure}[tpb!]
\centering
\begin{tabular}{cc}
\includegraphics[angle=0,width=8.5cm]{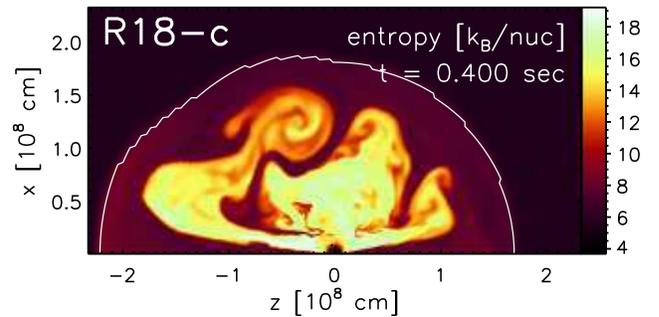} &  
\end{tabular}
\caption{Entropy distribution of the rotating Model R18-c 400\,ms
         after the start of the simulations. The white line marks the
         supernova shock. Note the two polar downflows. The rotation
         axis is oriented horizontally.}
\label{fig:rw-stot-evo}
\end{figure}

\begin{figure*}[tpb!]
\centering
\begin{tabular}{cc}
\includegraphics[width=8.5cm]{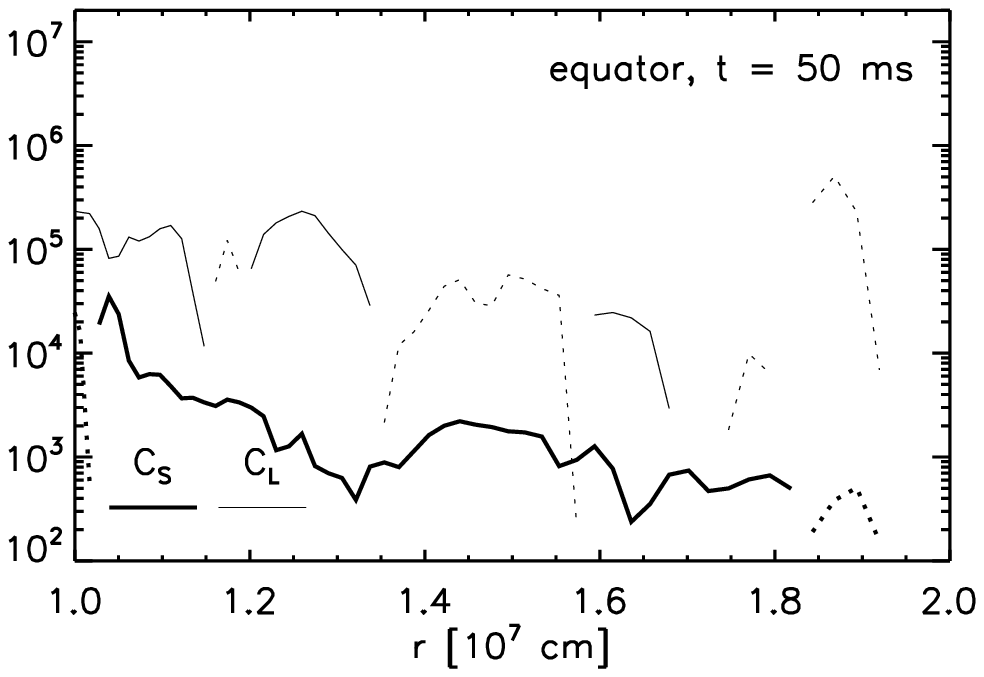}  &
\includegraphics[width=8.5cm]{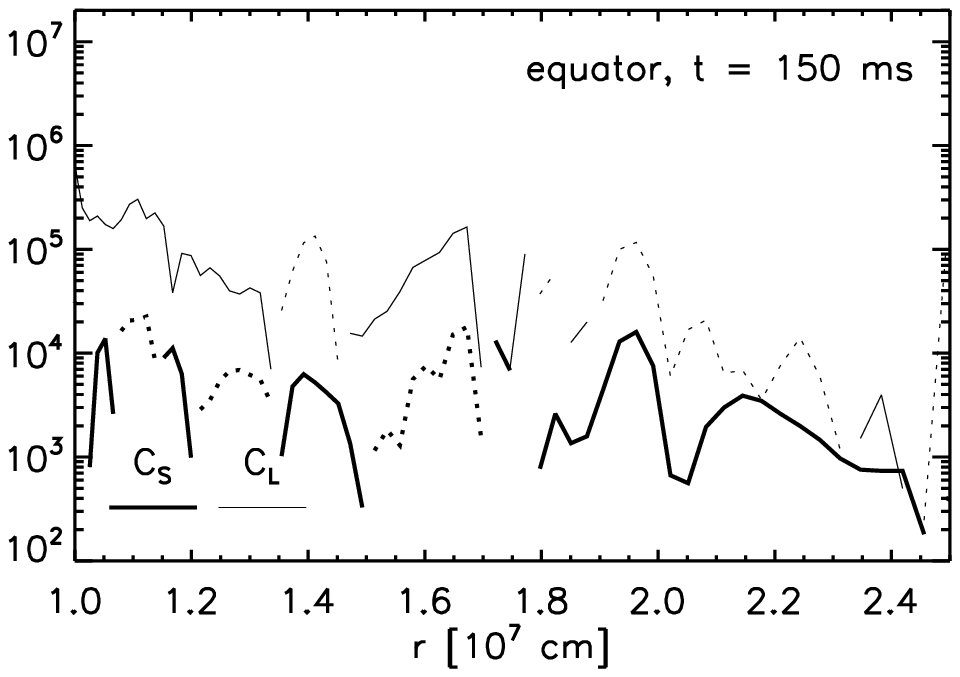} \\
\includegraphics[width=8.5cm]{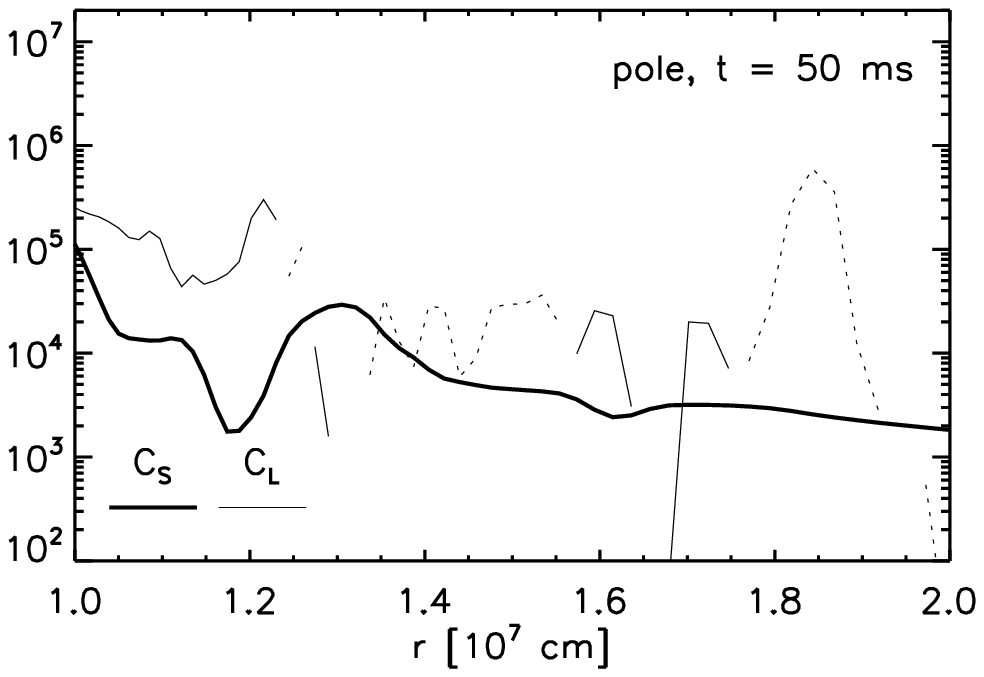}     &
\includegraphics[width=8.5cm]{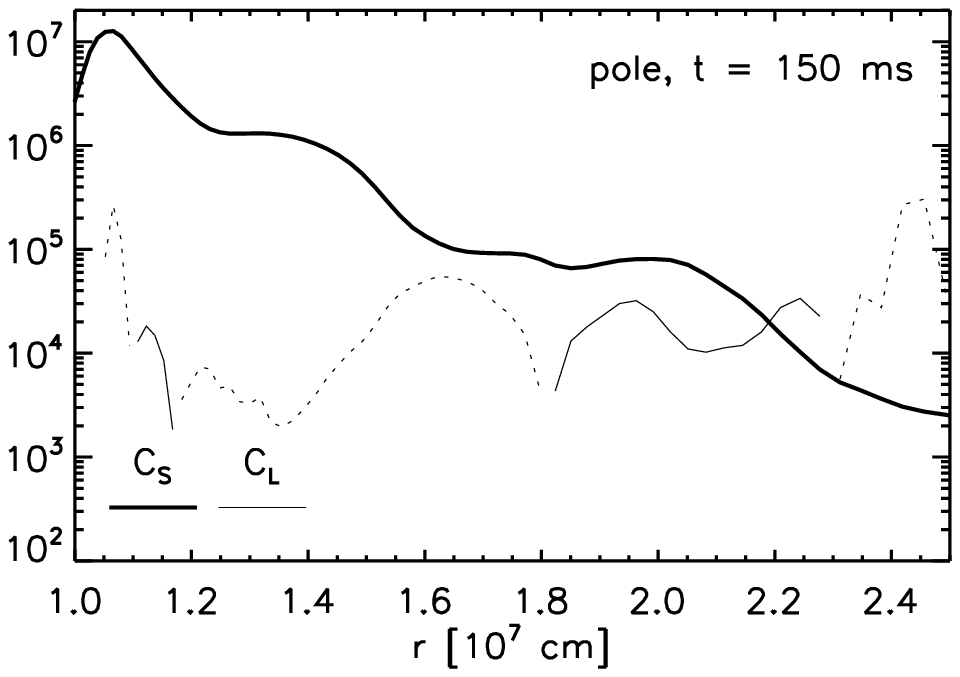}
\end{tabular}
\caption{Radial profiles of the Solberg-term, $\CS$, and of the
         Ledoux-term, $\CL$, (see Eq.~\ref{eq:hoi1}) for
         $\theta=5^{\circ}$ (``pole'') and $\theta=90^{\circ}$
         (``equator'') in Model R18-c. We show these quantities for
         $t=50\,$ms (left column) and $t=150\,$ms (right column). For
         regions in which $\CS$ or $\CL$ are negative, the absolute
         values are plotted as dotted lines. At $t=50\,$ms
         $|\CL|>|\CS|$ and unstable regions ($\CL+\CS<0$) are present
         for both latitudes. At a time of $150\,$ms the gradient
         $\djz^2/\dx$ has become sufficiently large to make
         $\CS>|\CL|$ at the pole, and thus to stabilise the flow,
         whereas in the equatorial region $|\CS|$ is still small.}
\label{fig:hoiland-profiles}
\end{figure*}

The rotating models evolve almost identically to the non-rotating ones
during the first 75\,ms after the start of the calculations.  This is
so because the Solberg-term, $\CS$, is negligible in this early
phase. The total angular momentum and the derivative of $j_z$ in the
postshock region are initially rather small
(Fig.~\ref{fig:jz_profiles}). However, the influence of the Solberg
term increases with time because there is a positive gradient of
$j_{z}$ upstream of the shock, and matter with increasingly large
specific angular momentum is advected into the postshock region (see
Figs.~\ref{fig:jz_profiles} and \ref{fig:jz-dist-150}). Therefore the
positive derivative of $j_z$ with $x$ grows within the postshock
flow. Note that, since we assume axisymmetry, there are no forces
(other than fictitious ones) acting in $\phi$ direction, and hence no
source terms for $j_{z}$ are present. The specific angular momentum of
a fluid element therefore remains constant, and $j_z$ is simply
carried along with the flow.

For $t>75\,$ms this causes the Solberg term to become sufficiently
large so that it affects the pattern of convection and thus leads to
differences compared to the non-rotating case: All the rotating models
develop downflows at both poles, whereas there is no preference for
the formation of polar downflows in the non-rotating models (compare
Fig.~\ref{fig:rw-stot-evo} with Figs.~\ref{fig:stot} and
\ref{fig:b18_rseed}). These polar downflows remain stable until they
are blown away from the vicinity of the neutron star by the
neutrino-driven wind. The stabilisation is caused by the positive
$x$-derivative of $j_z^2$ in the Solberg term, which is amplified by
the factor $1/x^3$ near the axis of rotation. Given a positive
derivative of $j_z^2$, a matter element pushed towards the axis feels
a larger centrifugal acceleration $a_c = j_z^2 / x^3$ than the
surrounding matter, and therefore moves back to its original position.
Analogously, a fluid element pushed away from the axis feels a
restoring force as well.  Thus, perturbations perpendicular to the
axis are suppressed and perturbations of a gas configuration in
rotational equilibrium can only grow parallel to the axis of rotation.

For $t>75\,$ms this stabilising effect of the positive angular
momentum derivative becomes sufficiently large to suppress convection
near the axis of rotation, i.e. to make $\CHone=\CS+\CL>0$ there. In
the rest of the postshock flow the Solberg term is negligible (because
of its dependence on $x^{-3}$) compared to the Ledoux term
(i.e. $|\CS| \ll |\CL|$) and convection is not affected much.  Radial
profiles of $\CS$ and $\CL$ illustrating this situation are shown in
Fig.~\ref{fig:hoiland-profiles}.

The fact that only polar downflows and no polar outflows form can also
be easily explained. Material inside a polar downflow always consists
of the lowest $j_z$-gas which is advected through the shock (see
Fig.~\ref{fig:jz-dist-150}). This guarantees a stable situation
because the angular momentum derivative with $x$ remains positive. In
contrast, a polar outflow, i.e. a rising polar bubble, would contain
postshock matter that would be rather well mixed, because a
convective plume encompasses matter from a larger range of latitudes.
Therefore such a polar bubble would not consist of gas with a lower
$j_z$ than the infalling material near the poles that surrounds such a
bubble. This situation would therefore be unstable due to the absence
of a positive derivative $\djz^2/\dx$.

\begin{figure}[tpb!]
\centering
\includegraphics[width=8cm]{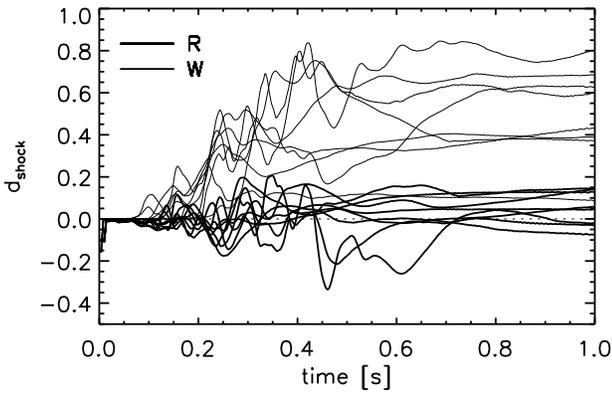}
\caption{Evolution of the shock deformation parameter $d_{\rm shock}$
        (see Eq.~\ref{eq:def_dshock}) for the rotating (R-series) and
        the non-rotating models (W-series). Positive and negative
        values of $d_{\rm shock}$ characterise oblate and prolate
        deformation of the shock, respectively.  }
\label{fig:rw-shock-def}
\end{figure}

Besides the differences in the pattern of convection another
morphological difference becomes evident: The rotating models remain
more spherical, whereas the non-rotating models in general develop a
clear prolate deformation (Fig.~\ref{fig:rw-shock-def}). This is
partly due to the polar downflows, which damp the shock expansion near
the poles. A second reason is the centrifugal acceleration of the
matter between neutron star and shock. Owing to the accumulation of
angular momentum behind the shock, the initially weak centrifugal
forces increase, and their radial components reach up to 20\% of the
gravitational acceleration. Consequently the shock is pushed out
farther in the equatorial region than in the non-rotating models. This
has interesting consequences for the explosion energy.

\subsection{Energetics}

In both rotating Models R12-c and R18-c that are listed in
Table~\ref{tab:restab_rw1218c} the explosion energies are higher and
the neutron star masses are correspondingly lower than in their
non-rotating counterparts, W12-c and W18-c, also listed in that
Table. In case of models R18-c and W18-c the energy difference amounts
to $\sim 20$\% (i.e. $0.2\times\foe$) and must be caused by rotational
effects. This difference builds up when the expanding and cooling
neutrino-heated matter in the gain layer recombines from free nucleons
to alpha particles (and partly to nuclei) and remains approximately
constant in the subsequent phase, in which the explosion energy
increases further due to the neutrino-driven wind (see
Fig.~\ref{fig:rw_eexp} and Appendix~\ref{app:eexp}). It is caused by
the larger equatorial shock radius in the rotating model R18-c and the
thus wider gain layer, which increases the recombining mass by
$0.013\,\Msol$ compared to the non-rotating case.

\subsection{Neutron star recoil}

What are the implications of the morphological differences between
rotating and non-rotating models for the neutron star kicks? In the
non-rotating case the highest recoil was obtained for Model B18-3, in
which a pronounced $l=1$ mode with a single polar downflow is
present. In the rotating case such a flow pattern cannot establish,
since we always obtain downflows at both poles. However, significant
asymmetries can still develop, since one of the polar accretion
funnels may be much stronger than the other, or a third downflow may
be dominating the mass distribution. High neutron star recoils are
thus not precluded, although we expect the mean and the maximum kicks
to be somewhat smaller than in the non-rotating case.

The results of our rather few simulations, which comprise only nine
rotating models (see Tables~\ref{tab:restab_r} and
\ref{tab:restab_rw1218c}), are in agreement with this expectation: The
largest neutron star recoil velocity obtained in the R-series of
models is $321\,$km/s, whereas it is $640\,$km/s in case of the
W-series (see Model W18-c in Table~\ref{tab:restab_rw1218c}). The
average kick velocities for the R- and W-type models are $151\,$km/s,
and $280\,$km/s, respectively. If one omits Model W18-c, the only
W-type model with a ``pure $l=1$ mode'', the average kick velocity of
the non-rotating models decreases to only $228\,$km/s, i.e. it is 50\%
larger than that of the rotating models. This is a relatively moderate
effect if one recalls that the initial angular velocity assumed in the
progenitor core of our calculations is clearly extreme compared to the
rotation rates obtained from the latest stellar evolution calculations
\citep{Heger+04}.

\subsection{Spin-kick alignment?}

Does rotation lead to an alignment of the kick direction with the
rotation axis (the so called ``spin-kick alignment'')? This question
cannot be conclusively answered on the basis of two-dimensional
axisymmetric simulations, because in this case the neutron star kick
is always along the rotation axis due to the assumed symmetry of the
calculations.

However in the context of our kick scenario also in the
three-dimensional case effects can be imagined which may lead to a
spin-kick alignment. On the one hand, the rotation axis is a preferred
physical direction of the system such that the development of global
anisotropies (e.g. polar accretion and outflow, bipolar oscillations)
might be favoured along this direction. On the other hand, if the
rotation period is smaller than the duration of the neutron star kick
by a one-sided, non-axial acceleration (in the corotating frame), then
any asymmetry will retain only its component parallel to the spin axis,
while the perpendicular component will be reduced or extinguished by
rotational averaging.

Our results seem to suggest that the first effect may be the more
important one. For the angular momentum present in our models there is
a tendency of anisotropies (e.g. of downflows) to develop preferably
aligned with the rotation axis. While for the relatively ``fast''
rotation of our models (in the sense discussed in
Sects.~\ref{sec:inimod} and \ref{sec:evoprot}), the second effect may
also contribute to produce spin-kick alignment, the influence of
rotational averaging will be weaker for slower and possibly more
realistic rotation. In case of ``slow'' rotation, i.e. for spin
periods of tens of milliseconds in the nascent neutron star and many
hundreds of milliseconds in the neutrino-heated convective postshock
layer (which are ten times or more larger than in our models),
rotation will be unable to enforce perfect alignment of the directions
of kick and spin.

Depending on the amount of angular momentum in the supernova core, the
hydrodynamic kick mechanism discussed in this paper therefore allows
for both possibilities, spin-kick alignment for rapid neutron star
rotation ($t_{\rm ns} \ll 1\,$s) and misalignment or incomplete
alignment for long rotation periods ($t_{\rm ns} \gtrsim$ some
$100\,$ms). This seems to be compatible with recent studies of
observational constraints on neutron star kicks for isolated pulsars
and for neutron stars in binary systems \citep{Wang+06}, although the
interpretation of observations is still ambiguous \citep{Johnston+05}.

\section{Robustness and long-time
         evolution of the neutron star recoils}
\label{sec:recoil_robustness}

We have seen above that rotation, even if it is noticeably faster than
in the most recent stellar evolution models, does not preclude neutron
star kicks of several hundred km/s. However, we have made a number of
approximations in our post-processing analysis and used assumptions in
our simulations whose impact on the neutron star recoil still needs
to be assessed. In addition, we have stopped most of our simulations at
a time of one second after core bounce, when the neutron star
acceleration was, in many cases, still high. Hence we need to comment
also on the later evolution of the kicks.  These issues are discussed
in the following.

\subsection{Anisotropic neutrino emission}
\label{sec:anisotropic_emission}

The neutron star recoil velocities, $\vzns$, that are listed in
Tables~\ref{tab:restab_b}--\ref{tab:restab_movns} are computed from
Eq.~(\ref{eq:vns_pgas}), i.e. they do not include the effects of
anisotropic neutrino emission. As we show in
Appendix~\ref{app:definitions}, anisotropic neutrino emission results
in a correction, $\vznsnu$, of the neutron star velocity which is
described by Eqs.~(\ref{eq:def_pnu}) and (\ref{eq:def_vnscorr}). In
Sect.~\ref{sec:recoil} we have already seen that this correction is
small for Models B12 and B18. This actually holds for most models.
Only in a few cases is $|\vznsnu/\vzns|>10\%$, and in most of these
cases the neutron stars have small recoil velocities (cf.
Tables~\ref{tab:restab_b}--\ref{tab:restab_movns}). The correction due
to anisotropic neutrino emission in general reduces the kick. This can
be understood from the fact that in most models a single prominent
accretion funnel is present. The neutron star recoil caused by gas
anisotropies is always directed towards this downflow, while the
neutrino emission associated with the ``hot spot'' created by the
downflow on the neutron star surface results in a ``neutrino-rocket
engine'' that kicks the neutron star in the opposite direction (this
circumstance was observed and discussed before by \citealt{Fryer04}).
However, the acceleration due to the neutrino emission remains small
because the anisotropy parameter of the accretion luminosity is
typically only a few per cent.

There are several reasons for that. On the one hand, the
neutrino-radiating tip of the accretion downstream is highly unstable
and its position varies with time, reducing the neutrino emission
anisotropy by temporal averaging. On the other hand there is a
projection factor of $\cos \theta$ of the downflow impact polar angle,
$\theta$, to be included due to the axial symmetry of our models. This
factor also reduces the kick. Finally, the time scale of neutrino
energy release from the accreted matter is typically significantly
longer (the cooling time scale is of order 10$\,$ms) than the time
scale that the gas remains compressed in the downflow tips (between
0.1 and 1$\,$ms) before it spreads around the neutron star surface.
Only very close to the lower end of the downdrafts the density of the
gas is so high ($\rho\la 10^{11}\,$g$\,$cm$^{-3}$) that the neutrino
emission is extremely large. During its violent impact on the NS
surface, the gas, however, overshoots equilibrium conditions.  Once
decelerated, it bounces back, reexpands immediately, and wraps around
the neutron star at radii considerably larger than the minimal radius
of impact. This is mainly due to the fact that the gas comes from far
out in the progenitor star and is shock heated during accretion. As a
consequence, its entropy is still considerably higher than the entropy
of the layers around and inside the neutrinosphere (remember that
neutrino cooling during the infall is too slow to cool the gas
efficiently). Therefore the gas floats and forms an essentially
spherical, high-entropy and low-density ($\rho\sim
10^{10}\,$g$\,$cm$^{-3}$) envelope that radiates neutrinos with
significantly lower rates than the dense tips of the impinging
downflows. A part of the gas is integrated in the cooling layer and in
response to the neutrino losses settles rather slowly on the NS, while
the other, higher-entropy part is added to the region outside of the
gain radius and is neutrino-heated until it is blown away again in the
neutrino-driven wind.  As a result, our models reveal that only at
most 10--15\% of the binding energy of the infalling gas in the
downflows are radiated highly anisotropically. A much larger part of
the released gravitational binding energy is not emitted in the
downflows but from the essentially spherical layer enwrapping the
nascent NS and settling on it\footnote{It should be noted that our
  transport approximation, which assumes that the transport equations
  in radial direction can be solved independently in all angular zones
  of the grid, has the tendency to overestimate the neutrino emission
  anisotropy compared to a fully multidimensional treatment. Therefore
  our ``neutrino recoil'' is likely to be an upper limit of the
  corresponding effect rather than an underestimation.}. Due to the
mass ejection in the wind, the total rate of energy loss in neutrinos
is actually significantly smaller than the rate of release of
gravitational binding energy corresponding to stationary accretion
with the mass infall rate through the downflow.

\subsection{Inertial mass of the neutron star}
\label{sec:ns_motion}

In most of our simulations we make the simplifying assumption that the
inertial mass of the neutron star is infinite, i.e. the consequences
of the neutron star motion are ignored during the hydrodynamic
simulation. This assumption is dropped in one set of models which is
listed in Table~\ref{tab:restab_movns}. In these simulations the
feedback effect of the neutron star motion is taken into account by
changing the frame of reference in every time step, thus allowing the
ejecta to move relative to the neutron star instead of following the
neutron star motion through the ambient gas (see Sect.~\ref{sec:grid}
and Appendices~\ref{app:definitions} and \ref{app:hydro_accel}).

Comparing the results obtained from both approaches for a sample of
about 30 simulations (which made use of the boundary parameters of
Models B12 and B18), one sees that any given model, all else being
equal, develops different explosion asymmetry and therefore NS kick,
although the explosion energy and time scale are very similar (see
Tables~\ref{tab:restab_b} and \ref{tab:restab_movns}).  The ensemble
distribution of kick velocities, however, shows little change, and in
particular neutron star velocities in excess of 400$\,$km/s after
1$\,$s of post-bounce evolution are found regardless of whether the
relative motion of the neutron star is included or not.

Inspecting our simulations with and without NS motion, we can actually
not discover any obvious differences caused by the moving NS (the
reader is invited to have a look at the movies for Models B12 and
B12-m6 which are provided as online material of this article).  We
think that there are a variety of reasons for that.  In the first
place, the neutron star acceleration and velocity are typically rather
small, in particular before and just after the explosion is launched
when the acceleration is still unsteady (see Figs.~\ref{fig:nsvel} and
\ref{fig:limcas_fastcon_v_ns}).  Secondly, the downflow deceleration
and impact on the NS surface are so extremely violent and create so
much sound wave and shock activity that the small effect of NS motion
cannot be discerned from other dynamical effects.  Thirdly, the
downflows and also the neutrino-driven wind at later stages are so
fast ($>\,$10\,000$\,$km/s) and their accelerations so high that the
neutron star motion even with hundreds of km/s (but still rather
modest acceleration) is only a small correction.

Since the explosions in our models are triggered by neutrino heating,
supported by violent hydrodynamic instabilities, we suspect that the
influence of the neutron star motion might just be masked and dwarfed
by other dynamics so that the explosion energy and time scale do not
reveal any visible dependence. On the other hand, the nonlinear growth
of the hydrodynamic instabilities in the shocked layer is so chaotic
that any small changes, independent of their detailed origin (e.g.,
different initial seed perturbations, different rounding errors on
different computers, different neutrino interactions, the moving
neutron star, etc.) lead to modifications of the mass and momentum
distributions at the end of our simulations. Taking into account the
NS motion by our transformation does not have any specific
consequences compared to other effects that influence randomness.

\subsection{Neutron star contraction and gravitational potential}
\label{sec:fastcon}

For practical reasons, all simulations listed in
Tables~\ref{tab:restab_b}--\ref{tab:restab_movns} and
Table~\ref{tab:restab_rw1218c} were performed with our ``standard''
prescription for the contraction of the neutron star core (see
Sect.~\ref{sec:hydro_bounds}), although the ``rapid contraction case''
also discussed in Sect.~\ref{sec:hydro_bounds} is potentially more
realistic. To study the corresponding differences, we take the
``high-perturbation'', non-rotating Model W12-c (see
Sect.~\ref{sec:rotation} and Table~\ref{tab:restab_rw1218c}) as a
reference case and perform an additional simulation, Model W12F-c, in
which we replace the slowly contracting inner boundary of Model W12-c
with the prescription for a rapidly contracting proto-neutron star.
Table~\ref{tab:restab_fastcon} compares some quantities characterising
the two models.

\begin{table*}
\begin{center}
  \caption{Important parameters of models W12-c and W12F-c.}
\label{tab:restab_fastcon}
\input{restab-fastcon}
\end{center}
\end{table*}

\begin{figure}[tpb!]
\centering
\includegraphics[angle=0,width=8.5cm]{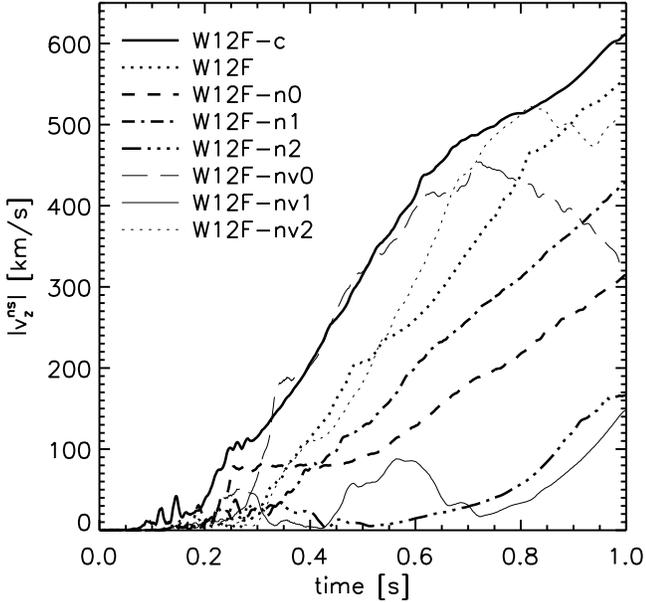}
\caption{Neutron star velocities (absolute values) as functions of
  time for Models W12F-c, W12F and several other models with fast
  neutron star contraction. In six out of eight models the neutron
  star moves faster than $300\,{\rm km/s}$ at $t=1\,$s.}
\label{fig:limcas_fastcon_v_ns}
\end{figure}

Model W12F-c explodes earlier and attains a higher explosion energy
than Model W12-c. This can be explained by the fact that for a
smaller inner boundary radius more gravitational energy is released,
and that for a shorter contraction time scale this release occurs
earlier (see also Appendix~\ref{app:eexp}).
With $\vzns(1\,{\rm s})=611\,\mathrm{km/s}$ the neutron star recoil
velocity of Model W12F-c is very high. Large kicks are also found 
in a set of simulations performed with rapid boundary contraction.
For testing this we consider for instance cases with
\begin{enumerate}
\item smaller initial random velocity perturbations of 0.1\% (Model
  W12F in Fig.~\ref{fig:limcas_fastcon_v_ns}),
\item a Newtonian gravitational potential and a constant central point
  mass chosen such that the same initial gravitational acceleration is
  obtained at a mass coordinate of $1.1~\Msol$ as in the models of
  \cite{Buras+03}, see Models W12F-n0, W12F-n1 and W12F-n2 in
  Fig.~\ref{fig:limcas_fastcon_v_ns},
\item a Newtonian gravitational potential and a varying central point
  mass, which is increased with time to reproduce the evolution of the
  gravitational acceleration at a mass coordinate of $1.1~\Msol$ in
  the models of \cite{Buras+03}, see Models W12F-nv, W12F-nv1 and
  W12F-nv2 in Fig.~\ref{fig:limcas_fastcon_v_ns}.
\end{enumerate}
All of these models have in common that they explode more quickly
than models with the standard boundary contraction. Yet, for all of
these variations we obtain at least one simulation with a neutron star
recoil velocity of more than $400\,\mathrm{km/s}$ at $t=1\,$s (see
Fig.~\ref{fig:limcas_fastcon_v_ns}). This demonstrates that a faster
neutron star contraction does not preclude high neutron star kicks and
in particular, it shows that it is {\em not the absolute value of the
time scale} for the onset of the explosion which matters. What matters
is the \emph{ratio} of the explosion time scale to the growth time
scale of low-mode anisotropies by hydrodynamic instabilities like
convection, the acoustic-vortex cycle or the SASI mechanism. With the
faster shrinking of the neutron star not only the explosion time scale
decreases, but also other important conditions change. In particular
the advection time scale in the postshock layer and the sound travel
time between shock and neutron star become shorter, because the faster
NS contraction initially leads to a smaller shock radius, too. Therefore
the velocities ahead and behind the stalled shock are higher and the
densities in the accretion layer are larger. Since the flow pattern
between shock and neutron star surface reacts and adjusts
on a hydrodynamic time scale, which is significantly shorter than
the contraction time scale of the neutron star, the growth of 
nonradial instabilities is accelerated in case of shorter dynamical
time scales in the accretion layer. Low-mode flow therefore develops 
faster (more details will be given in Scheck et al., in preparation).
Thus the faster NS contraction leads to larger global ejecta asymmetry,
in spite of faster explosions. Because of stronger neutrino heating by
the higher accretion luminosities the explosion time scales are
indeed similarly short as in the light-bulb studies of \cite{JM94,JM96}.
The previous ``burst-like'' light-bulb calculations 
were actually rather disfavorable for large global
anisotropies: Due to the high initial luminosities they produced
fast explosions, and because of the extended NS the growth of low-mode
nonradial instabilities was slow. What therefore finally matters is the
ratio of explosion time scale to low-mode growth time scale, and not
the absolute period of time in which the explosion develops.
The faster growth of non-radial instabilities 
can result in even larger values of the
anisotropy parameter $\alphag$ for ``rapid'' as compared to
``standard'' models with the same explosion energy. In other words,
the envelope in the $\alphag-\Eexp$ plane of
Fig.~\ref{fig:alpha-v_ns-e_exp} appears shifted towards larger values
of $\alphag$ for a faster contraction of the proto-neutron star, and
hence also the average recoil velocity (for a specified explosion
energy) increases.

In our largest sample of models sharing the same (slowly contracting)
boundary condition, i.e. the 18 B18-like models listed in
Tables~\ref{tab:restab_b} and \ref{tab:restab_movns}, only three
simulations develop neutron star recoil velocities of more than
$500\,{\rm km/s}$, and only seven produce neutron stars with more than
$300\,{\rm km/s}$ at 1 second.  In contrast, in just eight simulations
with rapid boundary contraction we obtain six models with neutron star
velocities of more than $300\,{\rm km/s}$ and three models with
neutron stars moving faster than $500\,{\rm km/s}$
(Fig.~\ref{fig:limcas_fastcon_v_ns}). Better statistics would require
more simulations, which should also be based on the same initial
model\footnote{The comparison between B and W models is viable,
however, because both progenitor models are quite similar.} and should
make use of the same gravitational potential.

We performed some of the simulations discussed above with
Newtonian gravity to demonstrate that the choice of the effective
relativistic potential in our models was not essential for our
results. We recall that only when we use the Newtonian gravitational
potential, momentum conservation can be expected analytically
(irrespective of numerical errors and independent of whether the point
mass is increased with time, or not). The results therefore show that
large neutron star recoil velocities are \emph{not} linked to any
violation of total momentum conservation associated with the use of
the effective relativistic potential (see the discussion in
Sect.~\ref{sec:recoil}).

\subsection{Long-time evolution of the neutron star kicks}
\label{sec:longterm}

In order to investigate how the neutron star recoil velocities evolve
beyond a time of one second after core bounce, we perform six
exemplary long-time simulations. For these we add 150 radial zones to
our grid and place the outer grid boundary at a larger radius of
$10^{10}\,$cm, which allows us to simulate the first $3$--$4\,$s of
the post-bounce evolution. In three of the simulations an infinite
inertial neutron star mass is assumed, while in the other models the
hydrodynamic feedback of the neutron star motion is taken into
account. Four of the six models are just continued from models which
we have computed up to a time of one second with our standard grid. We
map the corresponding data onto the larger grid at $t=750\,$ms and
extend the initial model profile from the old to the larger outer
boundary of the new grid.

\begin{figure}[tpb!]
\centering
\includegraphics[width=8.5cm]{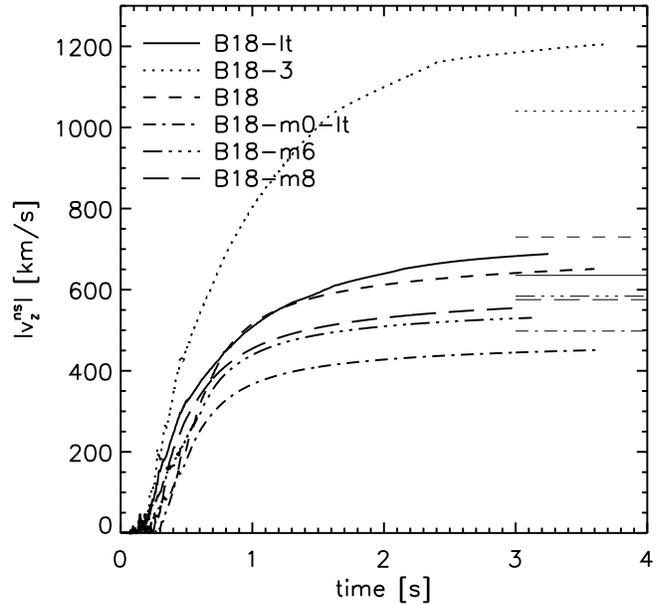}
\caption{Evolution of the neutron star velocities in six long-time
  simulations with the same boundary conditions as Model B18. After
  four seconds the acceleration has become very weak in all models and
  no significant further increase of the velocities is expected. For
  each model a thin horizontal line marks the extrapolated velocity
  value $v^{\infty}_{\rm ns}$ according to Eq.~\eqref{eq:vinfty},
  which is a rough estimate of the final neutron star velocity.}
\label{fig:vns_extrapolation}
\end{figure}

The evolution of the neutron star velocities for all of the long-time
simulations is displayed in Fig.~\ref{fig:vns_extrapolation}. The
neutron star of Model B18-3 is accelerated to more than
$1200\,\mathrm{km/s}$ within $3.7\,$s. This demonstrates that the
acceleration mechanism at work in our calculations has the potential
to explain even the highest observed pulsar velocities \citep[see
e.g.][]{Chatterjee+05}. The fact that Model B18-3 is the only one in
our sample that produces a neutron star with more than
$1000\,\mathrm{km/s}$ does not appear problematic to us. It may be a
matter of low-number statistics and might also change when more
extreme conditions are realized in models, e.g. by a faster
contraction of the neutron star than assumed in our standard set of
models. In this respect the sample of simulations plotted in
Fig.~\ref{fig:limcas_fastcon_v_ns} looks promising. In quite a number
of those the neutron stars have large velocities at one second and
also still high accelerations (see, e.g., Model W12F-c in
Table~\ref{tab:restab_fastcon}).

After $3$--$4$\,s the neutrino-driven wind has blown away all
downflows from the neutron star vicinity and has generated a nearly
spherically symmetric wind bubble around it. Therefore the neutron
star acceleration diminishes and the recoil velocities approach their
terminal values. The latter can be estimated by extrapolating the
velocities at $t=1\,$s, applying an average acceleration value
$\langle\azns\rangle$, as computed for the time interval between
$t=0.5\,$s and $1\,$s, over a time period $\Delta t_{\rm extrapol}$,
according to
\begin{equation}
v^{\infty}_{\rm ns} = \vzns(t=1\,{\rm s}) + \Delta t_{\rm extrapol}\,\times\,
                                                \langle\azns\rangle.
\label{eq:vinfty}
\end{equation}
The average acceleration $\langle\azns\rangle$ is introduced as a
time-average which is less sensitive to short-time variations and thus
allows for a more robust extrapolation of the velocities. The factor
$\Delta t_{\rm extrapol} = 0.35\,$s is ``calibrated'' by optimising
the estimates in case of the models of
Fig.~\ref{fig:vns_extrapolation}. The agreement of extrapolated and
computed terminal velocities is typically of the order of 10\%. In the
following section we use Eq.~\eqref{eq:vinfty} to estimate the final
neutron star velocities for all models listed in
Tables~\ref{tab:restab_b}--\ref{tab:restab_movns}. The basic findings
of our analysis do not depend on whether we use $\azns$ (the
acceleration values at the end of our simulations) or
$\langle\azns\rangle$ (the mean values in the last $0.5\,$s) for
extrapolating the velocities beyond the simulated period of one second
of evolution.

\section{Implications for the neutron star velocity distribution}
\label{sec:kick_distribution}

\begin{figure*}[tpb!]
\centering
\begin{tabular}{cc}
\includegraphics[width=8.5cm]{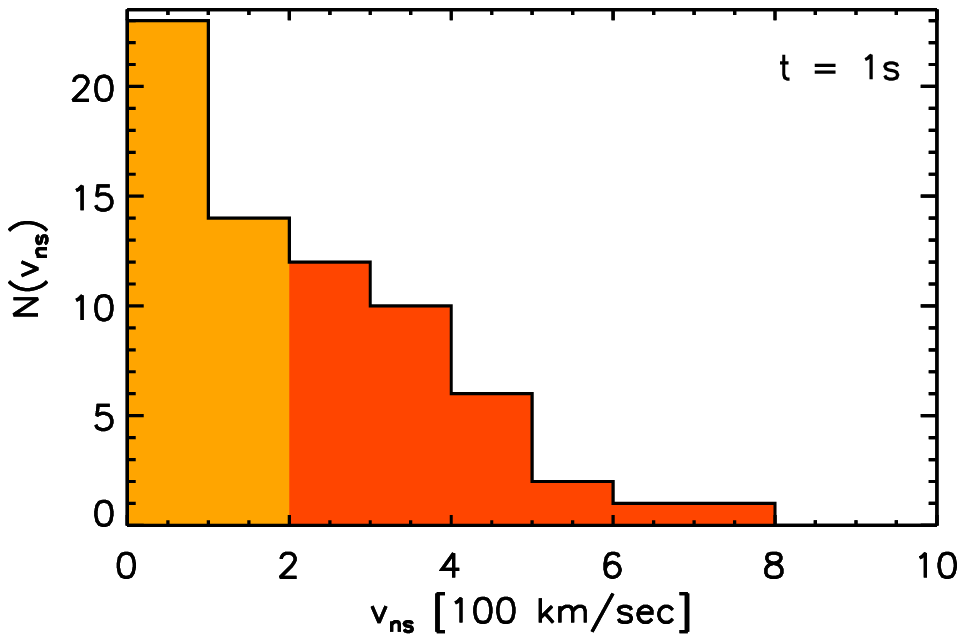}    &
\includegraphics[width=8.5cm]{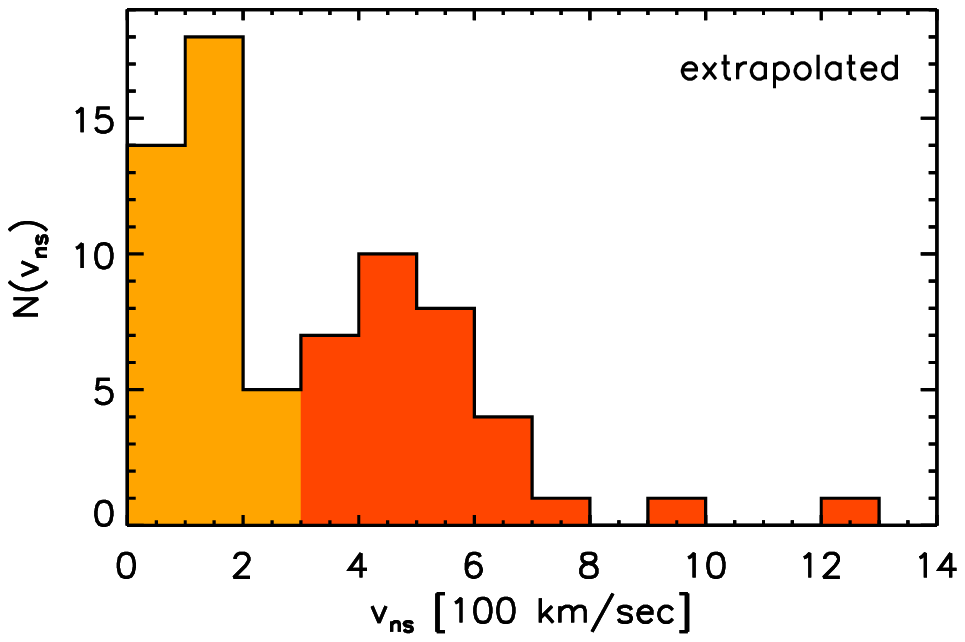}
\end{tabular}
\caption{Histograms of the neutron star velocity distribution for the
  70 models of Tables~\ref{tab:restab_b}--\ref{tab:restab_movns}. The
  left panel shows the velocity distribution at $t=1\,$s (solid black
  line). The darker shaded area corresponds to the fraction of models
  whose neutron stars are moving with more than $200\,\mathrm{km/s}$
  one second after bounce. The same models are displayed with dark
  shading also in the right panel, which shows the final velocity
  distribution as obtained by extrapolation with
  Eq.~\eqref{eq:vinfty}.}
\label{fig:vns_hist}
\end{figure*}

In Sect.~\ref{sec:correlations} we pointed out that
Fig.~\ref{fig:vzns_azns_eexp2}, showing the neutron star velocities
and accelerations at $t = 1\,$s, suggests the existence of two groups
of models. One group consists of cases with low velocities and on
average low acceleration, and the other group cases with high
velocities and significantly higher average acceleration. The latter
models are typically characterised by a strong $l = 1$ mode in the
flow pattern at the end of our simulations.

Provided the acceleration shows a trend of increasing more steeply
than linearly with the neutron star velocity, one can expect a growth
of the separation of both populations when the acceleration continues
over a longer period of time.  Thus a bimodal velocity distribution
will emerge, caused by the larger acceleration associated with the
presence of a dominant $l = 1$ mode in the models of the high-velocity
group.  To test this possibility, we extrapolate the neutron star
motions of all of our 70 models listed in
Tables~\ref{tab:restab_b}--\ref{tab:restab_movns} from one second to
the expected final conditions by applying Eq.~\eqref{eq:vinfty}.
Figure~\ref{fig:vns_hist} displays both the velocity distribution at
the end of the simulated evolution (at $t = 1\,$s; left panel) and the
terminal distribution (right panel).

A comparison of the panels in Fig.~\ref{fig:vns_hist} shows that most
neutron stars of the high-velocity and high-acceleration group (which
is indicated by the darker shading) accelerate to significantly higher
velocities on time scales longer than one second. In contrast, only
very few stars of the low-velocity group reach velocities in excess of
200$\,$km/s. As a consequence, a minimum develops in the extrapolated
distribution around 300$\,$km/s, separating clearly the two components
in velocity space.

We interpret this result as an interesting demonstration that the kick
mechanism discussed here is able to produce a bimodal distribution of
neutron star velocities simply due to the presence or absence of a
dominant $l=1$ mode in the spatial distribution of the supernova
ejecta. Invoking two different processes for neutron star acceleration
is not required. It is, however, unclear whether this may provide an
explanation of a possible bimodality in the observed velocity
distribution of pulsars. The existence of such a bimodality is not
only ambiguous, some authors finding hints
\citep[e.g.][]{CC98,Fryer+98,ACC02,Brisken+03}, while others favour a
one-component Maxwellian distribution
\citep[e.g.][]{LL94,HP97,Hobbs+05,Zou+05}. Also the parameters for the
two-component fits differ significantly between the publications.

Though our result is inspiring as well as tantalising, we refrain from
making a direct connection with observations. Such attempts are
hampered by the limitations of our analysis, which does not only
assume the extrapolation of Eq.~\eqref{eq:vinfty} to be valid for all
cases. Our analysis is also affected by our finding that the magnitude
of the neutron star kicks seems to depend on the neutron star
contraction (see Sect.~\ref{sec:fastcon}) that is mimicked in our
simulations by a moving inner boundary of the computational grid.
Moreover, our analysis is constrained to a set of 15$\,M_\odot$
stars\footnote{The employed progenitor models, however, exhibit large
  differences in core sizes and core density profiles, which actually
  may be considered as reflecting the variations over a broader range
  progenitor masses.}, while linking theory with observations would
require modelling explosions for a representative distribution of
supernova progenitors, making reasonable assumptions about the
progenitor dependence of the explosion energy and including the
effects from binary breakup. A large set of calculations would have to
account for the stochastic nature of the discussed neutron star
acceleration mechanism, thus establishing the distribution of kick
velocities as a function of the progenitor properties. One might have
the concern that in the combined data of all of these runs the minimum
visible in the velocity distribution of Fig.~\ref{fig:vns_hist} is
filled up. Finally, quantitatively meaningful calculations of neutron
star kicks will ultimately have to be obtained by three-dimensional
modelling.

\section{Summary and conclusions}
\label{sec:conclusions}

The aim of this work was an investigation of hydrodynamic
instabilities in the neutrino-heated postshock layer of
core-collapse supernovae and of the importance of such
instabilities for the development of explosion anisotropies
and neutron star kicks.

For this purpose we have presented a large number of (more than 70)
supernova simulations in two dimensions (i.e., assuming axisymmetry)
for different 15$\,M_\odot$ progenitor models, relying on the
viability of the neutrino-driven explosion mechanism. Since this
viability is still an open question and no explosions are obtained in
2D models with a detailed spectral treatment of neutrino transport for
stars more massive than about 11$\,M_\odot$
\citep[see][]{Buras+06,Buras+06b}, we triggered the explosions in our
simulations by replacing the contracting core of the nascent neutron
star by an inner boundary of the computational grid and assuming there
suitable neutrino luminosities from the neutron star core (see the
introduction in \citealt{Kifonidis+06} for a motivation and
justification of this procedure in the light of the results from
recent Boltzmann transport supernova simulations). The boundary was
placed at a Lagrangian mass coordinate of typically 1.1$\,M_\odot$,
where the neutrino optical depths were usually 10 or higher.  A
systematic variation of the core neutrino luminosities imposed at this
boundary allowed us to investigate the growth of hydrodynamic
instabilities and the development of the explosion in dependence of
the strength of the neutrino heating and thus of the size of the
explosion energy.

In contrast to previous work \citep{JM96,Kifonidis+03} the neutrino
luminosities of the neutron star core in the models presented here
were not assumed to decay exponentially, but -- in better agreement
with transport calculations for the whole neutron star -- were assumed
to remain (roughly) constant on a Lagrangian mass shell of $1.1~\Msol$
over hundreds of milliseconds after bounce. With this boundary
condition the approximative neutrino transport scheme developed for
the present study ensures a radial and temporal behaviour of the
neutrino luminosities and mean spectral energies as qualitatively also
found in more complete and fully consistent supernova models, i.e.
the core and accretion components of the neutrino emission are both
accounted for.

Our main results can be summarised as follows.
\begin{enumerate}

\item Random perturbations, by which we seed the growth of non-radial
  instabilities in our simulations, can grow from small initial
  amplitudes (between 0.1\% and some percent of the fluid velocity) to
  global asphericities by convective instability as well as the
  vortical-acoustic cycle \citep{Foglizzo01,Foglizzo02}, provided the
  time until the onset of rapid shock expansion is sufficiently long.
  Once the shock expansion gains momentum, the further growth of the
  instabilities, e.g. by the merging of smaller structures to larger
  ones, is quenched, and the flow pattern essentially freezes out. Not
  the absolute time until explosion matters in this context, but the
  ratio of the explosion time scale 
  to the typical growth time scale of the
  instability. A detailed investigation of the growth of different
  kinds of non-radial instabilities in the postshock flow and their
  competition will be published in a subsequent paper (Scheck et al.\
  2006, in preparation). The neutrino transport description and
  employed inner boundary condition for the transport used in this
  work ensured a sufficient delay of the shock acceleration, in
  contrast to the light-bulb parameters employed by \cite{JM96} and
  \cite{Kifonidis+03}.

\item The growth of the instabilities proceeds extremely nonlinearly
  and chaotically such that the final ejecta anisotropy turns out to
  be sensitive to the initial random pattern of the seed perturbations
  as well as small differences between numerical runs (connected,
  e.g., to small changes in the grid zoning, machine roundoff errors
  or small differences of the input physics). Despite the different
  ejecta geometry, however, integral parameters of the models such as
  the neutron star mass, explosion time scale or explosion energy,
  show little variability.

\item This is different for quantities, which depend on hemispheric
  asymmetries. The instabilities lead to symmetry breaking and the
  ejecta can attain a net linear momentum, balanced by the recoil
  absorbed by the neutron star. In practise, the momentum exchange was
  found to be mediated by gravitational as well as hydrodynamic
  forces. Typically the former are more important, but in cases where
  the neutron star accretes anisotropically over long periods of time,
  also hydrodynamic interaction can contribute significantly.  In our
  standard setup for the calculations, the neutron star is fixed (due
  to the use of the inner grid boundary) at the centre of the grid.
  Since it therefore does not start moving in spite of momentum gain,
  this situation can be considered as a case where the neutron star is
  assumed to have infinite inertial mass. In order to test whether
  this affects the results, we performed a number of runs by imposing
  the negative of the instantaneous neutron star velocity (as
  calculated from its attained momentum) on the ambient gas on the
  computational grid. This leads to a collective gas motion relative
  to the neutron star fixed at the grid centre and corresponds to a
  change of the frame of reference by applying a Galilei
  transformation after every hydrodynamics step. Of course, for any
  given model, all else being equal, the model with the transformation
  yields a different explosion asymmetry and a different neutron star
  kick (but still very similar explosion energy and time scale). But
  despite these differences of individual simulations, we could not
  detect any significant changes of the ensemble behaviour with
  respect to explosion parameters or magnitude and distribution of
  neutron star kick velocities.

\item Further tests also showed that the details of the neutrino
  treatment, the employed gravitational potential (i.e., performing
  the simulations with Newtonian gravity or an effective relativistic
  potential according to \citealt{Marek+06}), the assumed amplitude of
  initial perturbations or the assumed contraction of the inner grid
  boundary (which mimics the shrinking of the cooling nascent neutron
  star) do not have any qualitative influence on our results for the
  neutron star kicks. Quantitatively, we discovered indications
  (based on a limited set of computations, however) that a faster
  contraction of the forming neutron star -- which may correspond to a
  softer nuclear equation of state or more rapid cooling -- seems to
  favour higher neutron star kicks on average. This can be explained
  by a more rapid growth of low-mode non-radial instabilities, leading
  to larger values of the aniotropy parameter $\alphag$ for a given
  explosion energy.

\item While the neutrino flux imposed at the inner grid boundary was
  assumed to be isotropic in all of our simulations, the neutrino
  radiation at large distances from the neutron star could become
  anisotropic because of lateral differences in the neutrino emission
  and absorption. The biggest such differences are associated with
  long-lasting, narrow downflows through which the neutron star
  accretes gas anisotropically. The gas heats up strongly upon falling
  towards the neutron star surface and getting decelerated in shocks.
  We found, however, that the corresponding anisotropic neutrino
  emission produces a neutrino-mediated acceleration which accounts
  only for small corrections to the neutron star velocities produced
  by the asymmetric mass ejection.  These corrections rarely exceed
  10\%.  Both accelerations usually produce motion in opposite
  directions. The reason for this is that the neutron star receives a
  kick {\em towards} a downflow (which attracts the neutron star
  gravitationally or leads to a momentum deficit of the expanding
  ejecta shell on the side of the downflow), whereas the neutrino
  radiation is more intense in the hemisphere of the downflow and thus
  propels the neutron star in the other direction. Since the accretion
  luminosity is radiated near the neutron star surface, we do not
  think that our use of the inner boundary underestimates this effect.
  The inverse is more likely.  Our radial transport tends to
  overestimate the anisotropy of the outgoing radiation, because truly
  multi-dimensional transport would redistribute the locally emitted
  neutrinos more isotropically in all directions instead of favouring
  their radial propagation (see the discussion in \citealt{Livne+04}).

\item After one second of post-bounce evolution, which was the period
  of time we simulated for most models, we obtained maximum neutron
  star velocities up to 800$\,$km/s.  The models appear grouped in two
  populations, one in which the neutron stars move with less than
  200$\,$km/s and have low acceleration at $t = 1\,$s, and another
  one, roughly equally strong, where the stars have velocities higher
  than 200$\,$km/s and on average also higher accelerations (see
  Fig.~\ref{fig:vzns_azns_eexp2}, and the
  left panel of Fig.~\ref{fig:vns_hist}). The two groups differ by the
  absence or presence, respectively, of a strong or dominant $l=1$
  dipole mode in the gas distribution around the neutron star.  The
  simulated models cover roughly equally a range of explosion energies
  between about $0.3\times 10^{51}\,$erg and more than $1.5\times
  10^{51}\,$erg.  We could not detect any systematic variations of the
  typical magnitude or scatter of the kick velocities with the
  explosion energy. We also did not discover any obvious correlation
  of the kicks with the properties of the three considered
  15$\,M_\odot$ progenitor stars, which exhibited major differences in
  their core sizes and density structures. Rotation with a fairly high
  pre-collapse rate of 0.5$\,$rad/s in the iron core, which in view of
  the most recent stellar evolution models is probably unrealistically
  large for ordinary supernovae \citep[see][]{Heger+04}, lead to
  slightly higher explosion energies (due to a larger mass in the gain
  layer), a more spherical shock surface, and the presence of
  downflows at both poles of the rotating neutron star. This suggests
  a weaker contribution of an $l=1$ mode in this situation compared to
  the nonrotating models, where downflows in only one hemisphere are
  rather common. Although very extreme cases were missing in our
  fairly small sample of simulations with such rapid rotation, we
  nevertheless obtained kick velocities in excess of 300$\,$km/s
  (still rising at one second after bounce), and could not detect any
  bias towards the group with low velocities and low average
  acceleration.

\item The two populations in our velocity distribution at one second
  are certainly interesting in view of the possibility of a bimodality
  in the distribution of measured pulsar velocities, which however is
  still controversial. We therefore attempted to derive from our set of
  about 70 simulations the distribution at the time the neutron stars
  have reached their terminal velocities. In order to do that we
  continued some of our models until 3--4 seconds, at which time the
  accelerations have become very small. These models served for
  calibrating the typical period of time which a representative
  acceleration must be applied to extrapolate from the velocity at one
  second to the terminal values. The representative acceleration was
  taken as the average value between 0.5$\,$s and 1$\,$s after bounce,
  a choice which guaranteed that short-time fluctuations of the size
  and direction of the acceleration (which are rather frequent in case
  of low-energetic explosions) do not corrupt the extrapolation.
  Indeed the extrapolated velocity distribution revealed a clear
  bimodal structure with a minimum around 300$\,$km/s and a
  high-velocity component that extends up to 1300$\,$km/s (right panel
  of Fig.~\ref{fig:vns_hist}). This component consists of most of the
  neutron stars that belong to the high-velocity, high-acceleration
  group at one second.  Both components are similar in strength, but
  this may depend on the choices of parameters for the considered set
  of models.  The basic result of a bimodality, however, turned out to
  be very robust against variations of the exact way of extrapolation.
\end{enumerate}

Although the presence or absence of a pronounced $l=1$ mode in the
ejecta distribution offers a natural as well as suggestive way to
obtain a bimodality in the context of our hydrodynamic kick mechanism,
we refrain from claiming that our result is a strong support for the
existence of such a bimodality in the observed distribution of pulsar
velocities. There are too many uncertainties which might lead to a
filling of the minimum of our distribution. Not only do we assume that
our extrapolation law (Eq.~\ref{eq:vinfty}) can be applied with the
same value for the duration of the average acceleration to all models
of our sample, we also consider only a very constrained selection of
progenitor models, which is not representative for the true
distribution of supernova progenitors. Though our simulations do not
reveal systematic differences of the kicks in dependence of the
explosion energy or progenitor structure, we do not feel able to
exclude that a correlation of both over the range of supernova
progenitors could conspire such that the bimodality of
Fig.~\ref{fig:vns_hist} gets wiped out. 

The proposed hydrodynamic kick mechanism, however, leads to an
unambiguous prediction, which might be tested by future detailed
observations of supernova remnants: The measured neutron star velocity
should be directed opposite to the momentum of the gaseous supernova
ejecta. This is different from many theories which explain pulsar
kicks by anisotropic neutrino emission from the nascent neutron star.
In that case the direction of the acceleration can be independent of
ejecta asymmetries.

Apart from all the assumptions and approximations entering this work
and discussed in detail above, the biggest deficiency of the present
analysis is the fact that it is based on simulations which assume
axial symmetry with the polar axis being a coordinate singularity that
is impenetrable for the fluid flow. Currently it is neither clear to
which degree pronounced $l=1$ modes of the ejecta distribution and
long-lasting downflows of matter to the neutron star can develop in
the three-dimensional environment, and how common they are, although
first 3D simulations with the setup and input physics described here
are promising (they will be presented in a future publication, but see
\citealt{Janka+04a} for some results). Nor is it clear what the
distribution of neutron star recoil velocities from 3D models will be.
The large number of long-time simulations required by the stochastic
nature and long duration of the proposed hydrodynamic kick mechanism,
is currently out of reach due to its enormous demand of computer time.
Our results must therefore be considered as indicative but they are
far from providing definitive answers.

\begin{acknowledgements}
  We are grateful to T.~Plewa for his contributions to the early
  stages of this project and his continued interest in this work, to
  R.~Buras, M.~Rampp and S.~Bruenn for providing us with post-bounce
  models, to M.~Limongi and S.~Woosley for their progenitor models,
  and to J.~Niemeyer for valuable suggestions. We would especially
  like to thank an anonymous referee for his careful reading of the
  manuscript and his many insightful comments and suggestions, which
  helped us to improve our paper. Support by the
  Sonderforschungsbereich 375 on ``Astroparticle Physics'' of the
  Deutsche Forschungsgemeinschaft is acknowledged. We are also
  grateful to the Institute for Nuclear Theory at the University of
  Washington for its hospitality during several visits throughout the
  duration of this work. The computations were performed on the NEC
  SX-5/3C of the Rechenzentrum Garching (RZG) and the IBM p690
  clusters of the RZG and the John-von-Neumann Institute for Computing
  in J{\"u}lich.
\end{acknowledgements}

\bibliography{paper}

\appendix

\section{Post-processing of the simulations}
\label{app:definitions}

In the following we define and tabulate some interesting characteristic
quantities that were evaluated for our about 80 hydrodynamic models by
post-processing the data of the simulations. To keep the evaluation as
straightforward as possible we sometimes employ approximations which
we will detail below.

The inner boundary condition for the neutrinos is constrained by the
parameters $t_L$, $\Delta E^{\rm tot}_{\nu,\mathrm{core}}(t)$ and
$\Delta Y_{\rm e, core}(t)$. Their definitions are given in
Appendix~\ref{sec:app_neutrino_bounds}, while their actual values (at
$t=1$\,s in case of the time-dependent quantities) are listed in
Tables~\ref{tab:restab_b}--\ref{tab:restab_movns}. A characterising
value for the neutrino luminosities imposed at this inner boundary is
\begin{equation}
   \Lib(t) \equiv L_{e,\nue}(\rib,t) + L_{e,\nuebar}(\rib,t),
   \label{eq:L_innerbound}
\end{equation}
where $L_{e,\nue}$ and $L_{e,\nuebar}$ are the energy luminosities of
$\nue$ and $\nuebar$ defined in Appendix~\ref{app:transport_equation}.
Equation \eqref{eq:L_innerbound} neglects the contribution from
heavy-lepton neutrinos, whose interactions in the computational domain
are less important than those of $\nue$ and $\nuebar$, and who, in
particular, do not contribute to the neutrino heating behind the shock
at a significant level.

We also consider the sum of the $\nue$ and $\nuebar$ luminosities at a
radius of 500\,km,
\begin{equation}
   \Lfiveh(t) = L_{e,\nue}(r=500~{\rm km},t) + 
             L_{e,\nuebar}(r=500~{\rm km},t),
\end{equation}
and define the time average of this quantity in the time
interval $[0,\texp]$ as
\begin{equation}
 \left\langle L_{500} \right\rangle = \texp^{-1} \, \int_0^{\texp} \Lfiveh(t)\,\,\dt.
\end{equation}
The value of $\left\langle L_{500} \right\rangle$ represents
approximatively the $(\nue+\nuebar)$ luminosity that is responsible
for the energy deposition behind the supernova shock until the
explosion sets in at a post-bounce time $t=\texp$. Therefore the
difference between $\left\langle L_{500} \right\rangle$ and $\Lib$ can
be considered as a rough measure for the radial change of the neutrino
luminosities in contrast to their constancy in case of the light-bulb
scheme used by \cite{JM96}. Furthermore, we list the total energy in
$\nue$ and $\nuebar$ neutrinos that streams through a sphere with a
radius of 500\,km in the time interval $[0,t]$,
\begin{equation}
  \Delta E_{500}(t) = \int_0^{t}
                        (L_{e,\nue}+L_{e,\nuebar})(r=500\,\mathrm{km},t') \, \dt'.
\end{equation}

\begin{table*}[pt!]
\begin{center}
  \caption{Simulations based on the \cite{WPE88}/\cite{Bruenn93}
    post-bounce model. The luminosity time scale $t_L$ is 1\,s.
    Unless noted otherwise the inertial mass of the neutron star is
    assumed to be infinite for these and the simulations listed in the
    following tables, i.e. the neutron star takes up momentum but
    cannot move on the grid. For the definitions of the listed
    quantities see the main text. All time-dependent quantities are
    given at a time $t=1\,$s, when we terminated the
    simulations. Energies are given in units of $1\,\mathrm{B} =
    1\,\mathrm{Bethe} = \foe$.}
\label{tab:restab_b}
\input{restab-b}
\end{center}
\end{table*}

\begin{table*}
\begin{center}
  \caption{Simulations based on the \cite{Limongi+00}/Rampp
    post-bounce model. The luminosity time scale $t_{L}$ is 0.7\,s for
    these simulations. For more details, see the caption of
    Table~\ref{tab:restab_b}.}
\label{tab:restab_l}
\input{restab-l}
\end{center}
\end{table*}

\begin{table*}
\begin{center}
  \caption{Simulations based on the non-rotating
    \cite{WW95}/\cite{Buras+03} post-bounce model.  The luminosity
    time scale $t_{L}$ is 1\,s for these simulations. For more
    details, see the caption of Table~\ref{tab:restab_b}.}
\label{tab:restab_w}
\input{restab-w}
\end{center}
\end{table*}

\begin{table*}[pt!]
\begin{center}
  \caption{Simulations based on the rotating
    \cite{WW95}/\cite{Buras+03} post-bounce model. The luminosity
    time scale $t_{L}$ is 1\,s. For more details, see the caption of
    Table~\ref{tab:restab_b}.}
\label{tab:restab_r}
\input{restab-r}
\end{center}
\end{table*}

\begin{table*}
\begin{center}
  \caption{Simulations based on the \cite{WPE88}/\cite{Bruenn93}
    post-bounce model. The luminosity time scale $t_{L}$ is 1\,s. For
    more details, see the caption of Table~\ref{tab:restab_b}.
    Different from the models listed in all other tables, the recoil
    motion of the neutron star was accounted for in the simulations
    listed here (as described in Sect.~\ref{sec:grid} and
    Appendix~\ref{app:hydro_accel}).}
\label{tab:restab_movns}
\input{restab-movns}
\end{center}
\end{table*}

The explosion energy, $\eexp$, of a model is defined as the sum of the
total energy of all zones of the grid where this energy is positive,
i.e.
\begin{equation}
  \eexp = \sum_{e_{{\rm tot},i} > 0} e_{{\rm tot},i} \, \Delta m_i,
  \label{eq:def_eexp}
\end{equation}
where $i$ is the zone counter, $\Delta m_i$ the mass contained in zone
$i$, and the total specific energy $e_{\rm tot}$ is given by the sum
of the specific gravitational, kinetic, and internal energies,
\begin{equation}
e_{\rm tot} = e_{\rm grav} + \frac{1}{2}v^2 + e_{\rm int}.
  \label{eq:def_etot}
\end{equation}
For the sake of simplicity we use here the one-dimensional Newtonian
expression
\begin{equation}
  e_{\rm grav}(r) = -\frac{G M(r)}{r}
\end{equation}
to evaluate the gravitational energy, neglecting the relatively small
general relativistic corrections, which have been taken into account
in the simulations.

The explosion time scale, $\texp$, is defined as the time after the
start of the simulation when $\eexp$ exceeds $\rm 10^{48}\,erg$. It
turns out that the exact choice of this threshold value does not
matter very much. Other definitions of the explosion time scale (e.g.,
linked to the time when the expansion velocity of the shock exceeds a
certain value) do also not lead to qualitatively different results.

To characterise the deviation of the shape of the supernova shock from
a sphere we introduce a shock deformation parameter,
\begin{equation}
  d_{\rm shock} := \frac{\max\left(\Rs(\theta)\cos\theta\right) - \min\left(\Rs(\theta)\cos(\theta)\right)}
                    {2 \times \max\left(\Rs(\theta)\sin\theta\right)}-1,
\label{eq:def_dshock}
\end{equation}
where $\Rs(\theta)$ is the local shock radius as a function of polar
angle $\theta$.  The numerator and denominator in
Eq.~(\ref{eq:def_dshock}) are the maximum shock diameters in
projection on the symmetry axis and perpendicular to it, respectively.
A prolate deformation leads to a positive value of $d_{\rm shock}$, an
oblate deformation gives a negative value. Note that a linear shift of
the shock surface in the direction of the $z$-axis does not change
$d_{\rm shock}$.

The neutron star mass and the neutron star radius are considered to be
associated with a certain value of the density, $\rho_{\rm ns} =
10^{11}{\rm g~cm^{-3}}$. The neutron star radius, $\Rns$, is then
simply defined as the radius where the lateral average of the density
is equal to $\rho_{\rm ns}$, and the baryonic mass of the neutron star,
$\Mns$, is given by the sum of the central point mass and the mass
integral over all grid zones with densities $\geq \rho_{\rm ns}$.

In evaluating the neutron star recoil velocity, $\vnsvec$, we have to
distinguish between simulations in which we consider the neutron star
motion relative to the ejecta by changing the frame of reference after
each time step (see Sect.~\ref{sec:grid} and
Appendix~\ref{app:hydro_accel}), and simulations in which this motion
is not accounted for. In the first case no post-processing is
required, because the neutron star velocity is given at all times by
the accumulated effects of the Galilei transformations applied until
time $t$ or time step $m$,
\begin{equation}
  \vnsvec(t) = \sum_{n=1,\dots,m} \Delta \vec{v}^n_{\rm core},
\end{equation}
where $\Delta \vec{v}^n_{\rm core}$ is given by Eq.~(\ref{eq:vcore}).
In the second case, $\vnsvec$ is computed a posteriori, by making use
of linear momentum conservation. The total momentum of the system,
i.e. the sum of the neutron star momentum $\Pns=\Mns \vnsvec$ and the
momentum of the surrounding gas on the computational grid, $\Pgas$, is
initially zero (because all models that we consider are spherically
symmetric or equatorially and axially symmetric just after collapse).
Hence we have for all times
\begin{equation}
  \label{eq:vns_pgas}
  \vnsvec(t) = -\Pgas(t) / \Mns(t),
\end{equation}
and $\vnsvec(t)$ can be determined by evaluating the neutron star mass
and the momentum integral of the ejecta gas,
\begin{equation}
  \label{eq:def_pgas}
  \Pgas(t) = \int_{\Rns<r<\infty} \rho\,\vvec \; \dV.
\end{equation}
Here $\dV = r^2 \sin \theta \: \dr \: \dtheta \: \dphi$. Note that the
volume integral in Eq.~\eqref{eq:def_pgas} is limited by the outer
boundary of our Eulerian grid and that the momentum flux associated
with anisotropic mass flow over the grid boundary would have to be
taken into account.

Equation~(\ref{eq:vns_pgas}) may actually also be coined in terms of
an anisotropy parameter of the ejecta, $\alphag$ (see
\citealt{JM94,Herant95}). To accomplish this, we make use of the
following scalar quantity
\begin{equation}
  \Pej(t) := \int^{\Rs(\theta)}_{\Rns} \rho \left|\vvec\right| \; \dV,
\label{eq:def_pej}
\end{equation}
which has the dimension of a momentum. Then we can write the
anisotropy parameter as
\begin{equation}
  \alphag := |\Pgas| \; / \; \Pej,
\label{eq:def_alphag}
\end{equation}
and the absolute value of the neutron star velocity as
\begin{equation}
  \left| \vnsvec \right| = \alphag \; \Pej \; / \; \Mns.
\label{eq:rel_alphag_vns}
\end{equation}

The neutron star acceleration corresponding to the velocity change at
a given time is calculated by finite differences:
\begin{equation}
  \vec{a}_{\rm ns}^{(n)}  = \frac{\vec{v}_{\rm ns}^{(n+1)}-\vec{v}_{\rm ns}^{(n-1)}}{t^{(n+1)}-t^{(n-1)}}.
\end{equation}

In computing the recoil velocity according to Eqs.~(\ref{eq:vns_pgas})
and (\ref{eq:rel_alphag_vns}), we have so far neglected the fact that
the neutron star may also be accelerated by anisotropic neutrino
emission. While our core luminosities at the inner grid boundary are
assumed to be \emph{isotropic at all times} and no neutron star
acceleration can result from these, direction-dependent variations of
the thermodynamic variables in layers close to the neutron star
surface develop during the simulations and ultimately lead to
anisotropies of the neutrinospheric emission of neutrinos. In
particular, density inhomogeneities and local hot-spots (in
temperature) occur as a consequence of narrow accretion flows that
transport gas from the postshock layers to the neutron star, where
they are decelerated in shocks and radiate away energy in neutrinos.
The anisotropy of this neutrino emission can give rise to a ``neutrino
rocket effect'', whose magnitude can be estimated by considering the
integrated momentum of the escaping neutrinos.

For a transport scheme along radial rays like ours, the neutrino
momentum density has only a radial component and can thus be written
as (see also Appendix~\ref{app:transport})
\begin{equation}
p_{\nu} \,\vec{e}_r  = \frac{n_{\nu} \epsilon_{\nu}}{c}\,\vec{e}_r 
                   = \frac{F_{\nu}}{c^2}\,\vec{e}_r,
\end{equation}
where $F_{\nu}$ is the local neutrino energy flux and $\vec{e}_r$ the
unit vector in the radial direction. The integrated neutrino momentum
at time $t$ is then given by
\begin{align}
  \Pnu(t) & = \int_{\rib<r<\infty} p_{\nu} \,\vec{e}_r \,\dV
          \nonumber \\
          & = \int_{\rib<r<\rob} p_{\nu} \,\vec{e}_r \,\dV
           + \int_0^{t} {\rm d}t \oint_{r=\rob} p_{\nu} c \,\vec{e}_r \,\dS,
  \label{eq:def_pnu}
\end{align}
with the surface element $\dS = r^2 \sin \theta \: \dtheta \:
\dphi$. Here the surface integral accounts for the fact that a
significant amount of neutrino momentum may have left our grid through
the outer boundary by the time $t$. The momentum of the neutron star,
including now also the effect of anisotropic neutrino emission, is
\begin{equation}
\Pns = -\left( \Pgas + \Pnu \right),
\end{equation}
so that the neutron star velocity, corrected for the recoil by
anisotropic neutrino emission, can be written as
\begin{equation}
  \label{eq:def_vnscorr}
  \vnsveccorr = \vnsvec + \vnsvecnu =  -\Pgas / \Mns \, - \,\Pnu / \Mns.
\end{equation}

We finally note that for symmetry reasons $\Pgas$ and $\Pnu$, and thus
also $\Pns$ and $\vnsvec$, can have only a component parallel to the
symmetry axis, i.e. along the $z$-axis, in 2D axisymmetric
calculations. Equation~(\ref{eq:def_pgas}), for instance, therefore
reduces to
\begin{align}
  \label{eq:mombal2d}
  \Pzgas & = 2\pi \, \int_{\Rns}^{\infty} \!\dr
             \int_{0}^{\pi} \!\dtheta \, r^2 \sin \theta \,\, p_z(r,\theta) \nonumber\\
         & = 2\pi \, \int_{\Rns}^{\infty} \!\dr 
             \int_{0}^{\pi/2} \!\dtheta \, r^2 \sin \theta \,\, 
                   [ \, p_z(r,\theta) + p_z(r,\pi-\theta) \, ] \nonumber\\
         & =  P_{z,\mathrm{gas}}^{\rm N} + P_{z,\mathrm{gas}}^{\rm S}.
\end{align}
Here $p_z(r,\theta) = \rho \, ( v_r \cos \theta - v_{\theta} \sin
\theta)$ is the $z$-component of the momentum density of the gas, and
$P_{z,\mathrm{gas}}^{\rm N}$ and $P_{z,\mathrm{gas}}^{\rm S}$ are
introduced as the $z$-momenta of the gas in the northern and southern
hemispheres, respectively.

\section{Hydrodynamics in an accelerated frame of reference}
\label{app:hydro_accel}

In an inertial frame of reference the hydrodynamic equations are given
by
\begin{equation}
  \label{eqn:hydro_mass}
  \frac{\partial \rho}{\partial t} + \nabla \cdot (\rho \vvec) = 0,
\end{equation}
\begin{equation}
  \label{eqn:hydro_momentum}
 \rho \left( \frac{\partial \vvec}{\partial t} + (\vvec \cdot \nabla) \vvec \right) + \nabla {\cal P}
= \rho \gvec,
\end{equation}
\begin{equation}
  \label{eqn:hydro_energy}
 \frac{\partial \rho E}{\partial t} + \nabvec \cdot \left(\left(\rho
 \, E + {\cal P} \right)\,\vvec \right) = \vvec \cdot \rho \gvec,
\end{equation}
where $\rho$ is the density, $\vvec$ is the velocity, $\cal{P}$ is the
pressure, $\gvec$ is the gravitational acceleration and $E = \epsilon
+ v^2/2$ is the sum of internal energy, $\epsilon$, and kinetic energy,
$\epsilon_{\rm kin}$, per unit mass.

Let AF be a frame of reference that coincides with an inertial frame
IF at time $t=0$ and accelerates with a constant rate $a$ in
$z$-direction, $\avec=a\evec_z$. The Cartesian coordinates of both
frames are then related by
\begin{equation}
  \label{eqn:acc_coord_rel}
  (x',y',z',t') = (x,y,z-at^2/2,t)
\end{equation}
(primed quantities are used for the accelerated frame), which implies
that
\begin{equation}
  \label{eqn:acc_coord_der}
  \partial z'(x,y,z,t) / \partial t = -at \quad \mathrm{and} \quad \partial z(x',y',z',t') / \partial t' = at.
\end{equation}
For density, pressure, velocity, kinetic energy and gravitational
acceleration the following relations hold:
\begin{align}
  \label{eqn:acc_quan_rel}
  \rho'(x',y',z',t)     &= \rho(x,y,z,t),              \nonumber\\
  {\cal P}'(x',y',z',t) &= {\cal P}(x,y,z,t),          \nonumber\\
  \vvec'(x',y',z',t)    &= \vvec(x,y,z,t) - a t \evec_z \nonumber\\
  \epsilon_{\rm kin}'(x',y',z',t) &= \epsilon_{\rm kin}(x,y,z,t) - v_z^2/2 + (v_z-at)^2/2, \nonumber\\
  \vec{g}'(x',y',z',t)        &= \vec{g}(x,y,z,t) - a \evec_z .
\end{align}

From relations (\ref{eqn:acc_coord_rel})--(\ref{eqn:acc_quan_rel}),
it is easy to see that the equation of mass conservation
(\ref{eqn:hydro_mass}) does not change in the accelerated frame. The
momentum equation in this frame is
\begin{equation*}
  \label{eqn:hydro_momentum_acc}
  \rho' \left( \frac{\partial \vvec'}{\partial t} + (\vvec' \cdot \nabvec')\vvec' \right)
  + \nabvec' {\cal P}' - \rho\gvec'
  = -a t \rho \left( \frac{\partial v_x}{\partial z} \evec_x
                   + \frac{\partial v_y}{\partial z} \evec_y \right).
\end{equation*}
Note that in contrast to Eq.~(\ref{eqn:hydro_momentum}) there is now
an additional term on the right hand side, which affects the momentum
components perpendicular to the direction of acceleration. Thus for
instance the $x$-component of the time derivative of the velocity is
\begin{equation}
  \frac{\partial v'_x}{\partial t} =
  - \left( v'_x \frac{\partial v'_x}{\partial x'}
         + v'_y \frac{\partial v'_x}{\partial y'}
         + v'_z \frac{\partial v'_x}{\partial z'}
    \right)
  - \frac{1}{\rho'} \frac{\partial {\cal P}'}{\partial x'} + g'_x
  - at\frac{\partial v_x}{\partial z},
\end{equation}
where the additional (rightmost) term is negligible compared
to $v'_z \, ( \partial v'_x /\partial z')$, as long as $|at| \ll |v_z|$. 

Similarly, it can be shown that the additional terms arising in the
energy equation (\ref{eqn:hydro_energy}) for an accelerated frame of
reference are of order $t^2$ and can also be neglected, as long as
$|at| \ll |v_z|$ holds.

Within a typical time step $\Delta t$ of a supernova simulation (of
order $10^{-6}$\,s) the condition $|a\Delta t| \ll |v_z|$ is
satisfied, because the maximum neutron star accelerations are of
${\cal O}(10^8 \, \mathrm{cm/s^2})$, and hence $|a \Delta t| = {\cal
  O}(100 \, \mathrm{cm/s})$, which is much smaller than the relevant
velocities in the simulations, which are of ${\cal O}(10^6 {\rm
  cm/s})$.  Thus a solution of the inertial frame hydrodynamics
equations with the simple replacement $\vec{g} \rightarrow \vec{g'} =
\vec{g}-\vec{a}$ should yield an excellent approximation to the
solution of the hydrodynamic equations in the accelerated frame.

Unfortunately, in the present problem the neutron star acceleration,
and hence the instantaneous acceleration of the frame, $\vec{a}(t)$,
is not known a priori, because it is coupled to the solution of the
hydrodynamic problem during a considered time step.  Therefore we need
to make use of an operator-splitting approach, in which we first
ignore the acceleration of the frame of reference and simply solve the
inertial frame hydrodynamics equations (just using the gravitational
acceleration $\vec{g}$).  We can then compute the current value of
$\vec{a}(t)$, which is assumed to be constant over the time step,
using momentum conservation: The sum of the momenta of the neutron
star core, $\vec{P}_{\rm core}$, and the matter on the numerical grid,
$\vec{P}_{\rm grid}$, is conserved and initially zero, so that
$\Delta \vec{P}_{\rm core} = -\Delta \vec{P}_{\rm grid}$. We can then
use the relation
\begin{equation}
  \vec{a}(t^{n}) = - \frac{\vec{P}_{\rm grid}(t^{n}) -\vec{P}_{\rm grid}(t^{n-1})}
                         {M_{\rm core} \, \Delta t}
\label{eq:accel_det}
\end{equation}
to determine the acceleration in this time step. Finally we take the
effects of the global acceleration of our frame into account in a
second step, by adding
\begin{equation}
-\vec{a}(t^n) \,\Delta t \equiv -\Delta \vec{v}^n_{\rm core}
\label{eq:vcore}
\end{equation}
to the hydrodynamic velocity in each zone of the grid, in essence
performing a Galilei transformation to an instantaneous inertial frame
in which the neutron star is again at rest.

\section{Explosion energy}
\label{app:eexp}

The explosion energy of neutrino-driven supernovae consists of two
major contributions. The first is the recombination energy of the
matter in the gain layer at the onset of the explosion. This matter
consists of free nucleons and alpha particles at the time the
explosion starts. Almost all of this mass (except for some fraction in
the downflows, which is accreted onto the neutron star) ends up in a
dense shell behind the expanding shock. As the shock propagates
outward, the temperature in this expanding shell decreases and the
matter recombines to $\alpha$-particles and later to nuclei.

\begin{figure}[tpb!]
\centering
    \includegraphics[angle=0,width=8.5cm]{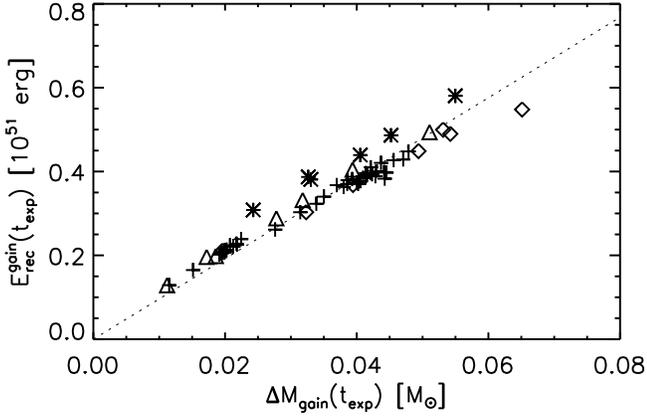}
    \caption{Available recombination energy, $E_{\rm rec}^{\rm gain}$,
      as a function of the mass in the gain layer, $\dMg$, at the time
      of explosion for the models of
      Tables~\ref{tab:restab_b}--\ref{tab:restab_movns}. The slope of
      this approximately linear relation corresponds to about 5\,MeV
      per baryon (dotted line).}
\label{fig:erec_mgain}
\end{figure}

Figure~\ref{fig:erec_mgain} displays the available recombination
energy of the matter in the gain layer at the time of the explosion,
\begin{equation}
  E_{\rm rec}^{\rm gain}(\texp) = \int_{\rm gain\;layer} \epsilon_{\rm rec}(r,\texp) \, \dV.
  \label{eq:def_erecgain}
\end{equation}
Here $\epsilon_{\rm rec}(r,t)$ denotes the density of recombination
energy available when matter consists of nucleons, $\alpha$-particles
and some mass fraction of heavy nuclei,
\begin{equation}
\epsilon_{\rm rec}(r,t) = B_{\rm h} \, n_{\rm h}^{\rm max}(r,t)
	         - \left( B_{\alpha} \, n_{\alpha}(r,t) +
                       B_{\rm h}  \, n_{\rm h}(r,t)  \right), 
\label{eq:def_epsrec}
\end{equation}
with $B_{\rm h}$ and $B_{\alpha}$ being the binding energies of a
representative heavy nucleus $(Z_{\rm h},A_{\rm h}=N_{\rm h}+Z_{\rm
  h})$ from the iron group (as assumed in our equation of state) and
of $\alpha$-particles, respectively, and $n_{\rm h}^{\rm max} =
\min(\,n_{\rm p}^{\rm tot}/Z_{\rm h},\,n_{\rm n}^{\rm tot}/N_{\rm
  h}\,)$ and $n_{\rm h}$ are the maximum and current number densities,
respectively, of this heavy nucleus when $n_{\rm p}^{\rm tot}$ and
$n_{\rm n}^{\rm tot}$ are the total (bound$+$free) number densities of
protons and neutrons.

Figure~\ref{fig:erec_mgain} shows that for all models of
Tables~\ref{tab:restab_b}--\ref{tab:restab_movns} $E_{\rm rec}^{\rm
  gain}(\texp) \approx N_{\rm b}^{\rm gain}(\texp) \times 5\,$MeV,
when $N_{\rm b}^{\rm gain}$ is the total number of baryons in the gain
layer. This means that due to the partial assembly of free n and p in
$\alpha$-particles at the time of explosion, about $5\,$MeV (instead
of $>8\,$MeV) remain available for being released by recombination
during the subsequent expansion and cooling.

\begin{figure}[tpb!]
\centering
    \includegraphics[angle=0,width=8.5cm]{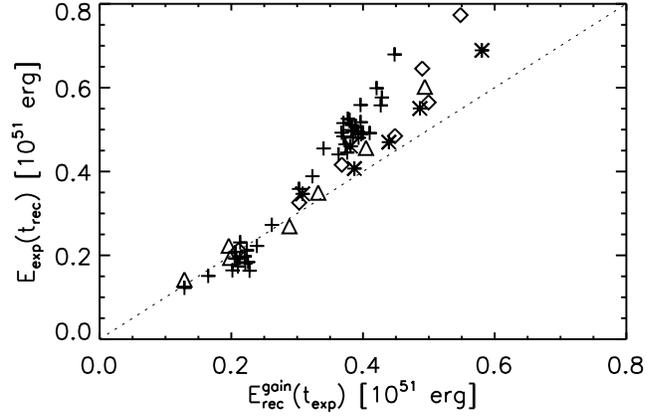}
    \caption{Explosion energy after the recombination of the ejecta,
      $\eexp(\trec)$, as a function of the available recombination
      energy in the gain layer at the onset of the explosion, $E_{\rm
        rec}^{\rm gain}(\texp)$, for the models of
      Tables~\ref{tab:restab_b}--\ref{tab:restab_movns}. For low
      explosion energies the two quantities agree well.}
\label{fig:eexp_erec}
\end{figure}

This recombination is essentially complete when the shock has reached
a radius of $3000\,$km (recombination to $\alpha$-particles happens
even much earlier). We denote this time by $\trec$.
Figure~\ref{fig:eexp_erec} demonstrates that the explosion energy at
time $\trec$, $\eexp(\trec)$, roughly equals the available
recombination energy, $E_{\rm rec}^{\rm gain}(\texp)$, at the onset of
the explosion. This means that neutrino heating essentially has the
effect of lifting the total energy of mass elements in the gain layer
close to zero (i.e., $\epsilon_{\rm tot} = \epsilon_{\rm kin} +
\epsilon_{\rm int} + \epsilon_{\rm grav} \approx 0$) and thus makes
this matter unbound and enables its expansion in the gravitational
potential of the forming neutron star. The excess energy of this
matter at time $\trec$, i.e. $\eexp(\trec)$, is provided by the
recombination of nucleons to $\alpha$-particles and finally to
iron-group nuclei. Only in case of higher explosion energies,
$\eexp(\trec)$ is clearly larger than $E_{\rm rec}^{\rm gain}(\texp)$
(Fig.~\ref{fig:eexp_erec}). In this case neutrino heating in the gain
layer is stronger and the heating time scale of the matter there
shorter than the expansion time scale when the shock begins to
accelerate outwards. Therefore neutrinos are able to deposit ``excess
energy'' in the ejecta before this matter has moved out of the region
of strong heating.

\begin{figure}[tpb!]
\centering
    \includegraphics[angle=0,width=8.5cm]{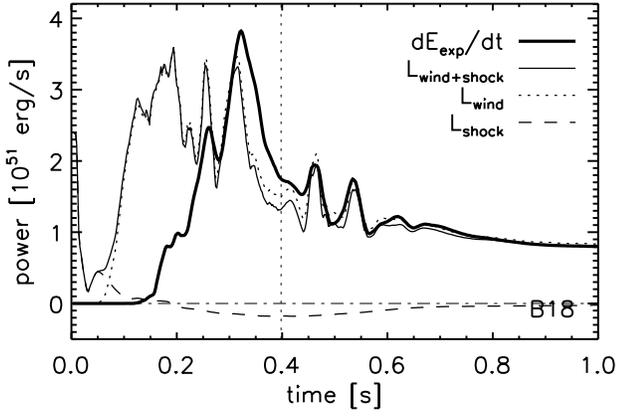}
    \caption{Evolution of the time derivative of the explosion energy
      (${\rm d}\eexp/\dt$, thick solid) for Model B18. Also shown are
      the wind power at a radius of 200\,km ($L_{\rm wind}$, dotted),
      the energy loss/gain rate at the shock by ${\cal P}\dV$ work and
      swept-up matter ($L_{\rm shock}$, dashed), and the sum of the
      latter two quantities ($L_{\rm wind+shock}$, thin solid).
      $L_{\rm wind+shock}$ agrees well with ${\rm d}\eexp/\dt$ for
      $t>\trec$ (right of the vertical line).}
\label{fig:pwind}
\end{figure}

The second contribution to the explosion energy comes from the
neutrino-driven baryonic wind which sets in after the surroundings of
the nascent neutron star have been cleaned from the initially heated
gas. Indeed this wind is an important energy source at ``late'' times.
To demonstrate this, we compare the time derivative of the explosion
energy, ${\rm d}\eexp/\dt$, with the wind power, $L_{\rm wind}$, and
the net energy loss/gain rate $L_{\rm shock}$ at the shock
(Fig.~\ref{fig:pwind}). The curve for ${\rm d}\eexp/\dt$ in
Fig.~\ref{fig:pwind} is calculated as the numerical derivative of the
energy integral
\begin{equation}
  \eexp(t) = \int_{V^+} \epsilon_{\rm tot}(r,t) \dV,
\end{equation}
where the integration is performed over the volume $V^+$, in which
the total energy $\epsilon_{\rm tot}(r,t)$ is positive (see also
Eq.~\ref{eq:def_eexp}). For $t>\trec$ this volume fills the region
between an inner boundary at $r \approx 200\,$km and the shock (except
for some parts of the accretion downflows, where $\epsilon_{\rm tot}$
may still be negative).

The explosion energy is subject to changes by ${\cal P}\dV$-work
performed at, and by energy fluxes through the boundaries of $V^+$, in
particular by the wind, whose power is given by the surface integral
\begin{equation}
  L_{\rm wind} = \oint_{r=200\,{\rm km}} 
( \epsilon_{\rm tot} + \epsilon_{\rm rec} + {\cal P} )
                               \, \max(v_r,0) \, {\rm dS}.
\label{eq:def_pwind}
\end{equation}
This expression takes into account the total energy ($\epsilon_{\rm
  tot} = \rho e_{\rm tot}$ with $e_{\rm tot}$ defined by
Eq.~\ref{eq:def_etot}) of the wind material streaming through the
inner boundary radius into $V^+$, the energy that will be set free at
larger radii by recombination (Eq.~\ref{eq:def_erecgain}), as well as
the work performed by pressure forces. Here we have neglected effects
due to downflows by omitting contributions to the surface integral
from zones with negative radial velocity.

The change of the explosion energy due to energy flow through the
outer boundary of $V^+$ (i.e., the shock), is given by the net energy
loss/gain rate
\begin{equation}
  L_{\rm shock} = \oint_{r=\Rs(\theta)^+} \!\! [ 
                (\epsilon_{\rm tot}+\epsilon_{\rm rec}+{\cal P}) \, v_r \, 
                + (\epsilon_{\rm tot}+\epsilon_{\rm rec}) \, 
                  \dot{R}_{\rm s} ] \, {\rm dS}.
\label{eq:def_pshock}
\end{equation}
The integration has to be performed over a surface located slightly
upstream of the shock. Compared to Eq.~(\ref{eq:def_pwind}) an
additional term arises here from the motion of the shock, which
propagates with a local velocity $\dot{R}_{\rm s}(\theta)$.

\begin{figure}[tpb!]
\centering
    \includegraphics[width=8.5cm]{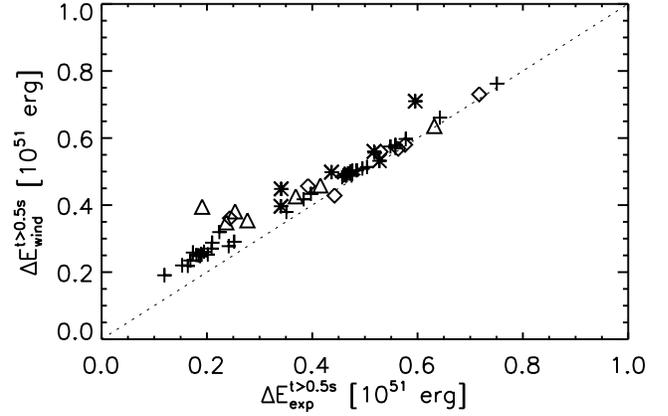}
    \caption{Relation between the increase of the explosion energy
      between $t=0.5\,$s and $t=1\,$s, $\Delta E_{\rm
        exp}^{t>0.5\,{\rm s}}$, and the integrated wind power during this
      time interval, $\Delta E_{\rm wind}^{t>0.5\,{\rm s}}$, for the
      models of Tables~\ref{tab:restab_b}--\ref{tab:restab_movns}.}
\label{fig:dexp05}
\end{figure}

Figure~\ref{fig:pwind} shows that these two terms explain the
evolution of ${\rm d}\eexp/\dt$ for $t>\trec$, i.e. 
\begin{equation}
  {\rm d}\eexp/\dt \approx L_{\rm wind} + L_{\rm shock}
\end{equation}
holds at late times, and the thin and thick solid lines in
Fig.~\ref{fig:pwind} almost coincide. Note also that $L_{\rm wind} \gg
|L_{\rm shock}|$.  This is true for all our models, and therefore the
increase of the explosion energy after about 0.5\,s post bounce is
(almost exclusively) associated with the time-integrated wind power
(see Fig.~\ref{fig:dexp05}).

\begin{figure}[tpb!]
\centering
    \includegraphics[width=8.5cm]{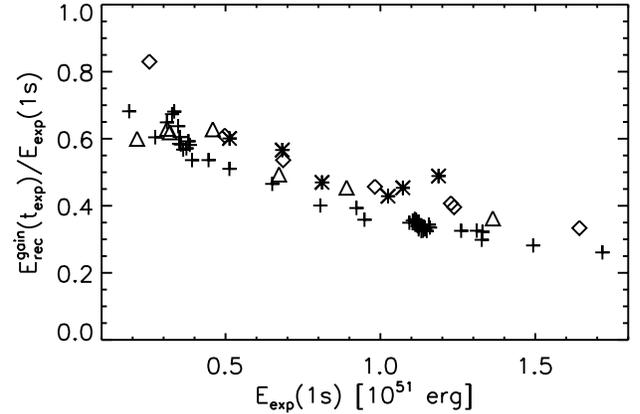}
    \caption{Ratio of the recombination contribution to the total
      explosion energy 1\,s after core bounce, $E_{\rm rec}^{\rm
        gain}(\texp)/\eexp(1\,\mathrm{s})$, as a function of
      $\eexp(1\,\mathrm{s})$ for the models of
      Tables~\ref{tab:restab_b}--\ref{tab:restab_movns}. For
      low-energy models the recombination contribution dominates,
      whereas for higher explosion energies the wind contribution
      becomes more important.}
\label{fig:recfrac}
\end{figure}

The relative importance of the two major constituents of the explosion
energy that we have discussed here, i.e., the nuclear recombination
energy of the matter in the gain layer and the integrated power of the
neutrino-driven wind, varies with the explosion energy. In our
``standard boundary contraction'' models the fraction of the explosion
energy provided by recombination drops from about 70\% for the
low-energy models to about 30\% for the model with
$\eexp(1\,\mathrm{s})\approx 1.5\times\foe$ (Fig.~\ref{fig:recfrac}).
This fraction declines because the wind power is proportional to a
higher power of the luminosity ($L_{\rm wind} \propto
L_{\nu}^{\alpha}$ with $\alpha \approx 3$; \citealt{Thompson+01}) and
although the mass in the gain layer at the onset of the explosion
scales linearly with the boundary luminosity (Fig.~\ref{fig:mgain}).

For the ``rapid boundary contraction'' cases the wind contribution is
even more important, e.g.  for Model W12F-c $E_{\rm rec}^{\rm
  gain}(\texp)/\eexp(1\,\mathrm{s}) \approx 0.2$, i.e.  about 80\% of
the explosion energy are generated by the neutrino-driven wind.  For a
fixed boundary luminosity the wind power is higher in this case than
for the ``standard boundary contraction'', because $L_{\rm wind}$
increases with decreasing neutron star radius \citep[see
e.g.][]{Thompson+01}.  However, $\dMg(\texp)$, and thus also $E_{\rm
  rec}^{\rm gain}(\texp)$, are similar for models with ``standard''
and ``rapid'' boundary contraction and the same $\Lib$. This is so because two
effects compensate each other roughly: On the one hand the density at a given
radius $r$ in the gain layer is lower for a faster contraction
($\rho^{\rm r}(r,t_{\rm exp}^{\rm r})<\rho^{\rm s}(r,t_{\rm exp}^{\rm
  s})$, where r and s denote the rapid and standard contraction cases,
respectively), but on the other hand also the gain radius is smaller
and thus located in a region of higher density, $\rho^{\rm r}(R_{\rm
  g}^{\rm r},t_{\rm exp}^{\rm r})>\rho^{\rm s}(R_{\rm g}^{\rm
  s},t_{\rm exp}^{\rm s})$.

\begin{figure}[tpb!]
\centering
\includegraphics[width=8cm]{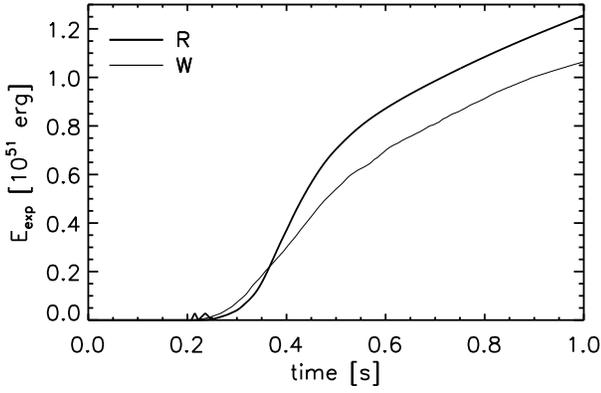}
\caption{Evolution of the explosion energy for Models R18-c and W18-c
         which are listed in Table~\ref{tab:restab_rw1218c}. The
         rotating model R18-c attains an explosion energy which is
         higher than in the non-rotating case W18-c due to a larger
         gain layer mass at the time of explosion.  Note also that the
         explosion energy in both cases is still increasing at
         $t=1$\,s.}
\label{fig:rw_eexp}
\end{figure}

Figure~\ref{fig:rw_eexp} indicates that the explosion energy is still
increasing at $t=1\,$s when we stopped most of our simulations. Yet,
with the subsequent drop of the core luminosity (we assume a
$t^{-3/2}$ behaviour at $t>t_L$, see Eq.~\ref{eq:libtimdep}) also the
wind power, which is proportional to $L_{\nu}^{\alpha}$ (see above),
must decline strongly. Therefore the explosion energy will grow only
moderately. In case of the long-time simulation B18-lt it rose from
$1.14\times\foe$ at $t=1\,$s to $1.43\times\foe$ at $t=2\,$s, and
reached $1.46\times\foe$ by the end of the simulation at $t=3.6\,$s.

\section{Neutrino transport}
\label{app:transport}

\subsection{Transport equation}
\label{app:transport_equation}

We start from the equation of radiation transport in spherical
symmetry
\begin{equation}
  \frac{1}{c} \ddt I + \mu \ddr I + \frac{1-\mu^2}{r} \ddmu I = S,
  \label{eq:transport1d}
\end{equation}
where $I = I(t,r,\epsilon,\mu)$ is the specific intensity,
$S=S(t,r,\epsilon,\mu)$ is the source function, $\epsilon$ is the
neutrino energy, $\mu = \cos \theta$ and $\theta$ is the angle between
radiation propagation and radial direction. Solid angle integration
yields the zeroth angular moment equation,
\begin{equation}
  \frac{1}{c} \ddt J + \frac{1}{r^2} \ddr (r^2 H) \\
              = S^{(0)} \equiv \frac{1}{2} \intdmu S
\end{equation}
with $\{J,H\}(t,r,\epsilon) := \frac{1}{2} \intdmu \mu^{\{0,1\}}
I(t,r,\epsilon,\mu)$. Integration over energy leads to
\begin{equation}
  \ddt E + \frac{1}{r^2} \ddr (r^2 F) = Q^+ - Q^-
  \label{eq:trans_inte}
\end{equation}
with $\{E,F\}(t,r) := 4 \pi \; \intdeps \{J/c,H\}(t,r,\epsilon)$ being
energy density and energy flux, respectively.  The source term has
been split in an emission rate $Q^+$ and an absorption rate $Q^- =
\kappa_{\rm a} c E$, which is proportional to the energy density. The
{\em flux factor} is defined as the ratio of flux to energy density,
\begin{equation}
  f(r,t) := F(r,t) \; / \; c E(r,t).
\label{eq:def_fluxf}
\end{equation}

In neutrino transport simulations solving the full Boltzmann equation
\citep[see e.g.][]{Buras+03,Buras+06,Buras+06b} this quantity shows
only little short-time variability during most phases of the supernova
evolution. Therefore $\partial f / \partial t = 0$ is an acceptably
good approximation.  With $L := 4\pi r^2 F = 4 \pi r^2 f c E$ one can
now rewrite Eq.~\eqref{eq:trans_inte} as
\begin{equation}
  \label{eq:transL}
  \ddt L + \ceff \ddr L = 4 \pi \; r^2 \; \ceff \; \{ Q^+ - Q^- \},
\end{equation}
where an effective speed of neutrino propagation has been introduced
as $\ceff := c f$. Provided $\ceff$ were known, the solution of
Eq.~\eqref{eq:transL} requires considerably less effort than the
numerical integration of Eq.~\eqref{eq:transport1d}. For vanishing
source terms $Q^+$ and $Q^-$ the neutrino energy or number density is
just advected along characteristics $r(t) = r_0 + \ceff \, t$.
Although $\ceff$ depends through $f(r,t)$ on the solution of the
transport problem (Eq.~\ref{eq:def_fluxf}), neutrino transport
calculations in the neutrino-decoupling layer of forming neutron stars
reveal that it can be well fitted by a $r$-dependent function which
depends on the steepness of the density profile (see
\citealt{Janka90,Janka_PhD}). Assuming further that the
(medium-dependent) coefficients $Q^+$ and $\kptl \equiv \kappa_{\rm
  a}/f = 4\pi r^2 Q^- / L$ are constant between two points $(r,t)$ and
$(\rstar,\tstar)$, which are connected by a characteristic line, i.e.,
\begin{equation}
  \rstar = r - \ceff \, (t-\tstar),
\end{equation}
Eq.~\eqref{eq:transL} can be integrated analytically to yield
\begin{eqnarray}
  \label{eq:transLsol}
  L(r,t) & = & L(\rstar,\tstar) \, \mathrm{e}^{-\kptl \ceff \, (t-\tstar)} \nonumber\\
         &   & + \frac{4\pi Q^+}{\kptl^3} \, \Big\{
                 [ 1 - \mathrm{e}^{-\kptl \ceff \, (t-\tstar)} ] \, [1 + (\kptl \rstar-1)^2] \nonumber\\
         &   &	 + \kptl \ceff \, (t-\tstar) [2 \kptl \rstar + \kptl \ceff \, (t-\tstar) - 2]
                                                \Big\},
\end{eqnarray}
where $L(r,t)$ and $L(\rstar,\tstar)$ are the luminosity values at
both ends of the characteristic line.

\begin{figure}[tpb!]
\centering
\includegraphics[width=4cm]{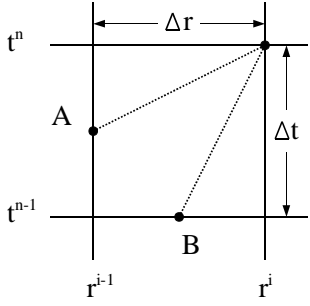}
\caption{The solution at $(r^i,t^n)$ is computed from the data at a
  point $(\rstar,\tstar)$ located on the same characteristic line.
  Depending on the grid spacing, $\Delta r$, the time step, $\Delta
  t$, and the effective speed of neutrino propagation, $\ceff$, either
  point A or point B must be used. The solution there can be obtained
  by interpolation in time or space, respectively.}
\label{fig:characteristics}
\end{figure}

We use Eq.~\eqref{eq:transLsol} to construct a numerical scheme to
solve Eq.~\eqref{eq:transL} in the general case: We assume that the
luminosity is known at the cell interfaces of a one-dimensional radial
grid for a time $t^{n-1}$, and that the cell-averaged values of the
quantities needed to compute the emission rate $Q^+$ and absorption
coefficient $\kptl$ are also known for that time. As a further
simplification we do not allow neutrinos to propagate in negative
radial direction (actually this is granted by defining a non-negative
function for the flux factor, see Sect.~\ref{sec:nudistfunc}).  Then
the luminosities at $t^n = t^{n-1} + \Delta t$ for each zone interface
(starting at the innermost zone) can be computed using
Eq.~\eqref{eq:transLsol}. In doing so we have to distinguish between
two cases (see Fig.~\ref{fig:characteristics}): If $\ceff \Delta t >
\Delta r$, we can use point A as the starting point of the
integration, $(\rstar,\tstar) = (r_{i-1},t_A)$.  The luminosity at
this point is derived from a linear interpolation between
$L(r_{i-1},t^{n-1})$ and $L(r_{i-1},t^n)$ (which is already known, as
we are integrating outwards). If $\ceff \Delta t \le \Delta r$, we use
point B, the luminosity at this point being given by a linear
interpolation between $L(r_{i-1},t^{n-1})$ and $L(r_{i},t^{n-1})$.

For time integration we use a predictor-corrector method: The
transport routine is called two times. In the first (predictor) step
the luminosities, emission rates and absorption coefficients of the
last time step $[L^{n-1}, Q^{n-1}, \kappa^{n-1}]$ are used to compute
preliminary values $(\tilde{Q}^n, \tilde{\kappa}^n)$ for the
neutrino-medium coupling at the next time level. In the second
(corrector) step the final values $[L^n, Q^n, \kappa^n]$ are
calculated using $[L^{n-1}, \frac{1}{2}(Q^{n-1}+\tilde{Q}^n),
\frac{1}{2}(\kappa^{n-1}+\tilde{\kappa}^n)]$ as input.

Equation \eqref{eq:transL} is solved not only for the energy
luminosity $L = L_e$, but also for the number luminosity $L_n = 4\pi
r^2 F_n = 4\pi r^2 f c n$ ($n$ is the particle density and $f$ is
assumed to be the same flux factor as for the energy transport).
Furthermore the equation has to be integrated for three neutrino
types, $\nue$, $\nuebar$, and $\nu_x$ (the latter denoting $\num$,
$\numbar$, $\nut$, and $\nutbar$, which are treated identically). In
the following we will use indices $\nu \in \{\nue,\nuebar,\nux\}$ and
$\alpha \in \{e,n\}$ to distinguish between these different
cases.

In the 2D case the neutrino transport is assumed to proceed only
radially, i.e. lateral components of the neutrino flux are ignored and
Eq.~\eqref{eq:transL} is integrated independently on different radial
``rays'', i.e. in radial direction for every lateral zone of the polar
coordinate grid.  Total luminosities of the star are obtained by
summing up the flux densities $L/4\pi r^2$ for all angular cells (at a
given radius $r$), appropriately weighting them with the corresponding
surface elements.

\subsection{Boundary conditions}
\label{sec:app_neutrino_bounds}

To integrate Eq.~\eqref{eq:transLsol} outwards, time-dependent boundary
conditions are required for the luminosities $L_{e,\nu_i}$ and
$L_{n,\nu_i}$, where $\nu_i = \nue,\nuebar,\nux$.  We assume
$L_{e,\nu_i}$ to be constant for a time interval $t_L$ (typically
1\,s), and to decay subsequently with a power-law dependence in time:
\begin{align}
L_{e,\nue}   (\rib,t) & = L_{\nu}^{\rm tot,0} \, K_{\nue}    \, h(t), \label{eq:lib1}\\
L_{e,\nuebar}(\rib,t) & = L_{\nu}^{\rm tot,0} \, K_{\nuebar} \, h(t),  \\
L_{e,\nux}   (\rib,t) & = L_{\nu}^{\rm tot,0} \, K_{\nux}    \, h(t), 
\end{align}
where
\begin{equation}
h(t) =
\begin{cases}
1.0           & \text{if $t\leq t_L$}, \\
(t_L/t)^{3/2} & \text{if $t > t_L$}.
\label{eq:libtimdep}
\end{cases}
\end{equation}
The constants $K_{\nu_i}$ denote the fractional contributions of the
individual luminosities to the total neutrino luminosity. They fulfill
the requirement
\begin{equation}
 K_{\nue} + K_{\nuebar} + 4 \, K_{\nux} = 1.
\label{eq:K_sum}
\end{equation}

\begin{figure}[tpb!]
\centering
\includegraphics[width=8.5cm]{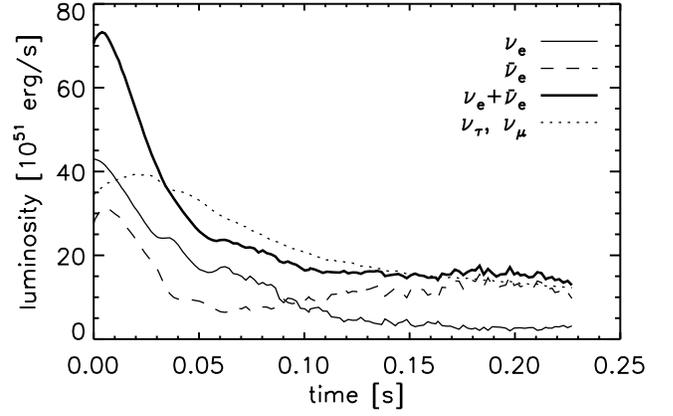}
\caption{Evolution after core bounce of the neutrino luminosities at a
  Lagrangian mass shell of $1.1\,\Msol$ from a supernova simulation
  with Boltzmann neutrino transport \citep{Buras+03}. After an initial
  phase of $~50\,$ms duration, the sum of the $\nue$ and $\nuebar$
  luminosities as well as the $\nu_{\mu}/\nu_{\tau}$ luminosities vary
  only slowly.}
\label{fig:boltz_lum}
\end{figure}

The functional form used in Eq.~(\ref{eq:libtimdep}) can be motivated
by the Boltzmann transport calculations of \cite{Buras+03}.  These
show that after a transient phase of $\sim 50$\,ms, which is short
compared to the explosion time scales of our simulations, the sum of
all luminosities is almost constant or varies only very weakly, at
least over the next $\sim 250$\,ms, for which data from the Boltzmann
transport simulations are available (see Fig.~\ref{fig:boltz_lum}).

According to Eqs.~(\ref{eq:lib1}--\ref{eq:libtimdep}) we need to
prescribe the time scale $t_L$ and the total initial luminosity
$L_{\nu}^{\rm tot,0}$. However, instead of choosing these two
quantities as basic parameters of our models, we prefer to prescribe
$t_L$ and the gravitational binding energy $\Delta
E^{\infty}_{\nu,\mathrm{core}}$ that is released by the neutron star
core asymptotically (i.e. for $t \rightarrow \infty$) via neutrino
emission. Introducing the energy that the core looses up to time $t$
\begin{equation}
  \Delta E^{\rm tot}_{\nu,\mathrm{core}}(t) = \int_0^{t}
         L_{\nu}^{\rm tot,0} \, h(t')  \, \dt',
\end{equation}
the following relations hold for the asymptotic energy loss
\begin{equation}
\Delta E^{\infty}_{\nu,\mathrm{core}} = 
\int_0^{\infty} L_{\nu}^{\rm tot,0} \, h(t) \, \mathrm{d}t =
3 \, \Delta E^{\rm tot}_{\nu,\mathrm{core}}(t_{L}) =
3 \, L_{\nu}^{\rm tot,0} \, t_L,
\end{equation}
i.e. our ansatz of Eq.~\eqref{eq:libtimdep} implies that $1/3$ of
$\Delta E^{\infty}_{\nu,\mathrm{core}}$ is radiated away within the
chosen time interval $t_L$ in neutrinos and antineutrinos of all
flavours.

We also prescribe the mean energies of neutrinos entering the
computational grid at the inner boundary. The corresponding values are
chosen to be $\langle\epsilon_{\nue}\rangle^{\rm ib} = 12 \, {\rm
MeV}$, $\langle\epsilon_{\nuebar}\rangle^{\rm ib} = 16 \, {\rm MeV}$,
and $\langle\epsilon_{\nux}\rangle^{\rm ib} = 20 \, {\rm MeV}$, and
kept constant during our simulations. Thereby also the number fluxes
$L_{n,\nu_i} = L_{e,\nu_i}/\langle\epsilon_{\nu_i}\rangle$ at $r=\Rib$
are defined.

The total lepton number lost by the neutron star
core until time $t$, normalised to the total baryon number $N_{\rm
  b,\,core}$ of the core, is given by
\begin{equation}
  \Delta Y_{\rm e,core}(t) = N_{\rm b,\,core}^{-1} \, \int_0^{t}
                         \left( L_{n,\nue}(\Rib,t')
                              - L_{n,\nuebar}(\Rib,t') \right) \dt'.
\label{eq:def_dYecoret}
\end{equation}
For $t=t_{\rm L}$ this yields
\begin{equation}
  \Delta Y_{\rm e,core}(t_L)
                  = \frac{L_{\nu}^{\rm tot,0} \, t_{\rm L}}{N_{\rm b,\,core}}
                    \left( \frac{K_{\nue}}{\langle\epsilon_{\nue}\rangle^{\rm ib}} 
                    - \frac{K_{\nuebar}}{\langle\epsilon_{\nuebar}\rangle^{\rm ib}} \right).
\end{equation}
We assume that the lepton number loss during time interval $t_{\rm L}$
is proportional to the energy loss during this time. Therefore we
choose $K_{\nu_i}=\mathrm{const}$, because
$\langle\epsilon_{\nu_i}\rangle^{\rm ib}=\mathrm{const}$, and set
$K_{\nue} = 0.2$, $K_{\nuebar} = 0.215$ for the calculations in this
paper. $K_{\nux}$ follows from Eq.~\eqref{eq:K_sum}.

\subsection{Neutrino distribution function}
\label{sec:nudistfunc}

To calculate the source terms in Eq.~\eqref{eq:transL} or
\eqref{eq:transLsol} we have to make an assumption about the neutrino
energy spectrum, i.e. about the energy dependency of the specific
intensity, which for particle energy and particle number is linked
with the particle distribution function $f_{D,\nu}$ in the following
way:
\begin{equation}
  I_{\nu,\{n,e\}}(t,r,\epsilon,\mu) = \\
  \left( \frac{\epsilon^{\{2,3\}}}{(hc)^3} \right) \; c \; f_{\mathrm{D},\nu}(t,r,\epsilon,\mu),
\end{equation}
where the exponent of $2$ applies for number transport and the
exponent of $3$ for energy transport, corresponding to the indices $n$
and $e$, respectively, of $I_\nu$.  We assume that
$f_{\mathrm{D},\nu}$ can be written as product of a Fermi-Dirac
distribution function,
\begin{equation}
  f_{\rm FD}(x,\eta) = \frac{1}{1 \; + \; \exp(x - \eta)},
\end{equation}
and an angle-dependent function $\gnu$,
\begin{equation}
  f_{D,\nu}(r,t,\epsilon,\mu) = g_{\nu}(r,t,\mu) \; f_{\rm FD}\left( \frac{\epsilon}{k_{\rm B}T_{\nu}(r,t)},\eta_{\nu}(r,t)\right),
\label{eq:def_fdnu}
\end{equation}
where in general the spectral temperature and degeneracy parameter,
$T_{\nu}$ and $\eta_{\nu}$, are different from the matter temperature
and equilibrium degeneracy parameter.

Furthermore we assume that $\eta_{\nu}$ is just a function of the
optical depth $\taunu$:
\begin{equation}
  \eta_{\nu}(\taunu) = \etaeqnu \, (1-\mathrm{e}^{-\taunu}) + \eta_{0,\nu} \, \mathrm{e}^{-\taunu},
  \label{eq:def_etaeff}
\end{equation}
where $\etaeqnu$ is the equilibrium degeneracy parameter and
$\eta_{0,\nu}$ is a chosen value as typically found in detailed
transport calculations for $\taunu \to 0$. The values for the
different neutrino types are (cf.
\citealt{Janka_PhD,Janka_Hillebrandt89,Myra_Burrows90,Keil+03}):
\begin{equation}
\begin{array}{lll}
  \eta_{\mathrm{eq},\nue} = (\mu_{\rm e} + \mu_{\rm p} - \mu_{\rm n}) /  \kb T, & \qquad & \eta_{0,\nue}    \equiv 3,  \\
  \eta_{\mathrm{eq},\nuebar} = - \eta_{\mathrm{eq},\nue},           & \qquad & \eta_{0,\nuebar} \equiv 2,  \\
  \eta_{\mathrm{eq},\nux} = 0,                               & \qquad & \eta_{0,\nux}    \equiv 0.
\end{array}
\end{equation}

With $\eta_{\nu}$ defined, $T_{\nu}$ can now be calculated from the
local average neutrino energy, which is computed from
$L_{e,\nu}$ and $L_{n,\nu}$ as
\begin{eqnarray}
  \label{eq:epsnuav}
  \langle\epsnu\rangle & = & L_{e,\nu} / L_{n,\nu}
                         = E_{\nu}(r,t) / n_{\nu}(r,t) \nonumber\\
           & = & \frac{ \intdeps \intdmu I_{e,\nu}(r,t,\epsilon,\mu) }
                      { \intdeps \intdmu I_{n,\nu}(r,t,\epsilon,\mu) } \nonumber\\
           & = & \kb T_{\nu} \, {\cal F}_3(\eta_{\nu}) / {\cal F}_2(\eta_{\nu}) 
\end{eqnarray}
with the Fermi integrals defined by
\begin{equation}
  {\cal F}_n(\eta) \equiv \int_0^{\infty} \mathrm{d}x  \; x^n \, f_{\rm FD}(x,\eta) 
\end{equation}
Thus the energy-dependent part of $f_{\mathrm{D},\nu}$ is fully defined. The
angle-dependent part is related to the flux factor by
\begin{equation}
  f_{\nu} = F_{\nu} / cE_{\nu} = \frac{\intdmu \mu I_{\nu}}{\intdmu I_{\nu}}
          = \frac{\intdmu \mu g_{\nu}(\mu)}{\intdmu g_{\nu}(\mu)} = \langle\mu_{\nu}\rangle.
\end{equation}
To solve Eq.~\eqref{eq:transL} only $\fnu$ is needed, not the
angle-dependent function $\gnu(\mu)$.  Far outside of the
neutrinosphere, $\fnu$ is approaching the {\em vacuum solution}. The
latter can be derived under the assumption that neutrinos are emitted
isotropically from the sharp surface of a sphere with radius $\Rnu$,
which is located at a distance $r$ from the observer. In this case the
flux factor is
\begin{equation}
  \fnuvac \, = \, {\textstyle \frac{1}{2}} \Big[ 1+\sqrt{1-(\Rnu/r)^2} \;  \Big].
\end{equation}
$\fnuvac$ approaches $1$ for $r \to \infty$ {\em (free streaming
  limit)} and $\fnuvac = 1/2$ at the neutrinosphere. In a more
realistic situation the neutrinosphere is not a sharp surface but a
layer with finite thickness in which neutrinos gradually decouple from
the stellar medium. In detailed transport calculations $\fnu(\Rnu)$ is
therefore found to be about $1/4$. \citep[see
e.g.][]{Janka_Hillebrandt89,Janka90}. How fast $\fnu$ approaches
$\fnuvac$ (with declining optical depth) depends on the steepness of
the density gradient at the neutrinosphere \citep{Janka90}. Inside the
neutrinosphere detailed transport calculations show that the flux
factor behaves roughly like $\fnu \propto \taunu^m$ with $m<0$.

Taking all this into account, the following function
constitutes a good approximation for the flux factors from
detailed transport calculations \citep{Janka_PhD,Janka90}:
\begin{equation}
  \fnu(\taunu) =
  \begin{cases}
	\frac{\displaystyle {\textstyle \frac{1}{2}}[1+D]}
         {\displaystyle 1+(1+D)(1-D^2)^{(n-1)/2}},
         & \quad \mathrm{if} \quad \taunu < \taunuo, \\
	1/4 \, (\taunu / \taunuo)^m,
         & \quad \mathrm{if} \quad \taunu > \taunuo.
  \end{cases}
  \label{eq:param_fluxf}
\end{equation}
Here $D = \sqrt{1-(\Rnu/r)^2}$, the neutrinosphere radius is defined
by $\tau_{\nu}(R_{\nu}) = \taunuo$ and we adopt $\taunuo = 1.1$. The
power-law index $m$ is chosen such that $\fnu(10) = 1/25$, and $n$ is
defined by a local power-law fit of the density profile around the
neutrinosphere, $\rho(r) \propto r^{-n}$. A higher value of $n$
therefore means a steeper density gradient.

\subsection{Neutrino reactions}

For calculating the neutrino-matter interaction rates the following
reactions are taken into account: charged-current processes with
neutrons (n) and protons (p),
\begin{align}
\label{eq:betaminus}
  \nue    + \mathrm{n} & \rightleftharpoons \mathrm{p} + \mathrm{e}^-, \\
  \nuebar + \mathrm{p} & \rightleftharpoons \mathrm{n} + \mathrm{e}^+,
\label{eq:betaplus}
\end{align}
thermal electron-positron (e$^{\pm}$) pair creation and annihilation,
\begin{equation}
  \mathrm{e}^+ + \mathrm{e}^- \rightleftharpoons \nu_i + \nubar_i \quad (i=\mathrm{e},\mu,\tau),
  \label{eq:pair_reacts}
\end{equation}
and neutrino scattering off nuclei (A), nucleons, and electrons and positrons,
\begin{align}
  \nu_i + \mathrm{A} & \rightleftharpoons \nu_i + \mathrm{A}, \\
  \nu_i + \left\{\begin{array}{c} \mathrm{n} \\ \mathrm{p} \end{array}\right\} & \rightleftharpoons \nu_i + \left\{\begin{array}{c} \mathrm{n} \\ \mathrm{p} \end{array}\right\}, \\
  \nu_i + \mathrm{e}^{\pm} & \rightleftharpoons \nu_i + \mathrm{e}^{\pm}.
\end{align}

\subsection{Optical depth}

Knowledge of the optical depth is necessary to evaluate
Eqs.~\eqref{eq:def_etaeff} and \eqref{eq:param_fluxf}. For this
purpose it is sufficient to compute $\tau_{\nu}$ approximately by
considering only the most relevant neutrino processes and assuming
that the neutrino spectrum is given by the spectrum for local
thermodynamic equilibrium. This means that instead of
Eq.~\eqref{eq:def_fdnu} we use 
\begin{equation}
  f_{{\rm D},\nu}^{\rm eq}(\epsnu,r) = f_{\rm FD}\left( \frac{\epsnu}{k_{\rm B}T(r)}, \eta_{\mathrm{eq},\nu}(r) \right)
\end{equation}
with $\eta_{\mathrm{eq},\nu}$ and $T$ instead of $\eta_{\nu}$ and
$T_{\nu}$.

The ``transport optical depth'' is defined as the integral
\begin{equation}
  \tautnu(r) = \int_r^\infty \dr' \langle\kaptnu\rangle(r')
  \label{eq:def_tautnu}
\end{equation}
of the energy-averaged ``transport opacity'' (i.e. the opacity which
is relevant for momentum transfer), $\langle\kaptnu\rangle(r)$ (see,
e.g., \citealt{Straumann89,Burrows_Thompson04}). In the following, all
neutrino interactions included in evaluating the opacity are
calculated without final-state lepton blocking, unless otherwise
stated.

The most important opacity-producing reactions are scattering off
nucleons (n,p) and nuclei ($Z_j,A_j$), where $j=1,2,\dots$ denotes the
considered nuclear species, and absorption by neutrons and protons in
case of $\nue$ and $\nuebar$, respectively. Thus one can write
\begin{equation}
  \langle\kaptnu\rangle = \langle\kpanu\rangle + \sum_{i \, \in \, \{{\rm n,p,}A_j\}} \langle\kaptnu^{{\rm s},i}\rangle.
\end{equation}
Here the (neutrino-flavour independent) scattering opacities are to
lowest order in neutrino energy over nucleon rest mass (i.e., without
effects of nucleon recoil, thermal motions, and weak-magnetism
corrections):
\begin{eqnarray}
  \label{eq:kappat_s}
  \langle\kaptnu^{\rm s,p}\rangle &=& \frac{1}{6} \left[ \frac{5}{4} \alpha^2 + (C_V-1)^2  \right]
                               \frac{\sigma_0}{(m_{\rm e} c^2)^2} \; \langle\epsnu^2\rangle \; n_{\rm p}, \\
  \langle\kaptnu^{\rm s,n}\rangle &=& \frac{ 5 \alpha^2 + 1}{24}
                               \frac{\sigma_0}{(m_{\rm e} c^2)^2} \; \langle\epsnu^2\rangle \; n_{\rm n}, \\
  \langle\kaptnu^{\mathrm{s},A_j}\rangle &=& \frac{1}{6} A_j^2 [C_A - 1 + \frac{Z_j}{A_j}(2-C_A-C_V)]^2 \nonumber\\
                           & & \times \, \frac{\sigma_0}{(m_{\rm e} c^2)^2} \; \langle\epsnu^2\rangle \; n_{A_j},
  \label{eq:kappat_sA}
\end{eqnarray}
for scattering off protons, neutrons, and nuclei with number densities
$n_{\mathrm{p}}$, $n_{\mathrm{n}}$, and $n_{A_j}$, respectively (see,
e.g., \citealt{Freedman+77,Straumann89,Burrows_Thompson04}).  The
absorption opacities for $\nue$ and $\nuebar$ are

\begin{eqnarray}
  \langle\kappa_{\nue}^{\mathrm{a}}\rangle &=&
    \frac{1}{4} ( 3 \alpha^2 + 1 ) \frac{\sigma_0}{(m_{\rm e} c^2)^2} \; n_{\rm n}\nonumber\\ 
	&& \times \; \left(        \langle\epsne^2\rangle
                         + 2\Delta  \langle\epsne\rangle
                         + \Delta^2   \right) \, \Theta(\langle\epsne\rangle), \\
  \langle\kappa_{\nuebar}^{\rm a}\rangle &=& 
    \frac{1}{4} ( 3 \alpha^2 + 1 ) \frac{\sigma_0}{(m_{\rm e} c^2)^2} \; n_{\rm p}\nonumber\\
	&& \times \; \left(  \langle\epsna^2\rangle^{\star}
                         + 2\Delta  \langle\epsna\rangle^{\star}
                         + \Delta^2 \langle\epsna^0\rangle^{\star}
                         \right).
  \label{eq:kappat_a}
\end{eqnarray}
\citep[see][]{Tubbs_Schramm75,Bruenn85}.

Here $\sigma_0 = 4 G_{\rm F}^2 m_{\rm e}^2 \hbar^2 / \pi c^2 = 1.76
\times 10^{-44}\,{\rm cm}^{2}$ (with the Fermi coupling constant
$G_{\rm F}$), $\Delta=1.2935$\,MeV is the rest mass difference of
neutrons and protons, $\alpha=1.254$, $C_A = \frac{1}{2}$, $C_V =
\frac{1}{2} + 2 \sin^2 \theta_{\rm W}$, and $\sin^2 \theta_{\rm W} =
0.23$.

In deriving Eqs.~\eqref{eq:kappat_s} -- \eqref{eq:kappat_a} (as well
as for all rates and source terms given below) the electron and
positron rest masses are ignored ($m_{\rm e}c^2 \ll \epsnu$) and
nucleons and nuclei are assumed to have infinite rest masses
($m_{\mathrm{n,p,}A_j}c^2 \gg \epsnu$) and to be nondegenerate.  For
electrons, phase space blocking is included in Eq.~\eqref{eq:kappat_a}
by the factor
\begin{equation}
  \Theta(\langle\epsnu\rangle) = 1-f_{\rm FD}\left( \frac{\langle\epsnu\rangle+\Delta}{k_{\rm B}T},\eta_{\rm e^-} \right),
\end{equation}
which accounts for the fact that a significant fraction of the
possible final electron states may be occupied. Phase space blocking
can be neglected in $\kappa_{\nuebar}^{\rm a}$
(Eq.~\ref{eq:kappat_a}), because the positrons are non-degenerate.

The neutrino energy moments are (generalising
Eq.~\ref{eq:epsnuav}) given by
\begin{equation}
  \langle\epsnu^n\rangle = (k_{\rm B} T_{\nu})^n \, \frac{ {\cal F}_{2+n}(\eta_{\nu}) }{ {\cal F}_2(\eta_{\nu}) },
  \label{eq:def_epnunav}
\end{equation}
\begin{equation}
  \langle\epsnu^n\rangle^{\star} = (k_{\rm B} T_{\nu})^n \, \frac{ {\cal F}_{2+n}(\eta_{\nu} - \Delta/k_{\rm B}T_{\nu}) }{ {\cal F}_2(\eta_{\nu}) },
  \label{eq:def_epnunavstar}
\end{equation}
and for evaluating Eqs.~\eqref{eq:kappat_s}--\eqref{eq:kappat_a} to
compute $\tau_{\mathrm{t},\nu}$ (Eq.~\ref{eq:def_tautnu}) for use in
Eq.~\eqref{eq:param_fluxf}, we take $\eta_{\nu} = \eta_{\mathrm{eq}, \nu}$
and $T_{\nu}=T$.

In contrast, Eq.~\eqref{eq:def_etaeff} is evaluated with the
``effective optical depth for equilibration'',
\begin{equation}
  \tau_{\nu}(r) = \int_r^\infty \dr' \langle\kappa_{\mathrm{eff},\nu}\rangle(r'),
\end{equation}
where the effective opacity is defined as
\begin{equation}
  \langle\kappa_{\mathrm{eff}, \nu}\rangle = \sqrt{
         \langle\kappa^{\mathrm{a}}_{\nu}\rangle \times
         \langle\kappa^{\mathrm{a}}_{\nu} + \kappa^{\mathrm{s}}_{\mathrm{t},\nu}\rangle }.
\label{eq:def_kappaeff}
\end{equation}
Here the spectrally averaged absorption opacity,
$\langle\kappa^{\mathrm{a}}_{\nu}\rangle$, is taken to include
neutrino-pair annihilation to $\mathrm{e}^{\pm}$-pairs
(Eq.~\ref{eq:pair_reacts}), which is assumed to be the most important
reaction for producing $\nu_x \bar{\nu}_x$ pairs. Both
$\langle\kappa^{\mathrm{a}}_{\nu}\rangle$ and
$\langle\kappa^{\mathrm{a}}_{\nu} +
\kappa^{\mathrm{s}}_{\mathrm{t},\nu}\rangle$ are evaluated for the
``true'' (not the local equilibrium) neutrino spectrum (i.e. for the
spectral temperature $T_{\nu}$ and the spectral degeneracy
$\eta_{\nu}$ instead of $T$ and $\eta_{\mathrm{eq},\nu}$) by employing
the source terms from the neutrino transport solution of the last time
step.

\subsection{Source terms}

Solving Eq.~\eqref{eq:transLsol} requires the knowledge of the
emission rates, $Q^+_{\nu_i}$, and absorption coefficients,
$\kptl_{\nu} = \kpanu / f_{\nu}$, which appear in this equation. Since
Eq.~\eqref{eq:transLsol} is used to determine the number fluxes,
$L_{n,\nu}$, and luminosities, $L_{e,\nu}$, of all neutrinos and
antineutrinos $\nu \in \{\nue,\nuebar,\nux\}$, the source terms need
to be calculated for the neutrino number as well as energy. In the
following, all these neutrino source terms are derived without taking
into account final-state lepton blocking, unless otherwise stated. As
in Eq.~\eqref{eq:def_kappaeff}, $\kpanu$ is defined to include the
contributions from the $\beta$-processes, Eqs.~(\ref{eq:betaminus})
and (\ref{eq:betaplus}) for $\nue$ and $\nuebar$ as well as those of
e$^+$e$^-$ pair annihilation (Eq.~\ref{eq:pair_reacts}). The
absorption coefficient $\kpanu$ can be computed from the corresponding
neutrino absorption rate by
\begin{equation}
  \kpanu = Q^-_{\nu} 4\pi r^2 f_{\nu} / L_{\nu}
  = (Q^{\rm a}_{\nu}+Q^{\rm ann}_{\nu\bar{\nu}}) \, 4\pi r^2 f_{\nu} / L_{\nu}.
\label{eq:def_kpanu}
\end{equation}
For the number transport the neutrino absorption and emission rates
(in units of number per cm$^3$ per second) by charged-current
$\beta$-reactions between leptons and nucleons can be written with our
approximations for the neutrino distribution function and the
appropriate statistical weights for the leptons as follows:

\begin{eqnarray}
  \label{eq:def_rae}
  \Rae & = & \sigma c \frac{L_{e,\nue} \, n_{\rm n}}{4\pi r^2 c f_{\nue}}
             \, \frac{ \langle\epsne^2\rangle + 2\Delta \langle\epsne\rangle + \Delta^2 }{\langle\epsne\rangle} 
             \, \Theta(\langle\epsne\rangle), \\
  \label{eq:def_raa}
  \Raa & = & \sigma c \frac{L_{e,\nuebar} \, n_{\rm p}}{4\pi r^2 c f_{\nuebar}}
             \, \frac{ \langle\epsna^2\rangle^{\star} + 2\Delta \langle\epsna\rangle^{\star} + \Delta^2 \langle\epsna^0\rangle^{\star}}{\langle\epsna\rangle}, \\
  \label{eq:def_ree}
  \Ree & = & \frac{1}{2} \sigma c \, n_{\rm p} \, n_{{\rm e}^-} \,
             [ \langle\epsem^2\rangle^{\star} + 2\Delta \langle\epsem\rangle^{\star} + \Delta^2 \langle\epsem^0\rangle^{\star} ], \\
  \label{eq:def_rea}
  \Rea & = & \frac{1}{2} \sigma c \, n_{\rm n} \, n_{{\rm e}^+} \,
             [ \langle\epsep^2\rangle + 2\Delta \langle\epsep\rangle + \Delta^2 ],
\end{eqnarray}
where $\sigma = \frac{1}{4}(3\alpha^2+1) \sigma_0 / (m_{\rm e}c^2)^2$
and the electron (positron) number density is
\begin{equation}
n_{\mathrm{e^\mp}} = {8\pi\over (hc)^3}\,(k_{\mathrm{B}})^3 
                                         {\cal F}_2(\pm\eta_{\mathrm{e}^-})\,. 
\end{equation}
The electron and positron energy moments are given by
\begin{align}
  \langle\epse^n\rangle = & (k_{\rm B}T)^n \, \frac{{\cal F}_{2+n}(\eta_{\mathrm{e}})}{{\cal F}_2(\eta_{\mathrm{e}})}, \\
  \langle\epse^n\rangle^{\star} = & (k_{\rm B}T)^n \, \frac{{\cal F}_{2+n}(\eta_{\mathrm{e}}-\Delta/k_{\rm B}T)}{{\cal F}_2(\eta_{\mathrm{e}})}.
\end{align}

The annihilation and production rates of neutrino number in
$\mathrm{e}^+\mathrm{e}^-$ pair reactions are given by (adapted from
\citealt{Schinder+87}; see also \citealt{Janka91} and
\citealt{Janka_PhD}, and references therein):
\begin{eqnarray}
\label{eq:rpairann}
  {\cal R}_{\nu}^{\rm ann} & = & \frac{\sigma_0 c}{(4\pi r^2 c)^2}
                      \frac{L_{n,\nu} L_{n,\nubar}}{\eavnu \eavnb}
                \, \Bigg\{ 
                      \frac{2}{9} \frac{\Phi(\fnu,\chinu)}{\fnu\fnubar}
                      \frac{\canu^2+\cvnu^2}{(m_{\rm e}c^2)^2}
                      \langle\epsnu\rangle \langle\epsnb\rangle \nonumber\\
               &   & + \frac{1}{6}\frac{1-\fnu\fnubar}{\fnu\fnubar}
                       (2\cvnu^2-\canu^2)
                       \Bigg\}, \\
\label{eq:rpairprod}
  {\cal R}_{\nu}^{\rm prod} & = & \frac{1}{18} \frac{\sigma_0 c}{(m_{\rm e}c^2)^2} 
                      n_{{\rm e}^-} n_{{\rm e}^+} \,\Bigg\{ (\canu^2+\cvnu^2)
                      \langle\epsem\rangle \langle\epsep\rangle \nonumber\\
                &   &  + \frac{3}{4} (m_{\rm e}c^2)^2 (2\cvnu^2-\canu^2) \Bigg\}.
\end{eqnarray}

These rates hold for neutrinos $\nu$ or antineutrinos $\nubar$ of all
flavours. In Eq.~\eqref{eq:def_kpanu}, $\cal{R}^{\rm a}_{\nu}$ and
$\cal{R}^{\rm ann}_{\nu}$ have to be used instead of $Q^{\rm a}_{\nu}$
and $Q^{\rm ann}_{\nu}$ for computing the absorption coefficient for
the number transport. In Eq.~\eqref{eq:rpairann}, $\Phi(\fnu,\chinu)$
is a geometrical factor,
\begin{equation}
  \Phi(\fnu,\chinu) = \frac{3}{4}\left[ 1 - 2\fnu\fnubar
      +\chinu\chinubar +\frac{1}{2}(1-\chinu)(1-\chinubar) \right],
\end{equation}
where we express the variable Eddington factor $\chinu$ in terms of
the flux factor $\fnu$ (Eq.~\ref{eq:def_fluxf}) using a statistical
form, which was derived by \cite{Minerbo78} on grounds of maximum
entropy considerations (for photons or nondegenerate neutrinos, as
assumed here):
\begin{equation}
  \chinu = \langle\munu^2\rangle = \frac{1}{3}
         + \frac{0.01932 \, \fnu + 0.2694 \, \fnu^2}
                {1 - 0.5953 \, \fnu + 0.02625 \, \fnu^2}.
\end{equation}

The weak coupling constants in Eqs.~\eqref{eq:rpairann} and
\eqref{eq:rpairprod} are given by
\begin{equation}
  \canu = 
  \begin{cases}
	\, +\frac{1}{2} & \text{for $\nu \, \in \, \{\nue,\nuebar\}$}, \\
	\, -\frac{1}{2} & \text{for $\nu \, \in \, \{\num,\numbar,\nut,\nutbar\}$},
  \end{cases}
\end{equation}
\begin{equation}
  \cvnu = 
  \begin{cases}
	\, +\frac{1}{2} + 2\sin^2\theta_{\rm W} & \text{for $\nu \, \in \, \{\nue,\nuebar\}$}, \\
	\, -\frac{1}{2} + 2\sin^2\theta_{\rm W} & \text{for $\nu \, \in \, \{\num,\numbar,\nut,\nutbar\}$}.
  \end{cases}
\end{equation}

The source term which describes the rate of change per unit of volume
in the evolution equation of the electron lepton number of the stellar
medium is
\begin{equation}
  Q_{\rm N} = \dot{Y}_{\rm e} \, n_{\rm b} = (\Rae-\Ree) - (\Raa-\Rea).
\label{eq:def_qn}
\end{equation}

The source terms which account for the absorption and emission of
energy through $\nue$ and $\nuebar$ are computed in analogy to
Eqs.~\eqref{eq:def_rae} -- \eqref{eq:def_rea} as
\begin{eqnarray}
  Q_{\nue}^{\rm a}   & = & \sigma c \frac{L_{e,\nue} \, n_{\rm n}}{4\pi r^2 c f_{\nue}}
             \, \frac{     \eavnet{3}
                       + 2\Delta \eavnet{2}
                       + \Delta^2 \eavne }{\eavne} \, \Theta(\eavne), \\
  Q_{\nuebar}^{\rm a} & = & \sigma c \frac{L_{e,\nuebar} \, n_{\rm p}}{4\pi r^2 c f_{\nuebar}} \nonumber\\
             & & \times \frac{      \eavnast{3}
                       + 3\Delta  \eavnast{2}
                       + 3\Delta^2 \eavnas
                       + \Delta^3 \eavnast{0} }{ \eavna}, \\
  Q_{\nue}^{\rm e}   & = & \frac{\sigma c}{2} \; n_{\rm p} \, n_{{\rm e}^-} 
             \left[        \eavemst{3}
                       + 2\Delta \eavemst{2}
                       + \Delta^2 \eavems \right], \\
  Q_{\nuebar}^{\rm e} & = & \frac{\sigma c}{2}  \; n_{\rm n} \, n_{{\rm e}^+}
             \left[         \eavept{3}
                       + 3\Delta  \eavept{2}
                       + 3\Delta^2 \eavep
                       + \Delta^3  \right].
\end{eqnarray}

The annihilation or production of energy in neutrinos ($\nu$) by
$\mathrm{e}^+\mathrm{e}^-$ pair reactions is given as
(\citealt{Janka91})
\begin{eqnarray}
\label{eq:qpairann}
  Q_{\nu}^{\rm ann} & = & \frac{\sigma_0 c}{(4\pi r^2 c)^2}
                      \frac{L_{e,\nu} L_{e,\nubar}}{\eavnu \eavnb}
                \, \Bigg\{ 
                      \frac{2}{9} \frac{\Phi(\fnu,\chinu)}{\fnu\fnubar}
                      \frac{\canu^2+\cvnu^2}{(m_{\rm e}c^2)^2}
                      \langle\epsnu^2\rangle \langle\epsnb\rangle \nonumber\\
               &   & + \frac{1}{6}\frac{1-\fnu\fnubar}{\fnu\fnubar}
                       (2\cvnu^2-\canu^2) \langle\epsnu^2\rangle
                       \Bigg\}, \\
\label{eq:qpairprod}
  Q_{\nu}^{\rm prod} & = & \frac{1}{36} \frac{\sigma_0 c}{(m_{\rm e}c^2)^2}
                          n_{{\rm e}^-} n_{{\rm e}^+} \nonumber\\
                   &   & \times \Bigg\{ 
           [\eavemt{2} \eavep + \eavept{2} \eavem] \,  (\canu^2+\cvnu^2) \nonumber\\
       && + \frac{3}{4} (m_ec^2)^2 [\eavem+\eavep] \,  (2\cvnu^2-\canu^2)
                                  \Bigg\}.
\end{eqnarray}

For annihilation of antineutrino ($\bar{\nu}$) energy,
$\langle\epsnu^2\rangle$ has to be replaced by
$\langle\epsnb^2\rangle$ and $\langle\epsnu\rangle$ has to be
exchanged with $\langle\epsnb\rangle$ in Eq.~\eqref{eq:qpairann},
while the production of $\nu$ and $\bar{\nu}$ was assumed to be
symmetric and both rates are given by Eq.~\eqref{eq:qpairprod}.

Also in scattering processes energy can be exchanged between neutrinos
and the stellar medium. For scattering off e$^-$ or e$^+$, using the
rates of \cite{Tubbs_Schramm75}, and ignoring electron phase space
blocking in the final reaction channels, the following spectrally
averaged expression for the energy transfer rate per unit of volume
can be derived \citep[see][]{Janka_PhD}:
\begin{eqnarray}
  \label{eq:qnuscat}
  Q_{\nu \mathrm{e}} & = & \frac{1}{12} (C_1+\frac{1}{6}C_2) \frac{\sigma_0 c}{(m_{\rm e}c^2)^2} n_{{\rm e}}
                        \frac{L_{e,\nu}}{4\pi r^2 c \fnu \eavnu} \nonumber\\
                  & \, & \Bigg\{
                \Big[ \eavnut{2}(\eave+\frac{3}{4}m_{\rm e}c^2) - \eavnu \eavet{2} \Big] \nonumber\\
                  &   & + \frac{3}{8} \frac{C_3}{C_1+\frac{1}{3}C_2} (m_{\rm e}c^2)^2
                          \Big[\eavnu-\frac{\eavet{2}}{\eave}\Big]
                            \Bigg\},
\end{eqnarray}
where e can be $\mathrm{e}^+$ or e$^-$ and $\nu$ stands for neutrinos
or antineutrinos of all flavours and the constants $C_1, C_2, C_3$ for
the different combinations are listed in Table~\ref{tab:coupcon}.  The
term $3m_{\rm e}c^2/4$ in the bracket results from a merge of the rate
expressions for the limits of relativistic and non-relativistic
electrons. In the latter case the neutrino-electron scattering cross
section is proportional to $\epsilon_\nu/(m_{\rm e}c^2)$ for
$\epsilon_\nu \gg m_{\rm e}c^2$ (cf.\ \citealt{Sehgal74}).

\begin{table}
\centering
\caption{Weak coupling constants for $\nu$ and $\bar{\nu}$ scattering off
e$^+$ or e$^-$ (cf. Eq.~\ref{eq:qnuscat}). $C_3^{\rm x}$ stands for
$C_3^x=(\CA-1)^2-(\CV-1)^2$, $\CA=\frac{1}{2}$, $\CV=\frac{1}{2}+2\sin^2\theta_{\rm W}$,
and $\nu_x$ can be $\nu_{\mu}$ or $\nu_{\tau}$.}
\begin{equation*}
  \begin{array}{lc|cccccc}
    \hline
    \hline
                         & \: & \: & C_1         &   \: & C_2         &   \: & C_3         \\ \hline
	\nue    \mathrm{e}^- & \: & \: & (\CV+\CA)^2 &   \: & (\CV-\CA)^2 &   \: & \CA^2-\CV^2 \\
	\nue    \mathrm{e}^+ & \: & \: & (\CV-\CA)^2 &   \: & (\CV+\CA)^2 &   \: & \CA^2-\CV^2 \\
	\nuebar \mathrm{e}^- & \: & \: & (\CV-\CA)^2 &   \: & (\CV+\CA)^2 &   \: & \CA^2-\CV^2 \\
	\nuebar \mathrm{e}^+ & \: & \: & (\CV+\CA)^2 &   \: & (\CV-\CA)^2 &   \: & \CA^2-\CV^2 \\
	\nux    \mathrm{e}^- & \: & \: & (\CV+\CA-2)^2 & \: & (\CV-\CA)^2 &   \: & C_3^x \\
	\nux    \mathrm{e}^+ & \: & \: & (\CV-\CA)^2 &   \: & (\CV+\CA-2)^2 & \: & C_3^x \\
	\nuxbar \mathrm{e}^- & \: & \: & (\CV-\CA)^2 &   \: & (\CV+\CA-2)^2 & \: & C_3^x \\
	\nuxbar \mathrm{e}^+ & \: & \: & (\CV+\CA-2)^2 & \: & (\CV-\CA)^2 &   \: & C_3^x
  \end{array}
\end{equation*}
\label{tab:coupcon}
\end{table}

Every transfer by neutrino-nucleon scattering, which is only ``nearly
conservative'', is taken into account following \cite{Tubbs79}. The
corresponding rate is \citep[see][]{Janka_PhD}:
\begin{align}
  Q_{\nu {\rm N}}  = &  \frac{1}{4} \frac{\sigma_0 c}{(m_{\rm e}c^2)^2}
                   {\cal C}_{\rm N} {\cal E}_{\rm N} \frac{n_{\rm N}}{m_{\rm N}c^2}
                   \, \{ \eavnut{4} - 6T\eavnut{3} \} \nonumber\\
                 & \times \frac{L_{e,\nu}}{4\pi r^2 c \fnu \eavnu}
\end{align}
with
\begin{equation}
{\cal C}_{\rm N} {\cal E}_{\rm N} =
  \begin{cases}
	\frac{2}{3}[(\CV-1)^2+\frac{5}{4}\alpha^2] & \text{for ${\rm N=p}$}, \nonumber\\
	\frac{1}{6}(1+5\alpha^2)                   & \text{for ${\rm N=n}$}.
  \end{cases}
\end{equation}
The symbol $\nu$ stands again for neutrinos and antineutrinos of all
flavours. Also scattering contributions are included in the energy
generation rate $Q^+$ and absorption coefficient $\kptl$ used in
Eq.~\eqref{eq:transLsol}. Considering scattering as an absorption
process followed immediately by an emission process, we add the net
energy exchange rates $Q_{\nu \mathrm{e}^-}$, $Q_{\nu \mathrm{e}^+}$,
$Q_{\nu \mathrm{p}}$ and $Q_{\nu \mathrm{n}}$ to $Q^-_{\nu}$ (used for
computing $\kptl_{\nu}$ in Eq.~\ref{eq:transLsol}) when the rates are
positive (i.e.  in case of energy transfer from neutrinos to the
stellar gas), and the absolute values of $Q_{\nu \mathrm{e}^-}$,
$Q_{\nu \mathrm{e}^+}$, $Q_{\nu \mathrm{p}}$ and $Q_{\nu \mathrm{n}}$
to $Q^+_{\nu}$ otherwise. The total neutrino energy source term to be
used in the gas energy equation including the contributions from
$\nue$ and $\nuebar$ absorption and emission, $\nu\bar{\nu}$ pair
production, and all scattering reactions is
\begin{eqnarray}
  Q_{\rm E} & = & \sum_{\nu \, \in \, \{\nue,\nuebar\}} \left(Q_{\nu}^{\rm a} - Q_{\nu}^{\rm e}\right)
                     + \sum_{\nu \, \in \, \{\nue,\num,\nut\}}
                       \left(Q_{\nu\nubar}^{\rm ann} - Q_{\nu\nubar}^{\rm prod}\right) \nonumber\\
          & + & \sum_{\nu \, \in \, \{\nue,\num,\nut,\nuebar,\numbar,\nutbar\}}
                       \left(Q^{\rm s}_{\nu {\rm e}^+} + Q^{\rm s}_{\nu {\rm e}^-} + Q^{\rm s}_{\nu {\rm p}} + Q^{\rm s}_{\nu {\rm n}}\right),
\label{eq:def_qe}
\end{eqnarray}
where $Q_{\nu\nubar}^{\rm ann} = Q_{\nu}^{\rm ann} + Q_{\nubar}^{\rm
  ann}$ and $Q_{\nu\nubar}^{\rm prod} = Q_{\nu}^{\rm prod} +
Q_{\nubar}^{\rm prod}$.

In practise, however, the lepton number source term $Q_{\rm N}$ as
well as the energy source term for the hydrodynamics part of the code
is not computed from Eq.~\eqref{eq:def_qn} and \eqref{eq:def_qe},
respectively, but from the luminosity change between points
$(r^i,t^n)$ and $(\rstar,\tstar)$ (cf.
Fig.~\ref{fig:characteristics}). The source terms $\tilde{Q}^{i}_{\rm
  N}$ and $\tilde{Q}^{i}_{\rm E}$ for a grid cell $i$ at time level
$t^n$ are then given by
\begin{equation}
  \tilde{Q}^{i}_{\rm N} = \frac{ L^{\mathrm{diff}}_{n}(r^i,t^n)
                               - L^{\mathrm{diff}}_{n}(r^{\star},t^{\star}) }
                               { \Delta V_i },
\label{eq:def_qtln}
\end{equation}
\begin{equation}
  \tilde{Q}^{i}_{\rm E} = \frac{ L^{\mathrm{tot}}_{e}(r^i,t^n)
                               - L^{\mathrm{tot}}_{e}(r^{\star},t^{\star}) }
                               { \Delta V_i },
\label{eq:def_qtle}
\end{equation}
where $\Delta V_i = \frac{4\pi}{3}(r_i^3-{r^{\star}}^3)$ is the part
of the cell volume crossed by the characteristic line between
$(r^i,t^n)$ and $(\rstar,\tstar)$, $L^{\mathrm{tot}}_{e}$ is the sum
of the luminosities of neutrinos and antineutrinos of all flavours,
and $L^{\mathrm{diff}}$ is the difference between the $\nue$ and
$\nuebar$ number fluxes, $L_{\nue,n}-L_{\nuebar,n}$.
Equations~\eqref{eq:def_qtln} and \eqref{eq:def_qtle} work well as a
description of the neutrino sources in the gas equations only, if the
neutrino fluxes do not exhibit a large degree of variability on the
radial and temporal scales of the $r$-$t$ cells. This, however, is
reasonably well fulfilled in the context considered in this paper.

Finally, the outgoing neutrino fluxes transfer also momentum to the
stellar fluid. To account for this, we include a momentum source term
$Q_{\rm M}$ which enters the Euler equation of the hydrodynamics
solver. It is sufficient to include only the most important reactions,
by which neutrinos transfer momentum, i.e. $\nue$ and $\nuebar$
absorption on n and p, respectively, and the scattering processes of
$\nu$ and $\bar{\nu}$ of all flavours off nucleons and nuclei (pair
processes and electron/positron scattering can be safely ignored). For
a neutrino or antineutrino $\nu$, the corresponding rate (in units of
erg/cm$^4$) is
\begin{equation}
  Q_{\rm M}^{\nu} = \frac{ L_{e,\nu} }{4 \pi r^2 c} \left(
    \frac{\langle\kpanu\epsilon_{\nu}\rangle}{\langle\epsilon_{\nu}\rangle}
    + \sum_{i \in \{\mathrm{p},\mathrm{n},A_j\}} \frac{\langle\kaptnu^{\mathrm{s},i}\epsilon_{\nu}\rangle}{\langle\epsilon_{\nu}\rangle}
  \right),
\end{equation}
where the first term in the sum is relevant only for $\nue$ and
$\nuebar$. The energy averages of the scattering transport opacities,
$\kappa^{\mathrm{s},i}_{\mathrm{t},\nu}$, and of the absorption
opacities, $\kpanu$, all weighted by the neutrino energy, are given by
\begin{align}
  \langle\kappa^{\mathrm{s,p}}_{\mathrm{t},\nu}\epsnu\rangle  = & 
  \frac{1}{6} \left[ \frac{5}{4} \alpha^2 +(\CV-1)^2\right]
  \frac{\sigma_0}{(m_{\rm e}c^2)^2} \langle\epsilon^3_{\nu}\rangle n_{\rm p},\\
  \langle\kappa^{\mathrm{s,n}}_{\mathrm{t},\nu}\epsnu\rangle  =  &
  \frac{5\alpha^2+1}{24} \frac{\sigma_0}{(m_{\rm e}c^2)^2}
  \langle\epsilon^3_{\nu}\rangle n_{\rm n},\\
  \langle\kappa^{\mathrm{s},A_j}_{\mathrm{t},\nu}\epsnu\rangle  = &
  \frac{1}{6} A_j^2 \left[ \CA-1+\frac{Z_j}{A_j}(2-\CA-\CV)\right]
  \frac{\sigma_0}{(m_{\rm e}c^2)^2} \langle\epsilon^3_{\nu}\rangle n_{A_j},\\
  \langle\kappa^{\rm a}_{\nue}\epsne\rangle = &             
  \frac{1}{4}(3\alpha^2+1) \frac{\sigma_0}{(m_{\rm e}c^2)^2}\\
  & \times \left( \langle\epsilon_{\nue}^3\rangle
         + 2\Delta \langle\epsilon_{\nue}^2\rangle
         +\Delta^2\langle\epsilon_{\nue}\rangle
  \right) \Theta(\langle\epsilon_{\nue}\rangle), \nonumber\\
  \langle\kappa^{\rm a}_{\nuebar}\epsna\rangle = &           
  \frac{1}{4}(3\alpha^2+1) \frac{\sigma_0}{(m_{\rm e}c^2)^2} \\
  & \times \left( \langle\epsilon_{\nuebar}^3\rangle^{\star}
         + 3\Delta \langle\epsilon_{\nuebar}^2\rangle^{\star}
         + 3\Delta^2\langle\epsilon_{\nuebar}\rangle^{\star}
         + \Delta^3\langle\epsilon_{\nuebar}^0\rangle^{\star} \nonumber
  \right).
\end{align}
The energy moments $\langle\epsilon_{\nu}^n\rangle$ and
$\langle\epsilon_{\nu}^n\rangle^{\star}$ are given in
Eqs.~\eqref{eq:def_epnunav} and \eqref{eq:def_epnunavstar}. They are
calculated using the nonequilibrium neutrino spectral parameters
$T_{\nu}$ and $\eta_{\nu}$. The momentum source term in the equation
of gas motion then reads
\begin{equation}
  Q_{\rm M} = \sum_{\nu \, \in \, \{\nue,\num,\nut, \nuebar,\numbar,\nutbar\}} Q_{\rm M}^{\nu}
\end{equation}
It was not included in the simulations presented in this paper, but
will be taken into account in future calculations.

The implementation of the source terms $\tilde{Q}_{\rm N}$,
$\tilde{Q}_{\rm E}$, and $Q_{\rm M}$ into the framework of our PPM
hydrodynamics code was discussed in detail by \cite{RJ02} and
\cite{Buras+06}.

We finish by pointing out that the approximative neutrino transport
scheme developed here employs two basic assumptions, which are radical
simplifications of the true situation:
\begin{enumerate}
\item In deriving Eq.~\eqref{eq:transLsol} from the transport equation
  we assumed that the flux factor $f(r,t)$ is a \emph{known} function,
  although it is actually dependent on the solution of the transport
  problem (see Eq.~\ref{eq:def_fluxf}). Equation~\eqref{eq:transLsol}
  certainly has the advantage of analytic simplicity, but also has a
  severe disadvantage: The source terms can be very large and the
  numerical use requires a very fine grid zoning at high optical
  depths. The cell size should fulfill the constraint that the optical
  depth of the cell stays around unity or less. Moreover, the
  implementation of the source terms in \eqref{eq:transLsol} and the
  medium sources (Eqs.~\ref{eq:def_qtln}, \ref{eq:def_qtle}) is not
  symmetric and the numerical scheme does not strictly conserve the
  total lepton number and total energy of neutrinos plus gas.
\item For treating the spectral dependence, we made the assumption
  that the neutrino phase space distribution function can be
  factorised into a product of an angle-dependent function $\gnu$ and
  an energy-dependent term, which we assume to be of Fermi-Dirac
  shape. This certainly constrains the spectral shape, but the
  factorisation also implies that the flux-factor is assumed not to be
  an energy-dependent quantity. This in turn means that the mean
  energy of the neutrinos flux, $\langle\epsnu\rangle_{\rm flux}
  \equiv L_{e,\nu}(r,t)/L_{n,\nu}(r,t)$ is identical with the mean
  energy of the local neutrino density, $\langle\epsnu\rangle_{\rm
    local} \equiv E_{\nu}(r,t)/n_{\nu}(r,t)$. This is certainly a
  problematic simplification in view of the fact that the neutrino
  interactions with the stellar medium are strongly energy-dependent.
\end{enumerate}

Nevertheless, the described neutrino transport treatment represents a
practical approximation which is able to reproduce basic features of
more detailed transport solutions and yields agreement with those even
beyond the purely qualitative level.

\end{document}

%% file: definitions.tex
\newcommand{\kb}{k_{\rm B}}

\newcommand{\Ye}{Y_{\rm e}}

\newcommand{\nubar}{\bar \nu} 
\newcommand{\nux}{\nu_x} 
\newcommand{\nuxbar}{\bar \nu_x} 
\newcommand{\nue}{\nu_{\rm e}} 
\newcommand{\nuebar}{{\bar \nu}_{\rm e}} 
\newcommand{\num}{\nu_\mu} 
\newcommand{\numbar}{{\bar \nu}_\mu} 
\newcommand{\nut}{\nu_\tau} 
\newcommand{\nutbar}{{\bar \nu}_\tau} 
\newcommand{\Msol}{M_{\odot}}

\newcommand{\rvec}{\vec{r}}
\newcommand{\vvec}{\vec{v}}
\newcommand{\evec}{\vec{e}}
\newcommand{\avec}{\vec{a}}
\newcommand{\gvec}{\vec{g}}
\newcommand{\nabvec}{\vec{\nabla}}

\newcommand{\ddr}{\frac{\partial}{\partial r}} 
\newcommand{\ddt}{\frac{\partial}{\partial t}} 
\newcommand{\ddmu}{\frac{\partial}{\partial \mu}} 
\newcommand{\intdmu}{\int_{-1}^{+1}{\rm d}\mu \;}
\newcommand{\intdeps}{\int_{0}^{\infty}{\rm d}\epsilon \;}

\newcommand{\kpanu}{\kappa^{\rm a}_{\nu}}

\newcommand{\ceff}{c_{\rm eff}}
\newcommand{\rstar}{r^{\star}}
\newcommand{\tstar}{t^{\star}}
\newcommand{\kptl}{\tilde{\kappa}}

\newcommand{\etaeqnu}{\eta_{{\rm eq},\nu}}

\newcommand{\epsnu}{\epsilon_{\nu}}
\newcommand{\epsnb}{\epsilon_{\nubar}}
\newcommand{\epsne}{\epsilon_{\nue}}
\newcommand{\epsna}{\epsilon_{\nuebar}}
\newcommand{\epse}{\epsilon_{\rm e}}
\newcommand{\epsem}{\epsilon_{{\rm e}^-}}
\newcommand{\epsep}{\epsilon_{{\rm e}^+}}

\newcommand{\eavnu}{\langle\epsnu\rangle}
\newcommand{\eavnb}{\langle\epsnb\rangle}
\newcommand{\eavne}{\langle\epsne\rangle}
\newcommand{\eavna}{\langle\epsna\rangle}
\newcommand{\eave}{\langle\epse\rangle}
\newcommand{\eavep}{\langle\epsep\rangle}
\newcommand{\eavem}{\langle\epsem\rangle}

\newcommand{\eavnut}[1]{\langle\epsnu^#1\rangle}

\newcommand{\eavnet}[1]{\langle\epsne^#1\rangle}

\newcommand{\eavet}[1]{\langle\epse^#1\rangle}
\newcommand{\eavept}[1]{\langle\epsep^#1\rangle}
\newcommand{\eavemt}[1]{\langle\epsem^#1\rangle}

\newcommand{\eavnas}{\langle\epsna\rangle^{\!\star}}

\newcommand{\eavems}{\langle\epsem\rangle^{\!\star}}

\newcommand{\eavnast}[1]{\langle\epsna^#1\rangle^{\!\star}}

\newcommand{\eavemst}[1]{\langle\epsem^#1\rangle^{\!\star}}

\newcommand{\taunu}{\tau_{\nu}}
\newcommand{\taunuo}{\tau_{\nu,1}}

\newcommand{\fnu}{f_{\nu}}
\newcommand{\fnubar}{f_{\bar{\nu}}}
\newcommand{\chinu}{\chi_{\nu}}
\newcommand{\chinubar}{\chi_{\bar{\nu}}}
\newcommand{\munu}{\mu_{\nu}}

\newcommand{\gnu}{g_{\nu}}
\newcommand{\Rnu}{R_{\nu}}

\newcommand{\fnuvac}{f_{\nu,{\rm vac}}}

\newcommand{\tautnu}{\tau_{{\rm t},\nu}}
\newcommand{\kaptnu}{\kappa_{{\rm t},\nu}}

\newcommand{\Rae}{{\cal R}^{\rm a}_{\nue}}
\newcommand{\Ree}{{\cal R}^{\rm e}_{\nue}}
\newcommand{\Raa}{{\cal R}^{\rm a}_{\nuebar}}
\newcommand{\Rea}{{\cal R}^{\rm e}_{\nuebar}}

\newcommand{\canu}{{\cal C}_{{\rm A}\nu}}
\newcommand{\cvnu}{{\cal C}_{{\rm V}\nu}}
\newcommand{\CA}{C_{\rm A}}
\newcommand{\CV}{C_{\rm V}}

\newcommand{\dr}{{{\rm d}r}} 
\newcommand{\dt}{{{\rm d}t}} 
 
\newcommand{\dtheta}{{{\rm d}\theta}} 
\newcommand{\dphi}{{{\rm d}\phi}}

\newcommand{\dV}{{{\rm d}V}} 

\newcommand{\dS}{{{\rm d}S}}

\newcommand{\Pzgas}{ P_{z,{\rm gas}}}
\newcommand{\Pgas}{\vec{P}_{\rm gas}}

\newcommand{\Pns}{\vec{P}_{\rm ns}}

\newcommand{\Pnu}{\vec{P}_{\rm \nu}}

\newcommand{\Pej}{ P_{\rm ej}}

\newcommand{\foe}{{10^{51}\,\mathrm{erg}}}

\newcommand{\eexp}{{E_{\rm exp}}}
\newcommand{\Eexp}{{E_{\rm exp}}}

\newcommand{\eexpn}{{E_{\rm exp,0}}} 
\newcommand{\texp}{{t_{\rm exp}}}
\newcommand{\trec}{{t_{\rm rec}}}

\newcommand{\dotMns}{{\dot M_{\rm ns}}}

\newcommand{\dYecoretone}{\Delta Y_{\rm e,core}}
  
\newcommand{\Mns}{{M_{\rm ns}}}

\newcommand{\dMg}{\Delta M_{\rm gain}}
\newcommand{\Rns}{{R_{\rm ns}}}  
\newcommand{\Rs}{{R_{\rm s}}}

\newcommand{\vzns}{{v_z^{\rm ns}}} 
\newcommand{\vznsnu}{{v_z^{\rm ns,\nu}}} 
\newcommand{\vnsvec}{{\vec{v}_{\rm ns}}} 
\newcommand{\vnsveccorr}{{\vec{v}_{\rm ns,corr}}} 
\newcommand{\vnsvecnu}{{\vec{v}_{\rm ns,\nu}}}

\newcommand{\ans}{{a_{\rm ns}}}
\newcommand{\azns}{{a_z^{\rm ns}}} 
 
\newcommand{\Lib}{{L_{\rm ib}}}
 
\newcommand{\Lfiveh}{{L_{\rm 500}}}

\newcommand{\rib}{{R_{\rm ib}}} 
\newcommand{\Rib}{{R_{\rm ib}}} 
\newcommand{\tib}{{t_{\rm ib}}} 
 
\newcommand{\rob}{{R_{\rm ob}}}

\newcommand{\alphag}{{\alpha_{\rm gas}}}

\newcommand{\dx}{{{\rm d}x}}

\newcommand{\djz}{{{\rm d}j_z}}
\newcommand{\CHone}{{\cal C}_{\rm H}}

\newcommand{\CS}{{\cal C}_{\rm S}}
\newcommand{\CL}{{\cal C}_{\rm L}}

\newcommand{\drhodS}{ \left( \frac{\partial \rho}{\partial S}   \right)_{P,Y_e}}
\newcommand{\drhodYe}{\left( \frac{\partial \rho}{\partial Y_e} \right)_{P,S}}

\def\apjl{ApJ}%
\def\apjs{ApJS}%
\def\ao{Appl.~Opt.}%
\def\aap{A\&A}%

%% file: restab-rw1218c.tex
\begin{tabular}{lrrrrrrrrrrrrr}
\hline
\hline
Model & $\Lib$ & $\Delta E^{\rm tot}_{\nu, \mathrm{core}}$ & $\dYecoretone$ & $\left\langle L_{500} \right\rangle$ & $\Delta E_{500}$ & $\eexp$ & $\texp$ & $\Mns$ & $v_z^{\mathrm{ns}}$ & $v_z^{\mathrm{ns},\nu}$ & $a_z^{\mathrm{ns}}$ & $\alphag$ & $d_{\rm shock}$ \\ 
 & [B/s] & [B] &  & [B/s] & [B] & [B] & [s] & [$\Msol$] & [km/s] & [km/s] & [$\mathrm{km/s}^2$] &  &  \\ 
\hline
W12-c &  29.7 &  71.5 &  0.11 &  68.7 &  57.1 &  0.40 & 0.301 &   1.535 &   44.4 &   54.0 &   85.4 & 0.03 &  0.63 \\ 
W18-c &  44.5 & 107.3 &  0.16 &  79.0 &  61.1 &  1.06 & 0.215 &   1.392 &  640.4 &   -8.5 &  444.4 & 0.21 &  0.08 \\ 
\hline
R12-c &  29.7 &  71.5 &  0.11 &  64.8 &  51.1 &  0.43 & 0.329 &   1.480 &   49.9 &   31.3 &  148.1 & 0.04 & -0.03 \\ 
R18-c &  44.5 & 107.3 &  0.16 &  75.5 &  58.1 &  1.26 & 0.236 &   1.345 &  166.1 &   -3.5 &  116.2 & 0.04 &  0.05 \\ 
\hline
\end{tabular}

%% file: restab-fastcon.tex
\begin{tabular}{lrrrrrrrrrrrrr}
\hline
\hline
Model & $\Lib$ & $\Delta E^{\rm tot}_{\nu, \mathrm{core}}$ & $\dYecoretone$ & $\left\langle L_{500} \right\rangle$ & $\Delta E_{500}$ & $\eexp$ & $\texp$ & $\Mns$ & $v_z^{\mathrm{ns}}$ & $v_z^{\mathrm{ns},\nu}$ & $a_z^{\mathrm{ns}}$ & $\alphag$ & $d_{\rm shock}$ \\ 
 & [B/s] & [B] &  & [B/s] & [B] & [B] & [s] & [$\Msol$] & [km/s] & [km/s] & [$\mathrm{km/s}^2$] &  &  \\ 
\hline
W12-c &  29.7 &  71.5 &  0.11 &  68.7 &  57.1 &  0.40 & 0.301 &   1.535 &   44.4 &   54.0 &   85.4 & 0.03 &  0.63 \\ 
W12F-c &  29.7 & 101.2 &  0.08 & 110.9 &  62.0 &  0.94 & 0.118 &   1.411 &  611.7 &   -1.9 &  580.6 & 0.21 &  0.31 \\ 
\hline
\end{tabular}

%% file: restab-b.tex
\begin{tabular}{lrrrrrrrrrrrrr}
\hline
\hline
Model & $\Lib$ & $\Delta E^{\rm tot}_{\nu, \mathrm{core}}$ & $\dYecoretone$ & $\left\langle L_{500} \right\rangle$ & $\Delta E_{500}$ & $\eexp$ & $\texp$ & $\Mns$ & $v_z^{\mathrm{ns}}$ & $v_z^{\mathrm{ns},\nu}$ & $a_z^{\mathrm{ns}}$ & $\alphag$ & $d_{\rm shock}$ \\ 
 & [B/s] & [B] &  & [B/s] & [B] & [B] & [s] & [$\Msol$] & [km/s] & [km/s] & [$\mathrm{km/s}^2$] &  &  \\ 
\hline
B10 &  24.7 &  59.6 &  0.09 &  57.1 &  45.9 &  0.19 & 0.294 &   1.426 & -164.1 &   44.4 & -180.2 & 0.24 &  0.67 \\ 
B11 &  27.2 &  65.5 &  0.10 &  58.8 &  46.3 &  0.27 & 0.280 &   1.401 &  -23.6 &    0.7 & -248.9 & 0.03 &  0.97 \\ 
B12 &  29.7 &  71.5 &  0.11 &  60.6 &  48.7 &  0.37 & 0.220 &   1.399 & -389.5 &   45.0 & -372.4 & 0.32 &  0.06 \\ 
B12-1 &  29.7 &  71.5 &  0.11 &  60.5 &  47.5 &  0.33 & 0.228 &   1.377 &   72.8 &   -4.7 &   47.9 & 0.07 &  0.22 \\ 
B12-2 &  29.7 &  71.5 &  0.11 &  60.9 &  48.5 &  0.39 & 0.212 &   1.391 &   85.8 &    9.7 &  345.7 & 0.07 &  0.82 \\ 
B12-3 &  29.7 &  71.5 &  0.11 &  60.9 &  46.5 &  0.38 & 0.207 &   1.369 &  242.0 &    2.0 &  464.3 & 0.18 &  0.97 \\ 
B12-4 &  29.7 &  71.5 &  0.11 &  61.1 &  47.7 &  0.35 & 0.216 &   1.385 & -115.1 &   20.4 & -154.2 & 0.10 &  0.51 \\ 
B12-5 &  29.7 &  71.5 &  0.11 &  61.0 &  47.8 &  0.33 & 0.211 &   1.387 & -206.9 &   11.6 & -483.1 & 0.19 &  0.52 \\ 
B13 &  32.2 &  77.5 &  0.12 &  62.4 &  49.6 &  0.45 & 0.188 &   1.378 & -355.3 &   32.0 & -408.0 & 0.25 &  0.36 \\ 
B14 &  34.6 &  83.4 &  0.13 &  63.6 &  49.6 &  0.51 & 0.198 &   1.345 & -128.0 &  -11.2 &  -66.7 & 0.07 &  0.40 \\ 
B15 &  37.1 &  89.4 &  0.14 &  65.3 &  50.3 &  0.65 & 0.162 &   1.318 &   36.1 &   -1.0 &   36.0 & 0.02 &  0.27 \\ 
B16 &  39.6 &  95.3 &  0.15 &  66.3 &  51.8 &  0.81 & 0.160 &   1.305 & -214.6 &   -2.6 & -334.4 & 0.08 &  0.57 \\ 
B17 &  42.1 & 101.3 &  0.15 &  67.6 &  53.3 &  0.95 & 0.146 &   1.289 &  -25.5 &   14.8 & -102.6 & 0.01 &  0.05 \\ 
B17-1 &  42.1 & 101.3 &  0.15 &  67.8 &  53.4 &  0.92 & 0.160 &   1.290 & -354.0 &    5.6 & -202.2 & 0.12 &  0.31 \\ 
B18 &  44.5 & 107.3 &  0.16 &  68.3 &  54.8 &  1.16 & 0.152 &   1.275 &  515.3 &    5.2 &  290.5 & 0.15 &  0.42 \\ 
B18-1 &  44.5 & 107.3 &  0.16 &  68.4 &  54.7 &  1.12 & 0.154 &   1.274 & -126.5 &   -0.8 &  -49.1 & 0.04 &  0.20 \\ 
B18-2 &  44.5 & 107.3 &  0.16 &  68.9 &  54.7 &  1.14 & 0.152 &   1.268 &   82.5 &   -5.2 &   16.5 & 0.02 &  0.07 \\ 
B18-3 &  44.5 & 107.3 &  0.16 &  68.8 &  57.1 &  1.15 & 0.142 &   1.305 &  798.8 &  -41.2 &  552.1 & 0.24 & -0.06 \\ 
B18-4 &  44.5 & 107.3 &  0.16 &  68.2 &  54.6 &  1.14 & 0.150 &   1.272 & -171.6 &    4.0 &   65.7 & 0.05 &  0.46 \\ 
B18-5 &  44.5 & 107.3 &  0.16 &  68.5 &  55.2 &  1.09 & 0.164 &   1.280 & -121.8 &   -0.9 &   15.4 & 0.04 & -0.02 \\ 
B18-6 &  44.5 & 107.3 &  0.16 &  68.7 &  55.4 &  1.11 & 0.160 &   1.283 &  502.1 &  -20.6 &  220.0 & 0.15 & -0.06 \\ 
B18-g1 &  44.5 & 107.3 &  0.16 &  68.7 &  54.5 &  1.12 & 0.142 &   1.269 &  -60.3 &    3.9 &  -55.4 & 0.02 &  0.06 \\ 
B18-g2 &  44.5 & 107.3 &  0.16 &  68.7 &  54.8 &  1.12 & 0.138 &   1.273 &  267.9 &   -8.1 &  126.7 & 0.08 &  0.28 \\ 
B18-g3 &  44.5 & 107.3 &  0.16 &  68.5 &  54.9 &  1.10 & 0.150 &   1.274 &   -7.4 &   -3.5 &    0.9 & 0.00 &  0.02 \\ 
B18-g4 &  44.5 & 107.3 &  0.16 &  68.7 &  54.5 &  1.16 & 0.132 &   1.270 & -416.8 &    1.7 & -150.9 & 0.11 &  0.37 \\ 
B19-g1 &  47.0 & 113.2 &  0.17 &  69.6 &  55.9 &  1.31 & 0.148 &   1.253 & -273.8 &    0.3 &  -96.7 & 0.07 &  0.41 \\ 
B19-g2 &  47.0 & 113.2 &  0.17 &  69.5 &  56.0 &  1.33 & 0.148 &   1.255 &  188.5 &    6.4 &   48.8 & 0.05 &  0.15 \\ 
B19-g3 &  47.0 & 113.2 &  0.17 &  70.0 &  56.6 &  1.26 & 0.132 &   1.263 &  366.6 &    1.1 &  183.7 & 0.10 &  0.13 \\ 
B19-g4 &  47.0 & 113.2 &  0.17 &  70.0 &  56.8 &  1.33 & 0.130 &   1.267 &  477.1 &  -18.3 &  195.6 & 0.12 & -0.02 \\ 
B20 &  49.5 & 119.2 &  0.18 &  71.0 &  57.3 &  1.49 & 0.128 &   1.238 &  133.2 &    5.6 &   52.6 & 0.03 &  0.40 \\ 
B21 &  51.9 & 125.1 &  0.19 &  72.1 &  58.5 &  1.72 & 0.122 &   1.222 &   30.6 &   -0.9 &  -20.2 & 0.01 &  0.24 \\ 
\hline
\end{tabular}

%% file: restab-l.tex
\begin{tabular}{lrrrrrrrrrrrrr}
\hline
\hline
Model & $\Lib$ & $\Delta E^{\rm tot}_{\nu, \mathrm{core}}$ & $\dYecoretone$ & $\left\langle L_{500} \right\rangle$ & $\Delta E_{500}$ & $\eexp$ & $\texp$ & $\Mns$ & $v_z^{\mathrm{ns}}$ & $v_z^{\mathrm{ns},\nu}$ & $a_z^{\mathrm{ns}}$ & $\alphag$ & $d_{\rm shock}$ \\ 
 & [B/s] & [B] &  & [B/s] & [B] & [B] & [s] & [$\Msol$] & [km/s] & [km/s] & [$\mathrm{km/s}^2$] &  &  \\ 
\hline
L12 &  42.4 &  94.6 &  0.13 &  90.7 &  70.7 &  0.51 & 0.321 &   1.677 &  278.5 &  -12.9 &  334.3 & 0.24 &  0.11 \\ 
L13 &  45.9 & 102.5 &  0.14 &  91.7 &  69.2 &  0.68 & 0.268 &   1.620 &  -92.6 &   -5.9 & -333.6 & 0.05 &  0.77 \\ 
L14 &  49.5 & 110.4 &  0.15 &  94.6 &  72.8 &  0.81 & 0.280 &   1.628 &  482.1 &  -22.0 &  297.1 & 0.26 &  0.31 \\ 
L15 &  53.0 & 118.3 &  0.17 &  96.2 &  75.2 &  1.02 & 0.266 &   1.617 & -239.5 &   -3.9 & -378.5 & 0.10 &  0.63 \\ 
L16 &  56.5 & 126.2 &  0.18 &  97.8 &  76.3 &  1.07 & 0.256 &   1.586 & -437.9 &   12.8 & -715.2 & 0.17 &  0.47 \\ 
L17 &  60.1 & 134.0 &  0.19 & 100.3 &  77.4 &  1.19 & 0.256 &   1.558 &  -24.7 &    5.5 &  -47.6 & 0.01 &  0.37 \\ 
\hline
\end{tabular}

%% file: restab-w.tex
\begin{tabular}{lrrrrrrrrrrrrr}
\hline
\hline
Model & $\Lib$ & $\Delta E^{\rm tot}_{\nu, \mathrm{core}}$ & $\dYecoretone$ & $\left\langle L_{500} \right\rangle$ & $\Delta E_{500}$ & $\eexp$ & $\texp$ & $\Mns$ & $v_z^{\mathrm{ns}}$ & $v_z^{\mathrm{ns},\nu}$ & $a_z^{\mathrm{ns}}$ & $\alphag$ & $d_{\rm shock}$ \\ 
 & [B/s] & [B] &  & [B/s] & [B] & [B] & [s] & [$\Msol$] & [km/s] & [km/s] & [$\mathrm{km/s}^2$] &  &  \\ 
\hline
W10 &  24.7 &  59.6 &  0.09 &  64.3 &  55.4 &  0.21 & 0.420 &   1.568 & -129.8 &   42.1 & -443.1 & 0.15 &  0.81 \\ 
W12 &  29.7 &  71.5 &  0.11 &  69.0 &  53.9 &  0.31 & 0.322 &   1.501 &  -97.7 &   -9.7 & -132.5 & 0.10 &  0.61 \\ 
W12-1 &  29.7 &  71.5 &  0.11 &  68.0 &  59.5 &  0.32 & 0.374 &   1.563 & -363.8 &   81.2 & -377.0 & 0.32 &  0.13 \\ 
W14 &  34.6 &  83.4 &  0.13 &  72.9 &  56.6 &  0.46 & 0.250 &   1.473 &  -62.0 &   -1.5 &   66.1 & 0.04 &  0.37 \\ 
W16 &  39.6 &  95.3 &  0.15 &  76.0 &  58.5 &  0.67 & 0.244 &   1.430 &  287.2 &   -5.5 &  464.2 & 0.14 &  0.68 \\ 
W18 &  44.5 & 107.3 &  0.16 &  79.3 &  61.5 &  0.89 & 0.226 &   1.401 & -283.6 &    4.2 & -290.1 & 0.11 &  0.44 \\ 
W20 &  49.5 & 119.2 &  0.18 &  82.0 &  63.5 &  1.36 & 0.216 &   1.354 & -377.3 &    0.6 & -277.0 & 0.10 &  0.39 \\ 
\hline
\end{tabular}

%% file: restab-r.tex
\begin{tabular}{lrrrrrrrrrrrrr}
\hline
\hline
Model & $\Lib$ & $\Delta E^{\rm tot}_{\nu, \mathrm{core}}$ & $\dYecoretone$ & $\left\langle L_{500} \right\rangle$ & $\Delta E_{500}$ & $\eexp$ & $\texp$ & $\Mns$ & $v_z^{\mathrm{ns}}$ & $v_z^{\mathrm{ns},\nu}$ & $a_z^{\mathrm{ns}}$ & $\alphag$ & $d_{\rm shock}$ \\ 
 & [B/s] & [B] &  & [B/s] & [B] & [B] & [s] & [$\Msol$] & [km/s] & [km/s] & [$\mathrm{km/s}^2$] &  &  \\ 
\hline
R10 &  24.7 &  59.6 &  0.09 &  59.9 &  48.8 &  0.25 & 0.418 &   1.521 &  -15.4 &  -14.3 & -118.7 & 0.02 & -0.02 \\ 
R12 &  29.7 &  71.5 &  0.11 &  64.6 &  49.9 &  0.50 & 0.316 &   1.461 & -235.8 &   17.5 & -203.4 & 0.16 &  0.15 \\ 
R14 &  34.6 &  83.4 &  0.13 &  69.2 &  52.4 &  0.69 & 0.264 &   1.420 &   88.4 &   14.6 &   86.9 & 0.04 &  0.15 \\ 
R16 &  39.6 &  95.3 &  0.14 &  71.9 &  56.0 &  0.98 & 0.256 &   1.396 &  321.2 &   -8.9 &  210.1 & 0.11 &  0.06 \\ 
R18 &  44.5 & 107.3 &  0.16 &  75.8 &  58.3 &  1.24 & 0.232 &   1.349 &   -4.8 &   -3.7 &  -26.7 & 0.00 & -0.07 \\ 
R18-g &  44.5 & 107.3 &  0.16 &  75.8 &  58.5 &  1.23 & 0.226 &   1.352 & -113.9 &    2.1 & -188.1 & 0.03 &  0.07 \\ 
R20 &  49.5 & 119.2 &  0.18 &  78.8 &  60.9 &  1.64 & 0.214 &   1.309 &  280.1 &    0.8 &  123.9 & 0.06 &  0.14 \\ 
\hline
\end{tabular}

%% file: restab-movns.tex
\begin{tabular}{lrrrrrrrrrrrrr}
\hline
\hline
Model & $\Lib$ & $\Delta E^{\rm tot}_{\nu, \mathrm{core}}$ & $\dYecoretone$ & $\left\langle L_{500} \right\rangle$ & $\Delta E_{500}$ & $\eexp$ & $\texp$ & $\Mns$ & $v_z^{\mathrm{ns}}$ & $v_z^{\mathrm{ns},\nu}$ & $a_z^{\mathrm{ns}}$ & $\alphag$ & $d_{\rm shock}$ \\ 
 & [B/s] & [B] &  & [B/s] & [B] & [B] & [s] & [$\Msol$] & [km/s] & [km/s] & [$\mathrm{km/s}^2$] &  &  \\ 
\hline
B12-m1 &  29.7 &  71.5 &  0.11 &  60.9 &  47.4 &  0.36 & 0.226 &   1.384 &  -56.8 &   -1.7 & -208.2 & 0.06 &  0.48 \\ 
B12-m2 &  29.7 &  71.5 &  0.11 &  60.9 &  47.7 &  0.31 & 0.222 &   1.385 & -100.0 &   19.1 &  -63.5 & 0.10 &  0.72 \\ 
B12-m3 &  29.7 &  71.5 &  0.11 &  61.2 &  47.8 &  0.38 & 0.210 &   1.388 &  272.6 &  -16.5 &   91.9 & 0.23 &  0.35 \\ 
B12-m4 &  29.7 &  71.5 &  0.11 &  60.9 &  47.0 &  0.35 & 0.209 &   1.378 & -104.3 &   -7.4 & -197.2 & 0.09 &  0.43 \\ 
B12-m5 &  29.7 &  71.5 &  0.11 &  60.8 &  47.9 &  0.35 & 0.219 &   1.389 &  365.6 &  -10.1 &  219.1 & 0.32 &  0.47 \\ 
B12-m6 &  29.7 &  71.5 &  0.11 &  60.7 &  48.4 &  0.36 & 0.229 &   1.395 & -334.1 &   42.4 & -462.9 & 0.30 &  0.26 \\ 
B18-m1 &  44.5 & 107.3 &  0.16 &  68.9 &  54.9 &  1.12 & 0.136 &   1.274 &   43.3 &   -4.8 & -108.8 & 0.02 &  0.12 \\ 
B18-m2 &  44.5 & 107.3 &  0.16 &  68.9 &  54.8 &  1.14 & 0.139 &   1.273 &  -86.8 &   -1.1 &  -31.1 & 0.03 &  0.20 \\ 
B18-m3 &  44.5 & 107.3 &  0.16 &  68.8 &  55.3 &  1.12 & 0.131 &   1.281 &   76.4 &   -8.8 &  -11.4 & 0.03 &  0.39 \\ 
B18-m4 &  44.5 & 107.3 &  0.16 &  68.5 &  54.9 &  1.14 & 0.150 &   1.274 & -118.7 &   14.5 & -156.4 & 0.05 &  0.13 \\ 
B18-m5 &  44.5 & 107.3 &  0.16 &  68.3 &  54.7 &  1.12 & 0.166 &   1.273 & -339.7 &   -4.5 & -152.4 & 0.13 & -0.06 \\ 
B18-m6 &  44.5 & 107.3 &  0.16 &  68.6 &  55.4 &  1.12 & 0.166 &   1.283 & -439.3 &   14.0 & -194.5 & 0.17 &  0.04 \\ 
B18-m7 &  44.5 & 107.3 &  0.16 &  68.8 &  54.7 &  1.12 & 0.138 &   1.272 &  109.2 &    8.6 &    2.1 & 0.04 &  0.38 \\ 
B18-m8 &  44.5 & 107.3 &  0.16 &  69.3 &  54.5 &  1.13 & 0.134 &   1.269 &  455.0 &   -4.1 &  187.4 & 0.17 &  0.05 \\ 
\hline
\end{tabular}